\documentclass[journal]{IEEEtran}
\usepackage{cite}
\usepackage{xcolor}
\usepackage{graphics}
\usepackage{amssymb}
\usepackage{makecell}
\usepackage{graphicx}
\usepackage{caption}
\usepackage{graphicx}
\usepackage{float} 
\usepackage{cuted}
\usepackage{fancybox}
\usepackage{bm}
\usepackage{amsmath}
\usepackage{subfigure}
\usepackage{algorithm}
\usepackage{algorithmic}
\usepackage{float}
\usepackage{multirow}
\usepackage{hyperref}
\hypersetup{
colorlinks=true,
linkcolor=blue,
citecolor=blue
}
\usepackage{booktabs}
\newcommand{\tabitem}{~~\llap{\textbullet}~~}
\begin{document}
	\title{A Tutorial on MIMO-OFDM ISAC:\\ From Far-Field to Near-Field}
	
	\author{\IEEEauthorblockN{Qianglong~Dai, Yong~Zeng,~\IEEEmembership{Fellow, IEEE}, Huizhi~Wang, Changsheng~You,~\IEEEmembership{Member, IEEE}, Chao~Zhou, Hongqiang~Cheng, Xiaoli~Xu,~\IEEEmembership{Member, IEEE}, Shi Jin,~\IEEEmembership{Fellow, IEEE},  A. Lee Swindlehurst,~\IEEEmembership{Life Fellow, IEEE}, \\ Yonina C. Eldar,~\IEEEmembership{Fellow, IEEE}, Robert Schober,~\IEEEmembership{Fellow, IEEE}, Rui Zhang,~\IEEEmembership{Fellow, IEEE} \\ and Xiaohu~You,~\IEEEmembership{Fellow, IEEE}}
   \thanks{This work was supported by the Natural Science Foundation for Distinguished Young Scholars of Jiangsu Province with grant BK20240070, by the Fundamental Research Funds for the Central Universities under grants 2242022k60004 and 3204002004A2, the ``Program for Innovative Talents and Entrepreneur in Jiangsu'' under grant 1104000402, and also by the Outstanding Projects of Overseas Returned Scholars of Nanjing under grant 1104000396. The work of A. Lee Swindlehurst was supported by the U.S. National Science Foundation under grant CCF-2322191. The work of Robert Schober was supported by the Federal Ministry for Research, Technology and Space (BMFTR) in Germany in the program of ``Souverän. Digital. Vernetzt.'' joint project xG-RIC (Project-ID 16KIS2432) and the Deutsche Forschungsgemeinschaft (DFG, German Research Foundation) under project SFB 1483 (Project-ID 442419336, EmpkinS).  \textit{(Corresponding author: Yong Zeng.)}}
    \thanks{Qianglong~Dai, Yong~Zeng, Xiaoli~Xu, Shi~Jin, and Xiaohu~You are with the National Mobile Communications Research Laboratory, Southeast University, Nanjing 210096, China. Yong~Zeng and Xiaohu~You are also with the Purple Mountain Laboratories, Nanjing 211111, China (e-mail: \{qldai, yong\_zeng, xiaolixu, shijin, xhyu\}@seu.edu.cn).}
    \thanks{Huizhi~Wang was with the National Mobile Communications Research Laboratory, Southeast University, Nanjing 210096, China. She is now with the School of Electrical and Electronic Engineering, Nanyang Technological University, 639798, Singapore (E-mail: HUIZHI001@e.ntu.edu.sg).}
    \thanks{Changsheng~You, Chao~Zhou, and Hongqiang~Cheng are with the Department of Electronic and Electrical Engineering, Southern University of Science and Technology, Shenzhen 518055, China (e-mail: youcs@sustech.edu.cn, \{zhouchao2024, chenghq2023\}@mail.sustech.edu.cn).}
    \thanks{A. Lee Swindlehurst is with the Department of Electrical Engineering \& Computer Science, University of California, Irvine, CA 92697, USA (e-mail: swindle@uci.edu).}
    \thanks{Yonina C. Eldar is with the Faculty of Electrical and Computer Engineering, Northeastern University, MA 02115, USA, and also with the Faculty of Mathematics and Computer Science, Weizmann Institute of Science, Rehovot 7610001, Israel (e-mail: yonina.eldar@weizmann.ac.il).}
    \thanks{Robert Schober is with the Institute for Digital Communications, FriedrichAlexander-University Erlangen-Nurnberg (FAU), 91054 Erlangen, Germany (e-mail: robert.schober@fau.de).}
    \thanks{Rui Zhang is with the Department of Electrical and Computer Engineering, National University of Singapore, Singapore 117583 (e-mail: elezhang@nus.edu.sg).}
    }

	\maketitle

\begin{abstract}
    Integrated sensing and communication (ISAC) is one of the key usage scenarios for future sixth-generation (6G) mobile communication networks, where communication and sensing (C\&S) services are simultaneously provided through shared wireless spectrum, signal processing modules, hardware, and network infrastructure. Such an integration is strengthened by the technology trends in 6G, such as denser network nodes, larger antenna arrays, wider bandwidths, higher frequency bands, and more efficient utilization of spectrum and hardware resources, which incentivize and empower enhanced sensing capabilities. Moreover, emerging applications such as Internet-of-Everything (IoE), autonomous ground and aerial vehicles, virtual reality/augmented reality (VR/AR), and connected intelligence have intensified the demands for both high-quality C\&S services, accelerating the development and implementation of ISAC in wireless networks. As in contemporary communication systems, orthogonal frequency-division multiplexing (OFDM) is expected to be the dominant waveform for ISAC, motivating the need for study of both the potential benefits and challenges of OFDM ISAC. Thus, this paper aims to provide a comprehensive tutorial overview of ISAC systems enabled by large-scale multi-input multi-output (MIMO) and OFDM technologies and discuss their fundamental principles, advantages, and enabling signal processing methods. To this end, a unified MIMO-OFDM ISAC system model is first introduced, followed by four frameworks for estimating parameters across the spatial, delay, and Doppler domains, including parallel one-domain, sequential one-domain, joint two-domain, and joint three-domain parameter estimation. Next, sensing algorithms and performance analysis are presented in detail for far-field scenarios where uniform plane wave (UPW) propagation is valid, followed by extensions to near-field scenarios where uniform spherical wave (USW) characteristics must be considered. Finally, the paper presents open challenges and outlines promising avenues for future research on MIMO-OFDM ISAC. 
\end{abstract}

\section{Introduction}

In June 2023, the International Telecommunication Union Radiocommunication Sector (ITU-R)  agreed on a draft of new recommendations regarding the framework and overall objectives for the future development of international mobile telecommunications (IMT) for 2030 and beyond, where 6G usage scenarios were addressed~\cite{ITU}. 
Among these novel scenarios, \emph{integrated sensing and communication} (ISAC) stands out as a transformative paradigm. In fact, the 3rd Generation Partnership Project (3GPP) Service and System Aspects working group had already initiated a feasibility study on ISAC in 2022~\cite{3GPP_ISAC}.
Generally speaking, ISAC refers to a new paradigm that integrates wireless communication and sensing (C\&S) capabilities into a unified system, by efficiently sharing wireless spectrum, signal processing modules, hardware, and network infrastructure, with potential joint design of waveforms and signal processing techniques~\cite{xiao2022waveform,liu2022survey,wei2023integrated,liu2022integrated,zhang2021enabling}.
The deep integration of C\&S functionalities facilitates concurrent transmission of communication data and sensing of environmental information, including both static background and moving targets, thus enhancing spectrum/energy/cost efficiency and achieving appealing cooperation gains~\cite{cui2023integrated,zhang2021enabling,yang2023multi}. Leveraging these advantages, ISAC is expected to find a wide range of applications such as high-accuracy localization and tracking~\cite{b38},  target classification/imaging, environmental monitoring, augmented human senses, autonomous vehicles \cite{autov}, and posture/gesture recognition~\cite{tong20226g}. 

ISAC originated from radar-communication coexistence (RCC) studies \cite{RCC}, which initially focused on interference mitigation for collocated radar and communication systems. With the evolution of wireless communication and digital radar technologies, such as multi-input multi-output (MIMO) \cite{paulraj1994increasing,fishler2004mimo} and orthogonal frequency division multiplexing (OFDM) \cite{b82}, the gap between C\&S has been gradually bridged. This technological convergence drove the emergence of dual functional radar communication (DFRC) systems \cite{hassanien2019dual,mealey1963method} that achieve higher spectral efficiency (SE) and hardware integration.
To enable the coexistence of radar and communication systems within the same frequency band, a portion of the sub-6 GHz bands was released for radar and communication~\cite{Shared_SA}, which promoted the evolution of DFRC towards ISAC. 
Afterwards, extensive research efforts have been devoted to optimizing ISAC performance by virtue of information theory~\cite{xiong2023fundamental,wei2023waveform,b8}, waveform design~\cite{dd5,dong2024communication,zhang2023integrated,dd9,dd10,AFDM,dd12}, signal processing techniques~\cite{lu2024random,chepuri2023integrated}, experiments and prototyping~\cite{zhang2024prototyping,zhang2024prototyping2}.

There are three typical design paradigms for ISAC, namely, communication-centric design, sensing-centric design, and C\&S co-design. In this paper, we focus on communication-centric designs that exploit communication waveforms to achieve sensing functions.
Leveraging globally and densely deployed wireless infrastructure and networks, communication-centric ISAC has emerged as one of the most promising approaches for 6G systems, providing significant advantages over separate C\&S designs, as follows.

\begin{itemize}
    \item \textbf{Efficient spectrum utilization:} The massive number of Internet-of-Things (IoT) devices and soaring demands for high-speed data services have imposed an unprecedented strain on the limited spectrum resources available for C\&S~\cite{feng2020joint}. In conventional DFRC systems, C\&S are performed in non-overlapping frequency bands, resulting in fragmented and poor utilization of spectrum resources. Communication-centric ISAC provides an efficient means to improve spectrum efficiency, since the spectrum pool can be flexibly allocated and even completely reused by C\&S tasks, as will be discussed in \mbox{Section~\ref{Sec:1-waveforms}}.

    \item \textbf{Pervasive sensing via wireless communication networks:} Pervasive sensing is an important ISAC vision for 6G that cannot be achieved by traditional radar sensing networks with limited coverage. This issue can be resolved by leveraging existing wireless infrastructure, including cellular base stations (BS) and backhaul links that are densely deployed. Working cooperatively, a collection of wireless networks can form a powerful and pervasive sensor network with ubiquitous coverage and connectivity.

    \item \textbf{Enhanced cost/energy efficiency:} Enhancing energy efficiency (EE) and cost efficiency (CE) is an important goal for 6G and ISAC. Pervasive sensing that relies on conventional standalone radars would require the deployment of a large amount of radar equipment, thus degrading EE/CE. Instead, exploiting existing communication systems to perform wireless sensing can significantly reduce the hardware cost and energy consumption for sensing, bringing great value to 6G.
\end{itemize}

 Among other communication-centric designs, OFDM ISAC has received significant attention from both academia and industry due to its effective C\&S performance. OFDM is the most popular waveform in contemporary communication systems, thanks to the following advantages.

 \begin{itemize}
     \item \textbf{High SE:} OFDM systems employ inverse fast Fourier transform (IFFT)/ fast Fourier transform (FFT) operations to achieve frequency domain transformations, dividing the channel bandwidth into multiple orthogonal subcarriers for parallel data transmission, thus enabling high SE.

     \item \textbf{Robustness against Inter-Symbol Interference (ISI):} The use of a cyclic prefix (CP) in OFDM efficiently mitigates the impact of multipath-induced ISI.

     \item \textbf{Easy equalization:} Since OFDM equalization is performed in the frequency domain, the receiver only needs to perform one-tap equalization on each subcarrier, making it much simpler and more efficient than time-domain equalization in single-carrier systems.

     \item \textbf{Flexible time-frequency resource allocation:} The number and spacing of subcarriers, together with the power and bit loading on each subcarrier can be dynamically adjusted based on available channel state information (CSI), thereby maximizing network throughput and minimizing error rates while ensuring desired performance for individual users.
 \end{itemize}
 
For target sensing, OFDM has also been considered to be a promising waveform with the following advantages.

\begin{itemize}
    \item \textbf{Array-manifold-like structure in the subcarrier and symbol domains:} After removing random information-bearing symbols, the remaining OFDM signal exhibits an array-manifold-like structure in both the subcarrier and symbol domains, enabling efficient estimation of the delay and Doppler frequency for the sensing targets, as will be shown in detail in \mbox{Section \ref{far-doa}}.

    \item \textbf{Efficient decoupled estimation of delay and Doppler:} The phase variations in OFDM channels induced by delay and Doppler in the subcarrier and symbol domains are independent. This allows for efficient decoupled estimation of delay and Doppler, where the data in the symbol domain can be used as snapshots for delay estimation and vice versa.

    \item \textbf{``Thumbtack-shaped" ambiguity function:} In both the subcarrier and symbol domains, the ambiguity function of OFDM signals with the information-bearing symbols removed exhibits a thumbtack shape, endowing OFDM waveforms with high delay and Doppler resolution, especially when the channel bandwidth is large and/or the coherent processing interval (CPI) is long.
\end{itemize}

\begin{figure*}[!t]
	\centering
	\includegraphics[width=0.95\linewidth]{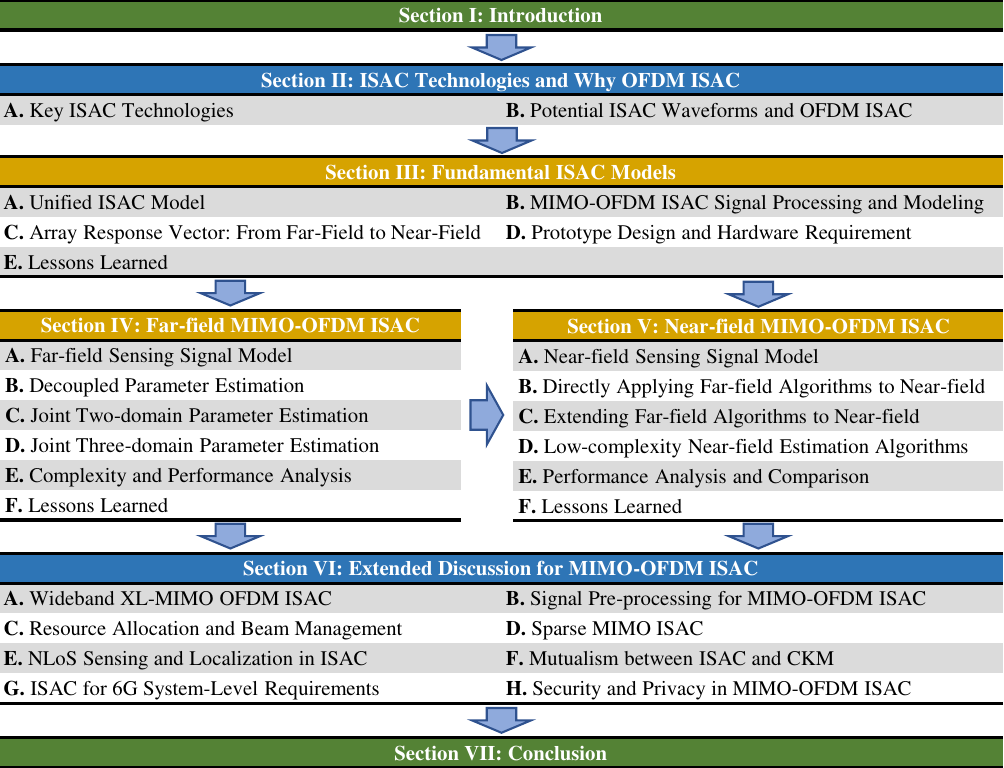}
	\caption{Organization of this paper.}
	\label{Astructure}
    \vspace{-0.4cm}
\end{figure*}

For the reasons listed above, and its compatibility with legacy communication systems, OFDM is expected to be the dominant waveform for ISAC. Numerous studies have demonstrated that sensing can be achieved by existing wireless standards using either OFDM pilot reference signals \cite{dd27,dd28} or the entire OFDM symbol \cite{dd29}, providing a hint of its potential for future 6G ISAC scenarios. However, the sensing capabilities of OFDM require further exploration. Some research has focused on optimizing MIMO-OFDM ISAC transmitter designs, such as waveforms \cite{wei2023integrated,liwave}, interference management \cite{clutter3}, power and time-frequency resource allocation \cite{9945983}, and beamforming \cite{l0}, but further investigation into efficient receiver signal processing techniques is still required. 

To this end, three key challenges must be addressed. First, a unified mathematical model for MIMO-OFDM ISAC is required to reveal the relationship between the system configuration and sensing performance. For instance, increasing the number of subcarriers and symbols can enhance the delay and Doppler resolution of OFDM waveforms, but this also increases the complexity of delay and Doppler estimation. Moreover, joint estimation of target parameters across the spatial, delay, and Doppler domains, significantly escalates signal processing complexity at the receiver, hindering real-time ISAC implementation. Finally, the uniform spherical wave (USW) characteristics in near-field scenarios invalidate traditional far-field algorithms designed based on the conventional uniform plane wave (UPW) model, posing new challenges in computational complexity and estimation performance for extending far-field methods to near-field scenarios. Therefore, a comprehensive study of mathematical models and signal processing techniques for MIMO-OFDM ISAC systems is of paramount importance.

Existing relevant overview papers~\cite{liu2022survey,liu2022integrated,zhang2021enabling,9540344} have primarily focused on the fundamental limits, waveform designs, application scenarios, and challenges of generic ISAC, without focusing on MIMO-OFDM designs. In \cite{wei2023integrated}, the authors considered ISAC signal design and optimization based on OFDM, but signal processing techniques only for far-field scenarios were considered, and joint processing across the time, frequency, and spatial domains was not considered. Unlike prior work, this paper aims to provide an in-depth tutorial overview of MIMO-OFDM ISAC that focuses on signal processing techniques at the receiver, extending from far-field to near-field scenarios. The main contributions of this work are summarized as follows.
 
\begin{itemize}
    \item First, based on the six basic ISAC modes discussed by 3GPP, we present a unified mathematical ISAC model and characterize the relationships among different ISAC transmitter and receiver designs. Based on this unified model, various input-output relationships and corresponding signal processing approaches for MIMO-OFDM ISAC systems are elaborated. Moreover, we propose four frameworks for estimating parameters across the spatial, delay, and Doppler domains, including parallel one-domain, sequential one-domain, joint two-domain and joint three-domain parameter estimation. Then, a comprehensive comparison of the four frameworks is provided.
    \item Next, we elaborate the four parameter estimation frameworks and corresponding estimation algorithms for far-field scenarios. To this end, the application and performance of various classical radar algorithms are discussed in the context of MIMO-OFDM ISAC, including inverse discrete Fourier transform/discrete Fourier transform (IDFT/DFT)-based methods, subspace-based super-resolution methods, and compressed sensing (CS) methods. Theoretical analyses and numerical results are presented to compare the computational complexity, estimation accuracy, and resolution of these algorithms.
    \item Finally, we present target sensing algorithms for more general near-field scenarios. In particular, we point out the challenges that arise when directly applying far-field based algorithms to near-field sensing problems. Subsequently, we extend the far-field algorithms to near-field sensing and study representative near-field algorithms with low complexity and super resolution. Performance comparisons and analyses of these algorithms are also conducted.
\end{itemize}

\begin{table}[!t]
\footnotesize
\renewcommand{\arraystretch}{1.5}
\centering
\caption{List of main notations.}
\label{notations}
\begin{tabular}{l|l}
\Xhline{1.4pt}
\makecell[c]{Notations} & \makecell[c]{Meanings}                                       \\ \Xhline{1.4pt}
$M_r$                 &   \begin{tabular}[c]{@{}c@{}}Number of antennas at the sensing receiver\end{tabular}           \\ 
$M_t$                 &   \begin{tabular}[c]{@{}c@{}}Number of antennas at the ISAC transmitter\end{tabular}           \\ 
$N$                  & Number of OFDM subcarriers          \\ 
$P$                  & Number of OFDM symbols              \\ 
$K$                  & Number of sensing targets            \\ 
$\mathbf{q}_k(r_k,\theta_k)$          & Location of the $k$-th target       \\ 
$r_k$          & Distance between the $k$-th target and Sen-RX       \\ 
$\theta_k$          & Angle from the $k$-th target to Sen-RX       \\ 
$\tau_k$      & Delay of the $k$-th target       \\ 
$\upsilon_k$      & Doppler frequency shift of the $k$-th target       \\ 
$\mathbf{x}(t)\in \mathbb{C}^{M_t\times1}$   &Signal transmitted by ISAC-TX \\ 
$\mathbf{y}_s(t)\in \mathbb{C}^{M_r\times1}$   & Signal received by Sen-RX \\ 
$\mathbf{Y}_s\in \mathbb{C}^{M_r\times N\times P}$    & Sensing signal after FFT demodulation  \\ 
$\overline{\mathbf{Y}}_s\in \mathbb{C}^{M_r\times N\times P}$ & \begin{tabular}[c]{@{}c@{}}Sensing signal after element-wise division\end{tabular}                   \\ 
\begin{tabular}[c]{@{}c@{}}$\mathbf{a}_{\rm R}(\theta)\in \mathbb{C}^{M_r\times1}$ \end{tabular} & \begin{tabular}[c]{@{}c@{}}Far-field steering vector in the spatial domain \end{tabular} \\ 
\begin{tabular}[c]{@{}c@{}}$\mathbf{a}_{\rm R}(r,\theta)\in \mathbb{C}^{M_r\times1}$ \end{tabular} & \begin{tabular}[c]{@{}c@{}}Near-field steering vector in the spatial domain \end{tabular} \\ 
\begin{tabular}[c]{@{}c@{}}$\mathbf{a}_{\tau}(\tau)\in \mathbb{C}^{N\times1}$  \end{tabular} & \begin{tabular}[c]{@{}c@{}} Steering vector in the delay domain \end{tabular} \\ 
\begin{tabular}[c]{@{}c@{}}$\mathbf{a}_{\upsilon}(\upsilon)\in \mathbb{C}^{P\times1}$ \end{tabular} & \begin{tabular}[c]{@{}c@{}}Steering vector in the Doppler domain\end{tabular} \\ 
\begin{tabular}[c]{@{}c@{}} $\mathbf{A}_{\theta}\in \mathbb{C}^{M_r\times K}$\end{tabular} & \begin{tabular}[c]{@{}c@{}}Far-field steering matrix in the spatial domain \end{tabular} \\ 
\begin{tabular}[c]{@{}c@{}} $\mathbf{A}_{\rm R}(r,\theta)\in \mathbb{C}^{M_r\times K}$\end{tabular} & \begin{tabular}[c]{@{}c@{}}Near-field steering matrix in the spatial domain \end{tabular} \\ 
\begin{tabular}[c]{@{}c@{}} $\mathbf{A}_{\tau}\in \mathbb{C}^{N\times K}$ \end{tabular} & \begin{tabular}[c]{@{}c@{}} Steering matrix in the delay domain \end{tabular} \\ 
\begin{tabular}[c]{@{}c@{}}$\mathbf{A}_{\upsilon}\in \mathbb{C}^{P\times K}$\end{tabular} & \begin{tabular}[c]{@{}c@{}}Steering matrix in the Doppler domain\end{tabular} \\ 

$\mathbf{X}_{\theta}\in \mathbb{C}^{M_r\times NP}$         & Data matrix for AoA estimation                                       \\ 
$\mathbf{X}_{\tau}\in \mathbb{C}^{N\times M_rP}$        & Data matrix for delay estimation                                       \\ 
$\mathbf{X}_{\upsilon}\in \mathbb{C}^{P\times M_rN}$         & Data matrix for Doppler estimation                                     \\ \Xhline{1.4pt}
\end{tabular}
\vspace{-0.5cm}
\end{table}

\begin{table}[t!]
\footnotesize
\renewcommand{\arraystretch}{1.25}
\centering
\caption{List of key acronyms.}
\label{Acronyms}
\begin{tabular}{l|l}
\Xhline{1.4pt}
\makecell[c]{Acronyms} & \makecell[c]{Definitions}                                       \\ \Xhline{1.4pt}
2D    & Two-dimensional \\ 
3D    & Three-dimensional \\ 
3GPP    & 3rd Generation Partnership Project  \\ 
6G    & Sixth-generation  \\ 
AoA     & Angle of arrival \\  
AoD     & Angle of departure \\
AWGN    & Additive white Gaussian noise \\
BS     & Base station \\
C\&S   & Communication and sensing \\
CS   & Compressed sensing \\
CP      & Cyclic prefix \\  
CPI     & Coherent processing interval \\ 
Com-RX    & Communication receiver \\ 
CKM    & Channel knowledge map \\
DFT    & Discrete Fourier transform \\
DFRC    & Dual functional radar communication \\  
ESPRIT  & \begin{tabular}[t]{@{}l@{}}Estimation of signal parameters via \\ rotational invariance technique\end{tabular}   \\ 
FFT     &  Fast Fourier transform\\
ISAC    & Integrated sensing and communication \\  
ISAC-TX    & ISAC transmitter \\ 
ISI     & Inter-symbol interference \\  
ICI     & Inter-carrier interference\\ 
IDFT     & Inverse discrete Fourier transform \\  
IFFT     & Inverse fast Fourier transform\\ 
LoS    &  Line-of-sight\\  
LS   &  Least squares\\  
MIMO    & Multi-input multi-output \\  
MUSIC   & Multiple signal classification \\  
MRC    &  Maximum-ratio combining\\  
MMSE    &  Minimum mean square error\\  
NLoS    &  Non-line-of-sight\\  
NOMA    &  Non-orthogonal multiple access\\  
OFDM    &  Orthogonal frequency division multiplexing\\  
OTFS    &  Orthogonal time frequency space\\ 
OMP     &  Orthogonal matching pursuit\\ 
PAPR    &  Peak-to-average power ratio\\ 
PM    &  Propagator method\\ 
RMSE    &  Root mean square error\\ 
SI      &  Self-interference \\ 
SE      &  Spectral efficiency\\     
Sen-RX  &  Sensing receiver \\ 
SNR     &  Signal-to-noise ratio \\ 
TLS    &  Total least square \\ 
UPW     &  Uniform plane wave\\  
USW     &  Uniform spherical wave\\  
ULA     &  Uniform linear array\\  
UE      &  User equipment\\     
XL-MIMO &  Extremely large-scale multi-input multi-output\\
ZF &  Zero-forcing\\
\Xhline{1.4pt}
\end{tabular}
\vspace{-0.6cm}
\end{table}

As outlined in \mbox{Fig.~\ref{Astructure}}, the rest of the paper is organized as follows. Section \ref{sec2} introduces the Key Performance Indicators (KPIs) and key technologies of ISAC, as well as the reasons for adopting OFDM waveforms. \mbox{Section \ref{sn1}} presents a unified ISAC framework and the corresponding MIMO-OFDM ISAC signal model. \mbox{Section \ref{FF_Sensing}} and \mbox{Section \ref{near-field}} detail parameter estimation methods for far-field and near-field target sensing, respectively. \mbox{Section \ref{openp}} provides directions worthy of further investigation, and \mbox{Section \ref{conclusion}} concludes this paper.

\textit{Notation:} Bold face lower- and upper-case letters denote vectors and matrices, respectively. The notations ${\left(\cdot\right)}^{T}$, ${\left(\cdot\right)}^{*}$, ${\left(\cdot\right)}^{H}$ denote transpose, conjugate, and conjugate transpose operations, respectively. The modulus of a complex scalar is given by $|a|$, $|\mathbf{a}|$ and $|\mathbf{A}|$ denote the element-wise modulus of vector $\mathbf{a}$ and matrix $\mathbf{A}$.  The transformation of vector $\mathbf{a}$ into a diagonal matrix is represented as $\mathrm{diag}\{\mathbf{a}\}$, and $[\mathbf{a}]_m$ denotes the $m$th element of vector $\mathbf{a}$. The operator $\mathrm{vec}\left(\mathbf{A}\right)$ converts matrix $\mathbf{A}$ into a column vector by sequentially stacking its columns. The average of all elements of $\mathbf{A}$ is obtained by $\mathrm{mean}\{\mathbf{A}\}$, and $\mathrm{angle}(a)$ represents the phase of a complex number $a$. The operators ${{\bf{A}}^{ \dagger}}$ and $\rm{det}(\mathbf{A})$ denote the pseudoinverse and determinant of matrix $\mathbf A$, respectively. The symbols $\otimes$ and $\odot$ denote the Kronecker product and Hadamard product, respectively. The rectangle pulse function is denoted as $\mathrm{rect}(t)=1, t\in \left[0,1\right]$. Notations frequently used in this paper are given in \mbox{Table~\ref{notations}}, and \mbox{Table~\ref{Acronyms}} summarizes key acronyms.

\section{ISAC Technologies and Why OFDM ISAC} \label{sec2}

\subsection{Key ISAC Technologies}

\subsubsection{Primary ISAC Tasks and KPIs}
To evaluate the performance of ISAC systems, several KPIs have been specified for C\&S services. The performance of communication tasks can be assessed from two key perspectives: efficiency and reliability. Efficiency is used to assess the amount of information successfully delivered from the transmitter to the receiver given limited resources.
Reliability refers to the ability of the system to withstand adverse communication environments, such as channel fading, interference, and noise. This metric focuses on signal quality, aiming to ensure that information is not corrupted or lost during transmission. On the other hand, sensing tasks can be generally classified into three categories, i.e., target detection, parameter estimation, and object imaging, with different KPIs.
Target detection refers to determining the presence or absence of targets obscured by clutter and interference~\cite{Kay_1998}.
Parameter estimation refers to the problem of extracting relevant parameters from target reflections. 
For unbiased estimators, the Cramér-Rao bound (CRB) characterizes the theoretical lower bound on mean squared error (MSE) between the actual parameters and their estimates~\cite{b8,9945983}. 
The number of sensing degrees of freedom (SDoF) is also an important KPI for parameter estimation, characterizing the maximum number of targets whose parameters can be estimated.
For multi-target sensing, resolution is an essential KPI that refers to the ability of the system to distinguish objects with similar parameter values.
Object imaging is a sensing task that aims to determine the type and shape of sensed objects. Imaging deviation is used as a KPI in this case to measure the accuracy of the imaging process~\cite{10078429}. In addition, image entropy and image contrast can be employed to evaluate the quality of the imaging results~\cite{8731684}.

\subsubsection{ISAC Transmitter}
The ISAC transmitter design involves optimizing both C\&S functions, including waveform design, resource allocation, antenna configuration, and beam management. Communication is typically based on the single-hop propagation between BS and user equipment (UE), whereas sensing is governed by the classical radar equation~\cite{beamforming7}. A dedicated waveform that can achieve a higher degree of integration between C\&S without interference is crucial. One straightforward solution is to schedule C\&S on orthogonal resource blocks, so that they do not interfere with each other~\cite{timedivision,fredivision2,spadivision}. However, orthogonal resource allocation usually suffers from poor SE and EE. Therefore, in order to maximize the integration gain between C\&S, it is more favorable to design a fully unified ISAC waveform.

Multi-antenna technology has evolved from MIMO in fourth-generation (4G) networks to massive MIMO in fifth-generation (5G) networks. For future 6G systems, several emerging technologies may further increase the number of available spatial DoFs, such as extremely large-scale MIMO (XL-MIMO)~\cite{b23,b39}, sparse MIMO~\cite{sparse1,sparse2}, modular MIMO~\cite{b37,b40}, and fluid~\cite{FAS}/movable antennas~\cite{move1,MAprototyping,groupMA,6DMA}. Millimeter-wave (mmWave) and terahertz (THz) frequency bands offer abundant spectral resources, where the large available bandwidth empowers both high-rate communications and high-resolution delay/distance sensing. Furthermore, the shorter wavelength allows for the deployment of more antenna elements within the same physical aperture, thereby achieving higher MIMO spatial multiplexing and beamforming gains. As array apertures increase, the beams of future XL-MIMO and mmWave/THz ISAC systems may be extremely narrow, increasing the challenges of beam management in achieving high gain and avoiding misalignment. In the literature, transmit beamformers are often designed to minimize the discrepancy between the resulting and targeted beampattern for sensing, while minimizing multi-user interference~\cite{beamforming1,beamforming2,beamforming3,beamformgin5}. To further address the complexity and cost incurred by fully digital MIMO, hybrid beamforming has been studied for enhancing ISAC performance in full-duplex systems~\cite{beamforming4,beammana1}. 

\subsubsection{ISAC Channel}
The mathematical models for C\&S channels differ significantly. Communication channel models are typically stochastic, focusing on signal transmission quality, while sensing channels are deterministic, incorporating target scattering. The transmitters and receivers in communication systems are usually assumed to be spatially separated, while the sensing transmitters and receivers can be collocated or separated, corresponding to ``monostatic sensing" or ``bistatic sensing", respectively. While multipath environments can enhance the spatial diversity of communication systems, radar sensing usually requires a line-of-sight (LoS) link for target detection. Channel fading models are also different for C\&S. For instance, large- and small-scale fading are typically used to model the propagation effects in communication channels, while for radar sensing, signal propagation is also characterized in terms of target scattering characteristics, such as the radar cross section (RCS). 

To integrate the acquisition of communication CSI with sensing services, several challenges need to be addressed. First, both monostatic and bistatic sensing need to be consistent with the communication system architecture. Second, the propagation characteristics exhibited by scatterers must be considered in the communication channel. In addition, precise channel models that can accurately characterize user/target mobility are needed for high-mobility applications. Recently, stochastic models and hybrid ISAC channel modeling have been proposed~\cite{channel2,channel4,channel3}. 

With ever-increasing array apertures and the use of mmWave/THz frequencies, the (radiative) near-field region will significantly expand, even up to several hundreds of meters. Thus, it is crucial to consider algorithm development and performance for ISAC systems that operate in near-field scenarios. In the near-field region, the more accurate USW model needs to be considered to characterize both the phase and amplitude variations across array elements~\cite{b39,b23}. The USW models for ISAC enable joint estimation of target angle and distance even with only spatial-domain array signal processing, thus reducing the requirements of high-precision synchronization. 

Furthermore, as the operating frequency increases, ISAC signals become more susceptible to blockages, which can significantly degrade C\&S performance. Fortunately, several techniques have been proposed to address this issue, such as using intelligent reflecting surfaces (IRSs)~\cite{ris2,ris3} and fully passive metal reflectors~\cite{metal} that establish virtual wireless links. Furthermore, one can also exploit prior knowledge of the local environment to enhance C\&S performance. For example, approaches based on a channel knowledge map (CKM) have been proposed to acquire CSI for achieving environment-aware C\&S~\cite{CKM1,CKMclutter}.

\subsubsection{ISAC Receiver}
Advanced signal processing techniques need to be developed for the ISAC receiver to achieve favorable C\&S performance, addressing issues such as transceiver synchronization, interference management, clutter suppression, and target parameter estimation. For monostatic sensing, the receiver is collocated with the ISAC transmitter, and thus it has full knowledge of the transmitted signals. As such, the sensing receiver does not need to perform a dedicated synchronization or demodulation of the communication symbols. However, self-interference occurs due to signal leakage between the transmitter and receiver, which may lead to non-linear distortion and difficulty in sensing weak targets~\cite{SI1,SI2}. For bistatic sensing, the ISAC receiver needs to estimate and compensate for clock bias, and the communication symbols need to be demodulated before performing sensing.

Typical parameters that are estimated include the angle of arrival (AoA), distance, and velocity of targets. Sensing receivers usually estimate the distance and velocity parameters by analyzing the propagation delays and Doppler shifts of the echo signals, respectively. For AoA estimation, multiple antennas are needed to provide spatial resolution. These estimation problems have been studied for many decades. For example, data-based algorithms such as Capon beamforming~\cite{capon} and amplitude and phase estimation (APES)~\cite{apes} can be used to improve the angle estimation accuracy by optimizing the beamforming gain at the receiver. Subspace-based techniques, such as multiple signal classification (MUSIC)~\cite{music} and estimation of signal parameters via rotational invariance techniques (ESPRIT)~\cite{esprit}, exploit the orthogonality between the signal subspace and the noise subspace to achieve super-resolution angle estimation. Compared to subspace-based algorithms, CS and sparse recovery methods exploit channel sparsity to reconstruct the transmitted signals and extract target parameters. However, sparse recovery inherently involves an NP-hard $\ell_0$-norm optimization, which is typically approximated using greedy algorithms like orthogonal matching pursuit (OMP)~\cite{OMP3}. ISAC techniques that use super-resolution sensing algorithms fall under the heading integrated super-resolution sensing and communication ($\rm{IS^2AC}$) \cite{zhang2023integrated,jingran_1,jingran_2}, and can provide highly accurate parameter estimates. However, such algorithms are known to be sensitive to array modeling errors, and they are costly to implement for large-scale arrays. These factors must be taken into account when considering them for ISAC scenarios.

\subsection{Potential ISAC Waveforms and OFDM ISAC}\label{Sec:1-waveforms}

To meet the diverse C\&S requirements of 6G ISAC applications, using an appropriate waveform is critical. For high communication performance with high data rates and low BER, waveforms should feature high SE, low peak-to-average power ratio (PAPR), robustness to Doppler effects and phase noise, resilience to nonlinear power amplifier noise, low signal processing complexity, and compatibility with legacy MIMO systems. For radar sensing, waveforms should enhance resolution, estimation accuracy, clutter rejection, and interference resistance by exhibiting low PAPR and favorable ambiguity function characteristics, such as narrow mainlobe width, low peak-to-sidelobe ratio (PSLR) or low integrated sidelobe ratio (ISLR). However, achieving all these attributes simultaneously is challenging, resulting in trade-offs between C\&S performance. Based on these trade-offs, one can classify waveform design approaches as either sensing-centric, communication-centric, or joint designs.

\begin{table*}[hbt]
	\renewcommand{\arraystretch}{1.2}
	\centering
	\caption{Comparison of advantages and limitations for various ISAC waveforms.}
	\label{waveform}
	\begin{tabular}{|ll|l|l|}
		\hline
		\multicolumn{2}{|c|}{\textbf{Wavefroms}}                            & \textbf{Pros}                                             & \textbf{Cons}                               \\ \hline
		\multicolumn{1}{|c|}{\multirow{15}{*}{\begin{tabular}[c]{@{}l@{}}Communication-\\ centric\end{tabular}}} & Single-carrier~\cite{dd3,dd4,b82,dd5,xiao2024integrated} & \begin{tabular}[c]{@{}l@{}} \tabitem Low PAPR\\ \tabitem Low energy consumption\end{tabular}                                                                                                        & \begin{tabular}[c]{@{}l@{}}\tabitem Low SE\\ \tabitem Sensitivity to ISI \\ \tabitem Compromised sensing performance\end{tabular}                                                                                      \\ \cline{2-4} 
		\multicolumn{1}{|c|}{}                                       & DDAM~\cite{xiao2024integrated,dd5}          & \begin{tabular}[c]{@{}l@{}} \tabitem Low PAPR \\ \tabitem Low guard interval overhead \\ \tabitem High SE \\ 
			\tabitem Low complexity  \end{tabular} & \begin{tabular}[c]{@{}l@{}} \tabitem Reliance on channel sparsity \end{tabular}                                                     \\ \cline{2-4} 
		\multicolumn{1}{|c|}{}                                       & OFDM~\cite{b82,zhang2023integrated}          & \begin{tabular}[c]{@{}l@{}} \tabitem High SE\\ \tabitem Robust against ISI\\ \tabitem Flexible time-frequency resource allocation \\ \tabitem Array-manifold-like structure in \\  \quad \ subcarrier and symbol domains\\ \tabitem Efficient decoupled estimation \\ \quad \  of delay and Doppler\\ \tabitem ``Thumbtack-shaped" ambiguity function\\ \tabitem  Backward compatibility \end{tabular} & \begin{tabular}[c]{@{}l@{}} \tabitem High PAPR\\ \tabitem Challenging self-interference cancellation \\ \tabitem Sensitivity to high Doppler spread\end{tabular}                                                     \\ \cline{2-4} 
		\multicolumn{1}{|c|}{}                      & OTFS\cite{dd9}/OCDM\cite{dd10}/AFDM\cite{AFDM}  & \begin{tabular}[c]{@{}l@{}} \tabitem Robust against Doppler spread\\ \tabitem High C\&S performance \end{tabular}                                                                     & \begin{tabular}[c]{@{}l@{}} \tabitem Complex signal processing\\ \tabitem High computational complexity\end{tabular}                        \\ \hline
		\multicolumn{1}{|c|}{\multirow{5}{*}{\begin{tabular}[c]{@{}l@{}}Sensing-centric\end{tabular}}}       & Plused \cite{dd11}         & \begin{tabular}[c]{@{}l@{}} \tabitem Simple for implementation \\ \tabitem Wide detection range\end{tabular}                                                                                              & \begin{tabular}[c]{@{}l@{}} \tabitem Low communication rate\\ \tabitem Short-range sensing blind spot \\ \tabitem Long-range ambiguities\end{tabular}                  \\ \cline{2-4} 
		\multicolumn{1}{|c|}{}                                       & Linear frequency modulation ~\cite{dd12}          & \begin{tabular}[c]{@{}l@{}} \tabitem High distance resolution\\ \tabitem Low PAPR\\ \tabitem Robust against Doppler spread\\ \tabitem No short-range sensing blind spot\end{tabular}                                        & \tabitem Low communication rate                                                                                                                                                     \\ \hline
		\multicolumn{1}{|c|}{\multirow{3}{*}{Joint design}}          & Multiplexing~\cite{xiao2022waveform,dd19,dd18,dd20}   & \begin{tabular}[c]{@{}l@{}} \tabitem Balance C\&S requirement\\ \tabitem Simple for implementation \\ \tabitem  Flexible resource allocation \end{tabular}                                                                                 & \begin{tabular}[c]{@{}l@{}} \tabitem Low radio resource utilization\\ \tabitem CDM raises the noise floor\\ \tabitem SDM limits C\&S coverage\end{tabular} \\ \cline{2-4} 
		\multicolumn{1}{|c|}{}                                       & Fusion~\cite{wei2023waveform,dd21,dd22,dd23,dd24,dd25,dd26,dd30}         & \tabitem High C\&S performance                                                                                                                                  & \begin{tabular}[c]{@{}l@{}} \tabitem Implementation challenges\\ \tabitem Complex optimization problems\end{tabular}                                                                          \\ \hline
	\end{tabular}
    \vspace{-0.2cm}
\end{table*}

In this paper, we focus on communication-centric ISAC systems. Communication-centric approaches leverage existing communication waveforms with various modifications to enable or enhance their sensing capability. However, the inherent randomness of communication waveforms typically degrades sensing performance due to issues like high PAPR, random autocorrelation properties, and reduced resolution \cite{b82}. Communication waveforms can be categorized as either single-carrier or multi-carrier.
Single-carrier waveforms feature low PAPR and minimal nonlinear distortion \cite{dd2}, making them suitable for scenarios with limited coverage but high communication quality requirements. Examples include single-carrier frequency division multiple access (SC-FDMA) \cite{dd3} methods, such as DFT-spread-OFDM (DFT-s-OFDM) \cite{b82,dd4}, widely used in the uplink of 4G and 5G systems. The authors of \cite{dd5} and \cite{xiao2024integrated} studied single-carrier ISAC using delay-Doppler alignment modulation (DDAM), which eliminates inter-symbol interference (ISI) through path-based beamforming and delay compensation.
On the other hand, multi-carrier waveforms enable increased communication rates and SE \cite{dd2}. OFDM \cite{b82,zhang2023integrated} is a quintessential example of multi-carrier waveforms and is ubiquitously used in 4G and 5G systems, featuring a ``thumbtack-shaped" ambiguity function that provides an inherent sensing capability. However, OFDM also suffers from inherent drawbacks such as high PAPR and sensitivity to Doppler spread, which result in increased hardware cost and degraded C\&S performance in high-mobility scenarios.

Several emerging multi-carrier waveforms, such as orthogonal time frequency space (OTFS) \cite{dd9}, orthogonal chirp division multiplexing (OCDM) \cite{dd10}, and affine frequency division multiplexing (AFDM) \cite{AFDM}, have been proposed to overcome the limitations of OFDM. Specifically, OTFS maps communication symbols to the delay-Doppler domain and extends them over the entire time-frequency domain via a two-dimensional (2D) transform. This approach fully exploits the sparsity of wireless channels, achieving high diversity gain and demonstrating more robust C\&S performance in high-mobility scenarios. Based on the Fourier orthogonal basis of OFDM, OCDM and AFDM introduce orthogonal chirp and affine chirp-like phase rotations, respectively, which shift signal orthogonality from the frequency domain to the joint time-frequency domain, thereby enhancing robustness against multipath fading and Doppler spread. Compared to OTFS and OCDM, AFDM can flexibly adjust two affine parameters to match time-frequency selective channels, ensuring orthogonality and full diversity gain, thus achieving better C\&S performance. However, these waveforms introduce additional 2D transforms, resulting in more complex transceiver architectures and increased computational complexity.

\mbox{Table~\ref{waveform}} presents the main types of ISAC waveforms along with their advantages and disadvantages. In summary, sensing-centric waveforms are limited by low communication rates. For joint designs, multiplexing methods, such as time division multiplexing (TDM) \cite{xiao2022waveform}, frequency division multiplexing (FDM) \cite{dd19}, code division multiplexing (CDM) \cite{dd18}, and space division multiplexing (SDM) \cite{dd20}, suffer from inefficient radio resource utilization, while fusion methods entail high implementation complexity and/or complex optimization problems. For communication-centric waveforms, emerging technologies like DDAM, OTFS, OCDM, and AFDM exhibit superior sensing performance in high-mobility scenarios but are often accompanied by complex signal processing and high computational demands. 

Consequently, OFDM remains the most likely waveform for future ISAC wireless networks. In the 3GPP proposals \cite{6GR,6GR1}, an agreement was reached that the CP-OFDM waveform as defined in the 5G new radio (NR) is supported as the basis for the 6G radio (6GR) downlink, the CP-OFDM and DFT-s-OFDM waveforms as defined in 5G NR are supported as the basis for the 6GR uplink. For MIMO-OFDM ISAC systems, the use of XL-MIMO and large time-frequency bandwidths significantly enhances the sensing performance, while simultaneously introducing increased parameter estimation complexity. This challenge becomes particularly pronounced when performing joint estimation across the spatial, delay, and Doppler domains, compounded by the additional complexity of near-field parameter estimation. Thus, in this paper, we present a tutorial overview of rigorous and unified models for MIMO-OFDM ISAC, methods and corresponding algorithms for parameter estimation across the three domains, and analyses of their performance and complexity. Furthermore, the extension of far-field methods to near-field scenarios will be elaborated. 

\section{Fundamental ISAC Models} \label{sn1}
\begin{figure}[!t]
  \centering
  \includegraphics[width=0.99\linewidth]{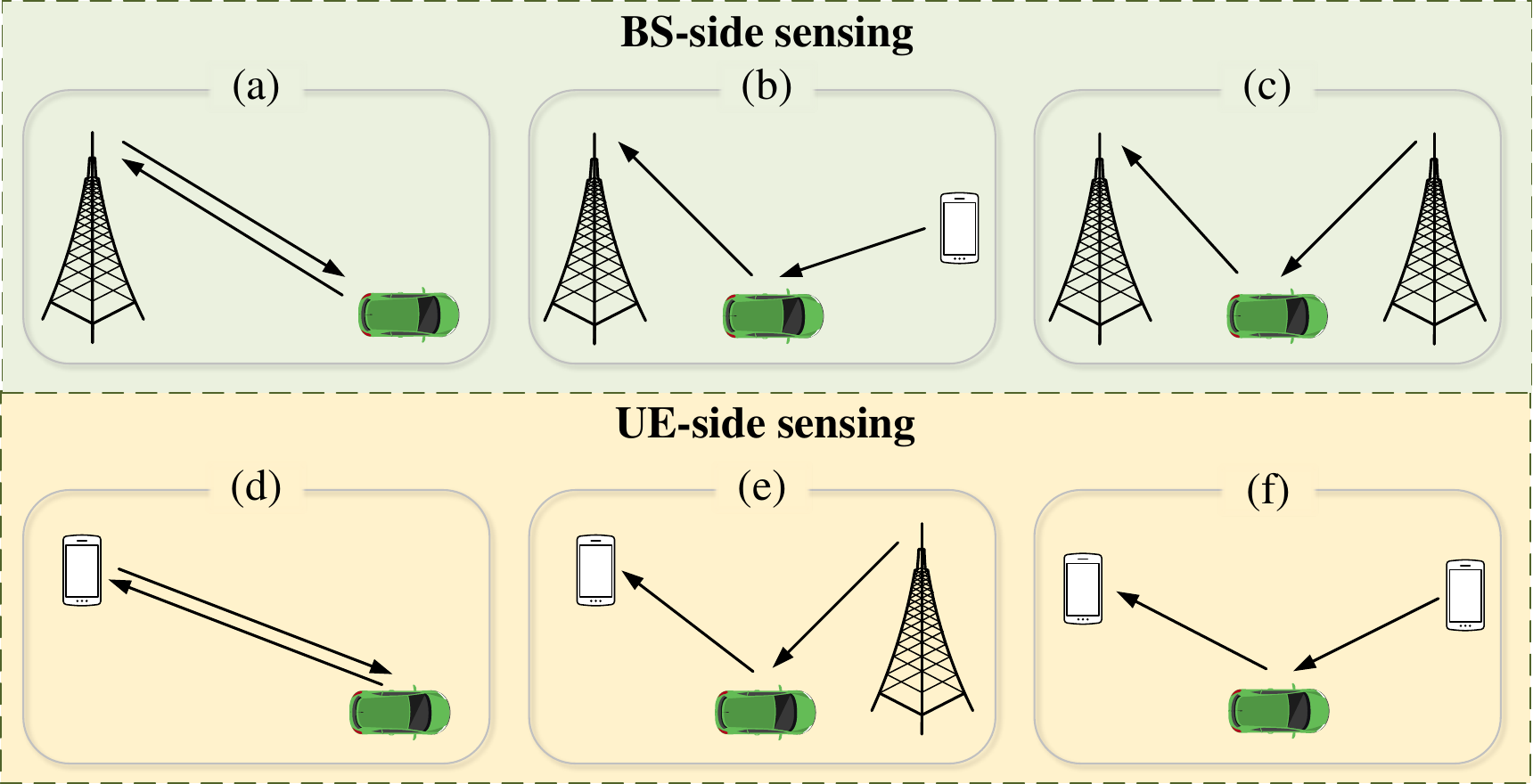}
  \caption{Six basic ISAC modes discussed by 3GPP, with BS-side sensing modes: (a) BS monostatic, (b) UE-to-BS bistatic, (c) BS-to-BS bistatic modes, and UE-side sensing modes: (d) UE monostatic, (e) BS-to-UE bistatic, (f) UE-to-UE bistatic modes.}
  \label{ISAC_modes}
    \vspace{-0.5cm}
\end{figure}
Considering the universal applicability of ISAC frameworks to all possible waveforms, we first introduce various ISAC modes, characterizing the relationships among transmitter and receiver signals and considering the required signal processing. As shown in \mbox{Fig.~\ref{ISAC_modes}}, 3GPP standardization bodies have discussed six basic ISAC modes \cite{dd31}, which can be divided into two categories depending on whether the sensing signal is received and processed at the BS or UE side. For BS-side sensing, as shown in \mbox{Fig.~\ref{ISAC_modes}} (a)-(c), ISAC signals could be sent by either the BS itself, a UE, or another BS, corresponding to the BS monostatic, UE-to-BS bistatic, and BS-to-BS bistatic modes, respectively. Similarly, for UE-side sensing, as shown in \mbox{Fig.~\ref{ISAC_modes}} (d)-(f), ISAC signals could be sent by either the UE itself, a BS, or another UE, corresponding to the UE monostatic, BS-to-UE bistatic, and UE-to-UE bistatic modes, respectively. In the following, we present the signal input-output relationships for C\&S based on a unified ISAC model that is applicable to all six modes above.
\subsection{Unified ISAC Model}
\begin{figure}[!t]
	\centering
	\includegraphics[width=0.9\linewidth]{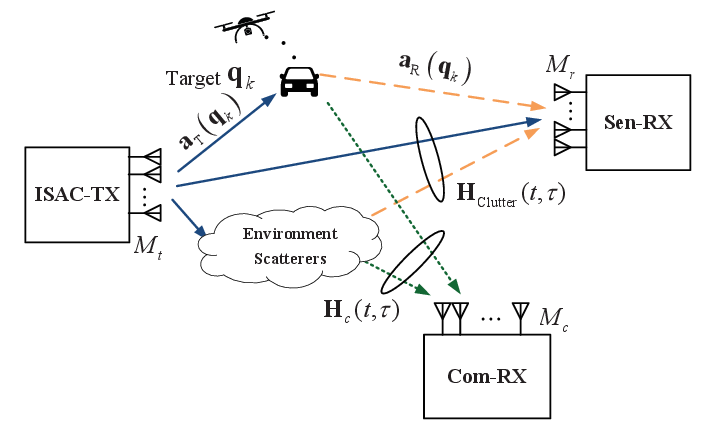}
	\caption{A generic ISAC setup that includes all the six modes shown in \mbox{Fig.~\ref{ISAC_modes}} as special cases.}
	\label{general model}
      \vspace{-0.6cm}
\end{figure}

\mbox{Fig.~\ref{general model}} shows a generic ISAC setup that includes the aforementioned six modes as special cases. It consists of three components, namely an ISAC transmitter (ISAC-TX), a communication receiver (Com-RX), and a sensing receiver (Sen-RX). The ISAC-TX transmits ISAC signals to the Com-RX to provide communication services, and to the Sen-RX for sensing $K$ targets whose locations are denoted as $\mathbf{q}_k, k=1,2,\cdots,K$. Denote the number of antennas at the ISAC-TX, Com-RX, and Sen-RX by $M_t$, $M_c$, and $M_r$, respectively. For sensing, we assume that each target has a LoS path with the ISAC-TX and the Sen-RX, while such an assumption is not needed for communication. The array response vector between the ISAC-TX and target $k$ is denoted as $\mathbf{a}_{\mathrm{T}}(\mathbf{q}_k) \in \mathbb{C}^{M_t \times 1}$, which in general is a function of the sensing target location $\mathbf{q}_k$. Similarly, the RX array response vector is denoted as $\mathbf{a}_{\mathrm{R}}(\mathbf{q}_k) \in \mathbb{C}^{M_r \times 1}$. Specific expressions for $\mathbf{a}_{\mathrm{T}}(\mathbf{q}_k)$ and $\mathbf{a}_{\mathrm{R}}(\mathbf{q}_k)$ in far-field and near-field scenarios will be provided in \mbox{Section~\ref{SO}}. 

Denote the ISAC signal transmitted by the ISAC-TX as $\mathbf{x}(t) \in \mathbb{C}^{M_t \times 1}$ and the communication channel impulse response between the ISAC-TX and Com-RX as $\mathbf{H}_c\left(t,\tau\right)\in\mathbb{C}^{M_c \times M_t}$, where $t$ and $\tau$ denote observation time and delay, respectively. Then, the communication signal $\mathbf{y}_c(t) \in \mathbb{C}^{M_c \times 1}$ received by the $M_c$-antenna Com-RX array is
\begin{equation}
	\setlength\abovedisplayskip{2pt}
	\setlength\belowdisplayskip{2pt}
	\begin{aligned}
		\mathbf{y}_c\left(t\right)=\mathbf{H}_c\left(t,\tau\right) \ast \mathbf{x}(t)+\mathbf{n}\left(t\right),
	\end{aligned}
	\label{d1}
\end{equation}
where $\ast$ denotes convolution, and $\mathbf{n}\bigl(t\bigr)\sim CN\bigl(0,\mathbf{\sigma}^2\mathbf{I}_{M_c}\bigr)$ is circularly symmetric additive white Gaussian noise (AWGN) with zero mean and variance $\sigma^2$. We assume communication signal processing methods consistent with traditional communication systems, and thus such methods will not be detailed here.

For target sensing, the ISAC signal sent by the ISAC-TX is reflected/scattered by the $K$ targets and other scatterers in the surrounding environment, and then received by the Sen-RX. A LoS link between the ISAC-TX and Sen-RX may also exist. Thus, the sensing signal $\mathbf{y}_s\left(t\right)\in\mathbb{C}^{M_r \times 1}$ received by the $M_r$-antenna Sen-RX array can be expressed as 
\begin{equation}
	\setlength\abovedisplayskip{2pt}
	\setlength\belowdisplayskip{2pt}
	\begin{aligned}
\mathbf{y}_{s}(t) =&\sum_{k=1}^K\alpha_k\mathbf{a}_{\mathrm{R}}\begin{pmatrix}\mathbf{q}_k\end{pmatrix}\mathbf{a}_{\mathrm{T}}^H\begin{pmatrix}\mathbf{q}_k\end{pmatrix}\mathbf{x}\begin{pmatrix}t-\tau_k-\tau_{\Delta}\end{pmatrix} \\
& \times e^{j2\pi\left(f_{\Delta}+\upsilon_k\right)t} +\mathbf{H}_{\mathrm{Clutter}}\left(t,\tau\right) \ast \mathbf{x}\left(t\right) +\tilde{\mathbf{z}}\left(t\right),
	\end{aligned}
	\label{d3}
\end{equation}
where $\alpha_{k}$, $\tau_{k}$, and $\upsilon_{k}$ denote the complex-valued channel coefficient, the propagation delay, and Doppler frequency of the path corresponding to the $k$-th sensing target, respectively, $\tau_{\Delta}$ and $f_{\Delta}$ are the symbol time offset (STO) and carrier frequency offset (CFO) caused by the difference between the reference clock and local oscillators (LO) at the ISAC-TX and Sen-RX. Note that we may assume $\tau_{\Delta}=0$ and $f_{\Delta}=0$ for monostatic ISAC modes, while $\tau_{\Delta} \neq 0$ and $f_{\Delta} \neq 0$ in general for bistatic ISAC modes. The term $\mathbf{H}_{\mathrm{Clutter}}\left(t,\tau\right)$ is the channel impulse response of the clutter, which consists of the signals scattered/reflected by objects irrelevant to the $K$ sensing targets and the LoS link between the ISAC-TX and Sen-RX. Note that for monostatic ISAC systems, the ISAC-TX and Sen-RX are collocated and operate in full-duplex mode to achieve maximum SE, making them vulnerable to self-interference (SI) due to imperfect isolation. The residual SI after the application of various cancellation techniques can also be considered as a component of the clutter. Finally, $\tilde{\mathbf{z}}\left(t\right)$ is the AWGN at the Sen-RX.

\begin{figure*}[!t]
\centering
\subfigcapskip=2pt 
\subfigure[BS monostatic ISAC mode.]{
\begin{minipage}[hbt]{0.33\linewidth}
\centering
\includegraphics[width=0.95\linewidth]{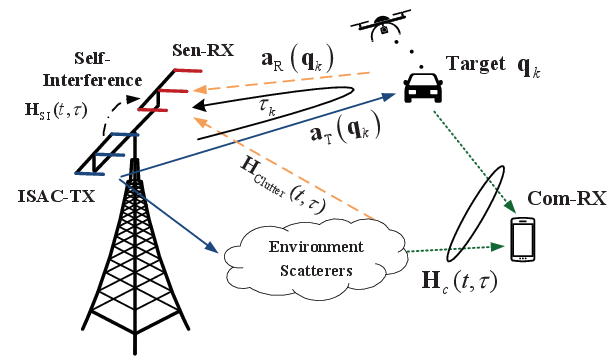}
\label{mono-down}
\end{minipage}
}\hspace{-3mm}%
\subfigure[UE-to-BS bistatic ISAC mode.]{
\begin{minipage}[hbt]{0.33\linewidth}
\centering
\includegraphics[width=0.95\linewidth]{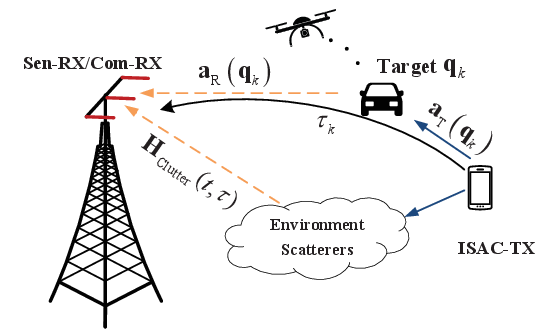}
\label{mono-up}
\end{minipage}%
}\hspace{-3mm}%
\subfigure[BS-to-BS bistatic ISAC mode.]{
\begin{minipage}[hbt]{0.33\linewidth}
\centering
\includegraphics[width=0.95\linewidth]{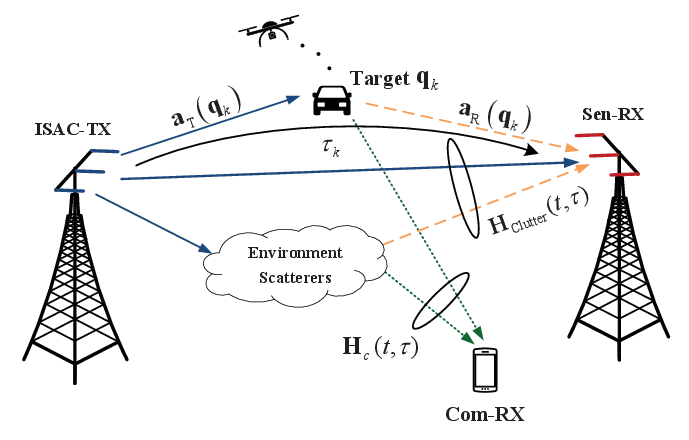}
\label{bi-down}
\end{minipage}%
}\hspace{-3mm}%
\centering
\caption{An illustration of the three BS-side sensing ISAC modes.}
\label{BSs}
\end{figure*}

\begin{table*}[!t]
\renewcommand{\arraystretch}{1.2}
\centering
\caption{The unified ISAC model and its special cases corresponding to the six basic modes.}
\label{gmc}
\begin{tabular}{|cc|ccc|ccc|}
\hline
\multicolumn{2}{|c|}{ISAC modes}                                                                                               & \multicolumn{3}{c|}{Transmitter and receiver}                                                                                                                                                                                           & \multicolumn{3}{c|}{Required processing}                                                                                                                                                          \\ \hline
\multicolumn{2}{|c|}{Unified model}                                                                                            & \multicolumn{1}{c|}{ISAC-TX} & \multicolumn{1}{c|}{Com-RX} & {Sen-RX} & \multicolumn{1}{c|}{SI cancellation} & \multicolumn{1}{c|}{Synchronization} & \begin{tabular}[c]{@{}c@{}}Clutter\\  rejection\end{tabular} \\ \hline
\multicolumn{1}{|c|}{\multirow{3}{*}{\begin{tabular}[c]{@{}c@{}}BS-side \\ sensing \end{tabular}}} & BS monostatic     & \multicolumn{1}{c|}{BS}                                                          & \multicolumn{1}{c|}{UE}                                                                & BS                                                          & \multicolumn{1}{c|}{\checkmark}                                                                    & \multicolumn{1}{c|}{}                & \checkmark                                                            \\ \cline{2-8} 
\multicolumn{1}{|c|}{}                                                                                     & UE-to-BS bistatic & \multicolumn{1}{c|}{UE}                                                          & \multicolumn{1}{c|}{BS}                                                                & BS                                                          & \multicolumn{1}{c|}{}                                                                     & \multicolumn{1}{c|}{\checkmark}               & \checkmark                                                           \\ \cline{2-8} 
\multicolumn{1}{|c|}{}                                                                                     & BS-to-BS bistatic       & \multicolumn{1}{c|}{BS-1}                                                        & \multicolumn{1}{c|}{UE}                                                                & BS-2                                                        & \multicolumn{1}{c|}{}                                                                     & \multicolumn{1}{c|}{\checkmark}               & \checkmark                                                            \\ \hline
\multicolumn{1}{|c|}{\multirow{3}{*}{\begin{tabular}[c]{@{}c@{}}UE-side \\ sensing \end{tabular}}} & UE monostatic     & \multicolumn{1}{c|}{UE}                                                          & \multicolumn{1}{c|}{BS}                                                                & UE                                                          & \multicolumn{1}{c|}{\checkmark}                                                                    & \multicolumn{1}{c|}{}                & \checkmark                                                            \\ \cline{2-8} 
\multicolumn{1}{|c|}{}                                                                                     & BS-to-UE bistatic & \multicolumn{1}{c|}{BS}                                                          & \multicolumn{1}{c|}{UE}                                                                & UE                                                          & \multicolumn{1}{c|}{}                                                                     & \multicolumn{1}{c|}{\checkmark}               & \checkmark                                                            \\ \cline{2-8} 
\multicolumn{1}{|c|}{}                                                                                     & UE-to-UE bistatic       & \multicolumn{1}{c|}{UE-1}                                                        & \multicolumn{1}{c|}{UE-2}                                                              & UE-2                                                        & \multicolumn{1}{c|}{}                                                                     & \multicolumn{1}{c|}{\checkmark}               & \checkmark                                                            \\ \hline
\end{tabular}
 \vspace{-0.3cm}
\end{table*}

The presence of clutter signals and SI makes it more challenging to distinguish the targets from the complex environment, and the STO and CFO will cause biases in the estimation of target parameters. Thus, SI cancellation, clutter rejection, and synchronization are essential for efficient ISAC signal processing, which will be discussed in \mbox{Section \ref{openp}}. After time and frequency synchronization, clutter rejection and SI cancellation, the resulting sensing signal $\mathbf{y}_s(t) \in \mathbb{C}^{M_r \times 1}$ in \mbox{(\ref{d3})} reduces to 
\begin{equation}
	\setlength\abovedisplayskip{2pt}
	\setlength\belowdisplayskip{2pt}
	\begin{aligned}
\mathbf{y}_{s}(t) =\sum_{k=1}^{K}\alpha_k\mathbf{a}_{\mathrm{R}}\begin{pmatrix}\mathbf{q}_k\end{pmatrix}\mathbf{a}_{\mathrm{T}}^H\begin{pmatrix}\mathbf{q}_k\end{pmatrix}\mathbf{x}\begin{pmatrix}t-\tau_k\end{pmatrix}e^{j2\pi\upsilon_kt}+\mathbf{z}(t),
	\end{aligned}
	\label{9}
\end{equation}
where the clutter rejection and SI cancellation residuals are typically considered as random noise and lumped together with the AWGN in $\mathbf{z}(t)$. The sensing signal in \mbox{(\ref{9})} is then exploited to estimate the parameters of the $K$ sensing targets, including the reflection coefficient $\alpha_k$, target location $\mathbf{q}_k$, propagation delay $\tau_k$, and Doppler frequency shift $\upsilon_k$.

Note that except for some differences in the number of antennas and computational capabilities, BS-side and UE-side sensing are symmetric in terms of signal modeling and processing. Therefore, in the following, we present detailed models and signal processing techniques for ISAC systems of the three BS-side sensing modes based on the unified ISAC model, while similar models can be derived for UE-side sensing as well. To this end, we first show that the generic ISAC setup in \mbox{Fig.~\ref{general model}} and the resulting signal in \mbox{(\ref{9})} can be reduced to the three BS-side sensing modes shown in \mbox{Figs.~\ref{ISAC_modes}} (a)-(c). 

\begin{itemize}

     \item \textbf{BS monostatic ISAC mode:} For BS monostatic ISAC mode, the ISAC-TX and Sen-RX shown in \mbox{Fig.~\ref{general model}} are collocated at the BS. To maximize the SE, the ISAC BS needs to employ in-band full-duplex operation for simultaneous downlink ISAC signal transmission and uplink sensing signal reception. Note that while collocated at the BS, we assume that the ISAC-TX and Sen-RX use different sets of antennas for ease of SI cancellation. The Com-RX in \mbox{Fig.~\ref{general model}} corresponds to a UE. In this case, the generic ISAC setup in \mbox{Fig.~\ref{general model}} reduces to the BS monostatic setup in \mbox{Fig.~\ref{mono-down}}. Furthermore, the ISAC signal $\mathbf{x}(t)$ in \mbox{(\ref{d1})}-\mbox{(\ref{9})} is transmitted by the BS, and is denoted as $\mathbf{x}_{\mathrm{BS}}(t)\in \mathbb{C}^{M_t\times 1}$.

    \item \textbf{UE-to-BS bistatic ISAC mode:} For UE-to-BS bistatic ISAC mode, the ISAC-TX in \mbox{Fig.~\ref{general model}} corresponds to a UE, while the Com-RX and Sen-RX are collocated at the BS, as shown in \mbox{Fig.~\ref{mono-up}}. Note that unlike the BS monostatic mode in \mbox{Fig.~\ref{mono-down}}, since only signal reception needs to be performed at the BS in \mbox{Fig.~\ref{mono-up}}, the Com-RX and Sen-RX may use the same set of antennas. However, synchronization is essential due to the use of different local oscillators (LO) and reference clocks by the ISAC BS and the UE. The ISAC signal $\mathbf{x}(t)$ in \mbox{(\ref{d1})}-\mbox{(\ref{9})} is transmitted by the UE, and is denoted as $\mathbf{x}_{\mathrm{UE}}(t)\in \mathbb{C}^{M_t \times 1}$. 

     \item \textbf{BS-to-BS bistatic ISAC mode:} For BS-to-BS bistatic ISAC mode, the ISAC-TX in \mbox{Fig.~\ref{general model}} corresponds to a BS, while the Com-RX and Sen-RX occur at a UE and another BS, respectively, as shown in \mbox{Fig.~\ref{bi-down}}. Note that unlike the UE-to-BS bistatic mode in \mbox{Fig.~\ref{mono-up}}, although the BS bistatic mode also requires synchronization, it can be achieved through a wired connection between the two ISAC BSs. Furthermore, the ISAC signal $\mathbf{x}(t)$ in \mbox{(\ref{d1})}-\mbox{(\ref{9})} is broadcast by the transmit BS, and is denoted as $\mathbf{x}_{\mathrm{BS}}(t)\in \mathbb{C}^{M_t\times 1}$.
\end{itemize}

\begin{figure*}[!t]
  \centering
  \includegraphics[width=0.85\linewidth]{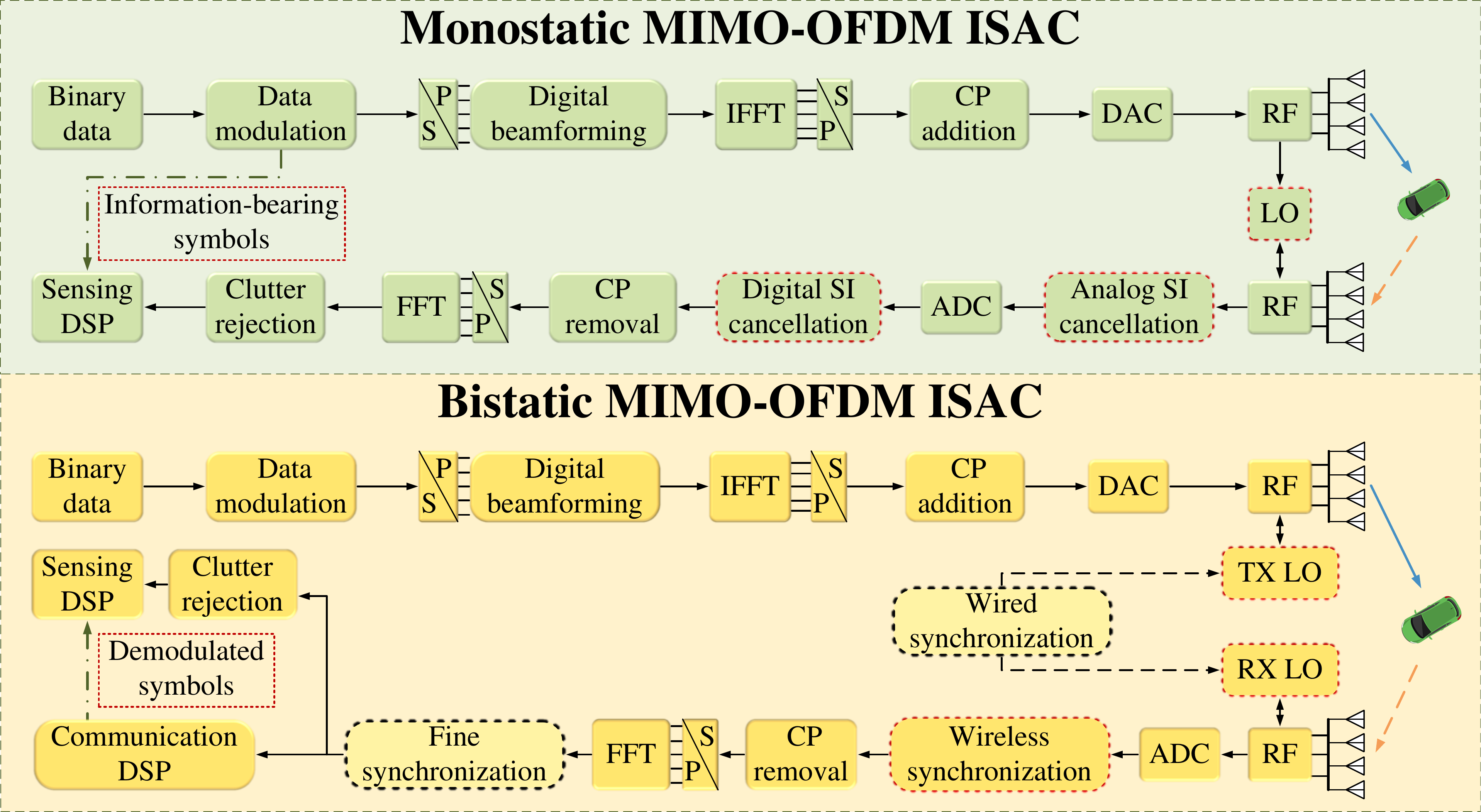}
  \caption{Primary signal processing operations for monostatic and bistatic MIMO-OFDM ISAC systems.}
  \label{OFDMp}
  \vspace{-0.4cm}
\end{figure*}

The corresponding ISAC-TX, Com-RX, and Sen-RX of the unified model in \mbox{Fig.~\ref{general model}}, as well as the required processing for the six ISAC modes, are summarized in \mbox{Table~\ref{gmc}}.

In the following, we focus on ISAC systems implemented with OFDM waveforms, which is the dominant waveform in 4G and 5G communication systems, and will continue to play the same role for 6G. In OFDM-ISAC systems, target sensing can be performed based on pilots only or on the fully reconstructed transmitted OFDM data frame. The former approach can be easily implemented, but its sensing performance is limited by the availability of time-frequency resources allocated to pilots. The latter can fully exploit all of the time-frequency resources for target sensing using known (for monostatic sensing) or demodulated communication symbols (for bistatic sensing), at the expense of increased complexity. In this paper, we employ the OFDM frame-based approach for sensing tasks using all time-frequency resources allocated to both pilots and communication data. 

\subsection{MIMO-OFDM ISAC Signal Processing and Modeling} \label{SO}

For MIMO-OFDM ISAC, the ISAC-TX transmits OFDM signals over $N$ subcarriers and $P$ OFDM symbols. The subcarrier spacing and the OFDM symbol duration without the CP are denoted by $\Delta f$ and $T$, respectively, where $T=\frac{1}{\Delta f}$. The system bandwidth is thus $B=N\Delta f$. Let $T_{\mathrm{cp}}$ denote the CP duration, so the OFDM symbol duration including CP is $T_s=T+T_{\mathrm{cp}}$. To avoid ISI for communication, the CP duration should be no smaller than the multi-path delay spread $\tau_{\mathrm{com}}$ of the communication channel $\mathbf{H}_c(t, \tau)$, since the Com-RX only begins receiving communication signals upon receiving its first path. However, for target sensing, the CP duration should be larger than the propagation delay $\tau_{\mathrm{sen}}$ corresponding to the ISI-free sensing distance \cite{ISI}, which is different from the communications case since the Sen-RX begins to receive signals as soon as the ISAC-TX transmits, so that no nearby targets are missed. Thus, in ISAC systems, the CP should be set as $T_{\mathrm{cp}}\geq \tau_{\max}$, where $\tau_{\max}=\max\{\tau_{\mathrm{com}},\tau_{\mathrm{sen}}\}$. The OFDM signal $\mathbf{x}(t)\in \mathbb{C}^{M_t \times 1}$ transmitted by the $M_t$-antenna ISAC-TX array is given by
\begin{equation}
	\setlength\abovedisplayskip{2pt}
	\setlength\belowdisplayskip{2pt}
	\begin{aligned}
		\mathbf{x}(t)=\sum_{p=0}^{P-1}\sum_{n=0}^{N-1}\mathbf{w}_{n,p} b_{n,p}e^{j2\pi n\Delta f(t-pT_s-T_{\mathrm{cp}})}\mathrm{rect}\left(\frac{t-pT_s}{T_s}\right),
	\end{aligned}
	\label{nISAC-TXsm}
\end{equation}
where $\mathbf{w}_{n,p}\in\mathbb{C}^{M_t\times 1}$ and $b_{n,p}$ are the transmit beamforming vector and the transmitted symbol on the $n$-th subcarrier of the $p$-th OFDM symbol, respectively. Contrary to traditional OFDM radar, where all subcarriers and symbols can be exploited to sense targets, a user typically cannot be allocated all of the time-frequency resources for communications. Thus, $\mathbf{w}_{n,p}$ depends on the users' resource allocation and beam management. In addition, if a user occupies the $(n,p)$-th resource element (RE), $b_{n,p}\neq0$ is the information-bearing symbol; otherwise, $b_{n,p} = 0$. Later, in \mbox{Section~\ref{RABM}}, we will further discuss this issue. 

Substituting \mbox{(\ref{nISAC-TXsm})} into \mbox{(\ref{9})}, the resulting sensing signal $\mathbf{y}_s(t) \in \mathbb{C}^{M_r \times 1}$ for radar processing at the Sen-RX is
\begin{align}	\label{OSM}
	\setlength\abovedisplayskip{2pt}
	\setlength\belowdisplayskip{2pt}
        \mathbf{y}_s(t)=&\sum_{k=1}^K\sum_{p=0}^{P-1}\sum_{n=0}^{N-1}\alpha_{k}^{n,p}\mathbf{a}_{\mathrm{R}}(\mathbf{q}_k)b_{n,p}e^{j2\pi n\Delta f(t-\tau_k-pT_s-T_{\mathrm{cp}})} \nonumber \\
        &\qquad\quad\times e^{j2\pi \upsilon_{k}t}\mathrm{rect}\left(\frac{t-\tau_k-pT_s}{T_s}\right)+\mathbf{z}(t),
\end{align}
where $\alpha_k^{n,p}=\alpha_{k}\mathbf{a}_{\mathrm{T}}^H(\mathbf{q}_k)\mathbf{w}_{n,p}$ is the equivalent channel coefficient including the impact of transmit beamforming.
Then, the signal in \mbox{(\ref{OSM})} is partitioned into $P$ blocks, each of duration $T_s$. After discarding the CP, the $p$-th ($p=0,\cdots,P-1$) block is given by
\begin{equation}
	\setlength\abovedisplayskip{2pt}
	\setlength\belowdisplayskip{2pt}
	\begin{aligned}
        \mathbf{y}_s^{p}\left(t\right)=\mathbf{y}_s\left(t+pT_s+T_{\mathrm{cp}}\right)\mathrm{rect}\Bigg(\frac{t}{T}\Bigg), \  t\in [0,T).
        \end{aligned}
	\label{25}
\end{equation}

\begin{figure*}[!t]
  \centering
  \includegraphics[width=0.95\linewidth]{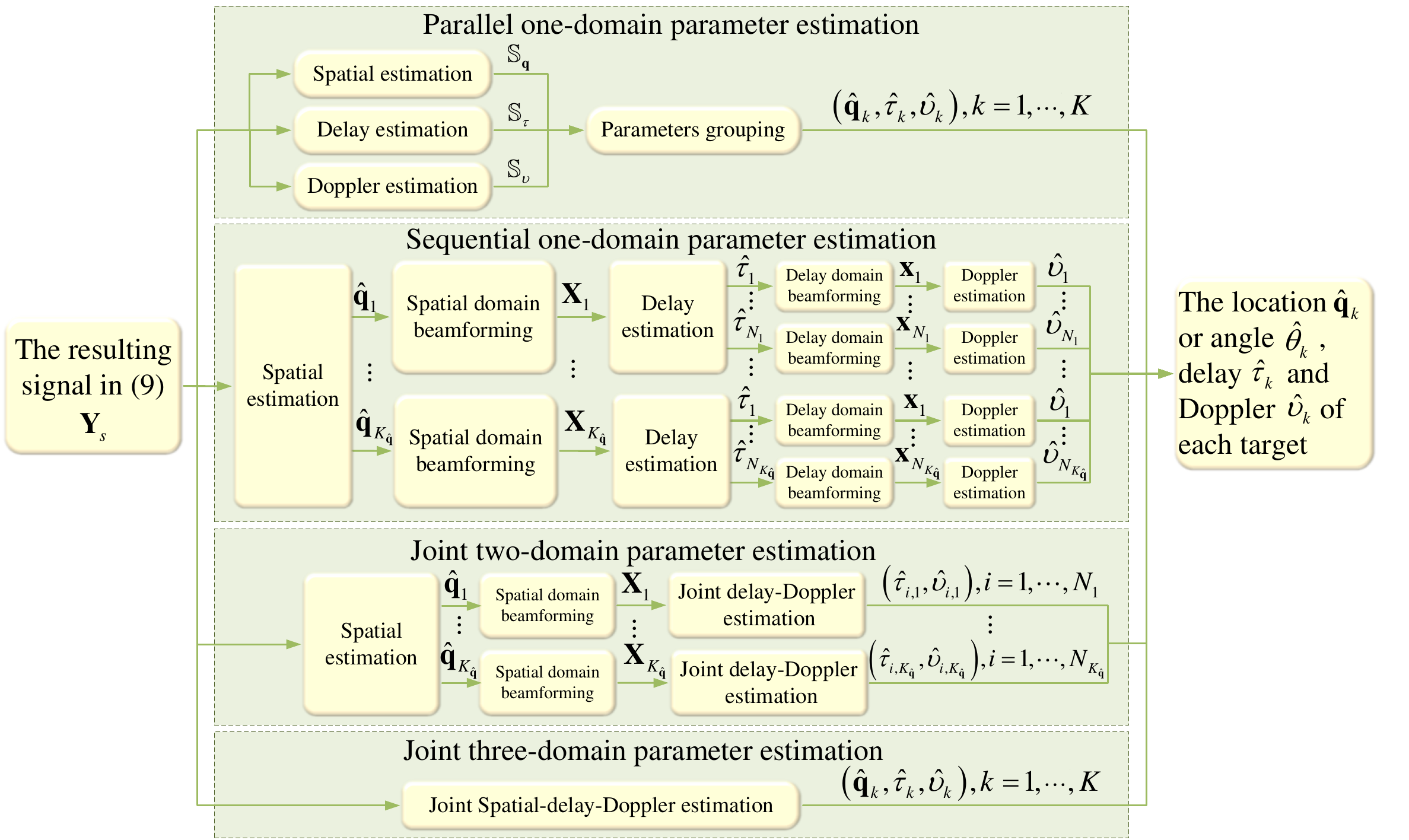}
  \caption{The four parameter estimation frameworks in MIMO-OFDM ISAC.}
  \label{PE}
    \vspace{-0.4cm}
\end{figure*}

The Sen-RX samples the sensing signal with sampling rate $B=N\Delta f$. To avoid inter-carrier interference (ICI) caused by large Doppler frequency shifts, the subcarrier spacing is typically assumed to be much larger than the Doppler frequency shift, i.e.,  $\Delta f\gg |\upsilon_{k}|, \forall k=1,2,\cdots,K$. Under this assumption, the $q$-th ($q=0,1,\cdots,N-1$) sample of the $p$-th block can be expressed as
\begin{equation}
	\setlength\abovedisplayskip{2pt}
	\setlength\belowdisplayskip{2pt}
	\begin{aligned}
        \mathbf{y}_{s}^{p}[q]=&\mathbf{y}_{s}^{p}\left(q/B\right)\\
         =&\sum_{k=1}^K\sum_{n=0}^{N-1}\alpha_{k}^{n,p}\mathbf{a}_{\mathrm{R}}(\mathbf{q}_k)b_{n,p}e^{j2\pi n\Delta f\left(q/B-\tau_{k}\right)} \\
        &\qquad \qquad \times e^{j2\pi\upsilon_{k}\left(q/B+pT_{s}+T_{\mathrm{cp}}\right)}+\mathbf{z}^{p}\left[q\right] \\
\approx&\sum_{k=1}^K\sum_{n=0}^{N-1}\overline{\alpha}_{k}^{n,p}\mathbf{a}_{\mathrm{R}}(\mathbf{q}_k)b_{n,p}e^{j2\pi nq/N}e^{-j2\pi n\Delta f\tau_{k}} \\
&\qquad \qquad \qquad\qquad \times e^{j2\pi pT_{s}\upsilon_{k}}+\mathbf{z}^{p}[q],
        \end{aligned}
	\label{nRXSMBS}
\end{equation}
where $\overline{\alpha}_{k}^{n,p}=\alpha_{k}^{n,p}e^{j2\pi\upsilon_{k}T_{\mathrm{cp}}}$, $\mathbf{z}^{p}[q]$ is the resulting noise at the $p$-th block. Note that in \mbox{(\ref{nRXSMBS})}, we have used the fact that $B=N\Delta f$ and $|v_kq/B|<|v_kN/B|=|v_k|/\Delta f$, so that $e^{j2\pi\upsilon_{k}q/B}\approx 1$ for $|\upsilon_{k}|/ \Delta f\ll 1$.

After performing an $N$-point DFT for each of the $P$ blocks in \mbox{(\ref{nRXSMBS})}, we obtain the following frequency-domain signal for the $n$-th $(n=0,1,\cdots,N-1)$ subcarrier of the $p$-th $(p=0,1,\cdots,P-1)$ OFDM block $\bar{\mathbf{y}}_s^p[n]\in \mathbb{C}^{M_r\times 1}$ as
\begin{equation}
	\setlength\abovedisplayskip{2pt}
	\setlength\belowdisplayskip{2pt}
	\begin{aligned}
        &\bar{\mathbf{y}}_s^p[n]=\frac{1}{N}\sum_{q=0}^{N-1}\mathbf{y}_{s}^{p}[q]e^{-j2\pi nq/N} \\
        &=b_{n,p}\sum_{k=1}^K\overline{\alpha}_{k}^{n,p} \mathbf{a}_{\mathrm{R}}(\mathbf{q}_k) e^{-j2\pi n\Delta f\tau_k}e^{j2\pi pT_s\upsilon_{k}} +\bar{\mathbf{z}}^p[n],
        \end{aligned}
	\label{d161}
\end{equation}
where $\bar{\mathbf{z}}^p[n]$ is the resulting AWGN in the frequency domain. As a result, we obtain a total of $M_r\times N\times P$ signal values across the spatial, subcarrier, and symbol domains, which can be organized as a tensor $\mathbf{Y}_s \in \mathbb{C}^{M_r \times N\times P}$ by setting
\begin{equation}
	\setlength\abovedisplayskip{2pt}
	\setlength\belowdisplayskip{2pt}
	\begin{aligned}
        \mathbf{Y}_s(:,n,p)=\bar{\mathbf{y}}_s^p[n].
        \end{aligned}
	\label{16}
\end{equation}

The main signal processing operations associated with MIMO-OFDM ISAC systems for the monostatic and bistatic modes are illustrated in \mbox{Fig.~\ref{OFDMp}}. The operations for the monostatic and bistatic ISAC modes mainly differ in three aspects. First, in the monostatic mode, the ISAC-TX and Sen-RX use the same LO and reference clock, avoiding the need for synchronization with no STO or CFO in this case. By contrast, the bistatic mode requires different LOs and reference clocks, necessitating synchronization at the RX. Second, unlike the bistatic case, the monostatic ISAC-TX and Sen-RX operate in full-duplex mode and require SI cancellation. Finally, the information-bearing symbols transmitted by the ISAC-TX are available at the monostatic Sen-RX since they are collocated. However, in the bistatic case, only pilot data is available, and the information-bearing symbols need to be demodulated before target sensing. Synchronization between basestations in bistatic systems can be achieved via both wireless and wired connections, and fine synchronization can be achieved after FFT demodulation to further reduce the residual STO and CFO in scenarios requiring higher accuracy.

Parameter estimation is one of the most important sensing tasks in ISAC systems, as it underpins functions like target localization, tracking, and imaging. Therefore, in the following, we will focus on estimation of the location $\mathbf{q}_k$ (or the angle $\theta_k$ in the far-field case), the propagation delay $\tau_k$, and the Doppler frequency shift $\upsilon_k$ of the sensing targets.
Generally speaking, the target parameters are estimated from the spatial, delay, and Doppler domain data. Inspired by traditional joint parameter estimation methods \cite{2DPE,3DPE}, we propose four frameworks to estimate target parameters across the three domains. As shown in \mbox{Fig.~\ref{PE}}, depending on whether the parameters are estimated jointly or separately in each domain, the four frameworks are parallel one-domain, sequential one-domain, joint two-domain, and joint three-domain parameter estimation. The parallel one-domain approach separately estimates the sets of target locations $\mathbb{S}_{\mathbf{q}}=\{\mathbf{q}_1,\cdots,\mathbf{q}_K\}$ or angles $\mathbb{S}_{\theta}=\{{\theta}_1,\cdots,{\theta}_K\}$, delays $\mathbb{S}_{\tau}=\{\tau_1,\cdots,\tau_K\}$ and Doppler shifts $\mathbb{S}_{\upsilon}=\{\upsilon_1,\cdots,\upsilon_K\}$, and then associates the estimated parameters into groups $(\mathbf{q}_k/\theta_k,\tau_k,\upsilon_k)$ for each of the $K$ sensing targets. This approach has a relatively low computational complexity, but its sensing performance is limited by potential errors in the parameter grouping.

On the other hand, the sequential one-domain approach performs spatial, delay, and Doppler domain estimation sequentially. For example, spatial domain estimation can be applied first to determine the $K_{\hat{\mathbf{q}}}$ distinguishable locations/angles. When the spatial resolution is limited, it may not be possible to distinguish all $K$ targets in the spatial domain alone. Thus, in general, $K_{\hat{\mathbf{q}}}\leq K$. After location/angle estimation, beamforming can be applied to separate the $K_{\hat{\mathbf{q}}}$ signals in the spatial domain, denoted as $\mathbf{X}_{k_{\hat{\mathbf{q}}}}, k_{\hat{\mathbf{q}}}=1,\cdots,K_{\hat{\mathbf{q}}}$. Subsequently, delay and Doppler domain processing can be applied. Note that the order in which the different types of parameters are estimated depends on the resolution and target separation in the three domains. The sequential one-domain estimation method does not require parameter grouping, but its complexity is greater than parallel one-domain processing. 

Similar to the sequential one-domain method, joint two-domain parameter estimation first determines the parameter in one domain, and then exploits beamforming methods to separate the signals of different targets, but it jointly estimates the parameters in the remaining two domains. \mbox{Fig.~\ref{PE}} illustrates the special case of joint processing of the delay-Doppler domains after location/angle estimation has been performed, while other combinations are also possible. This method can achieve higher estimation accuracy, but its complexity is much higher than that of the decoupled estimation methods. Last, joint three-domain parameter estimation refers to joint estimation of all parameters. This approach can achieve the highest estimation accuracy but also incurs the highest computational complexity. 

The details of each method will be discussed in \mbox{Sections~\ref{FF_Sensing}} and \mbox{\ref{near-field}} for the far-field and near-field cases, respectively. Note that the TX and RX steering vectors $\mathbf{a}_{\mathrm{T}}(\mathbf{q}_k)$ and $\mathbf{a}_{\mathrm{R}}(\mathbf{q}_k)$ are functions of target location $\mathbf{q}_k$, which in general depends on both the angle of departure (AoD)/AoA $\theta_k$ and distance $r_k$ between the target and the reference antenna. In the following, we present expressions for the RX steering vectors for far-field and near-field scenarios. Similar results can be obtained for the TX array response. 

\subsection{Array Response Vector: From Far-Field to Near-Field}
\begin{figure}[!t]	
	\subfigure[Uniform plane wave.] 
	{
		\begin{minipage}{4.5cm}
			\hspace{-4ex}
			\includegraphics[scale=0.4]{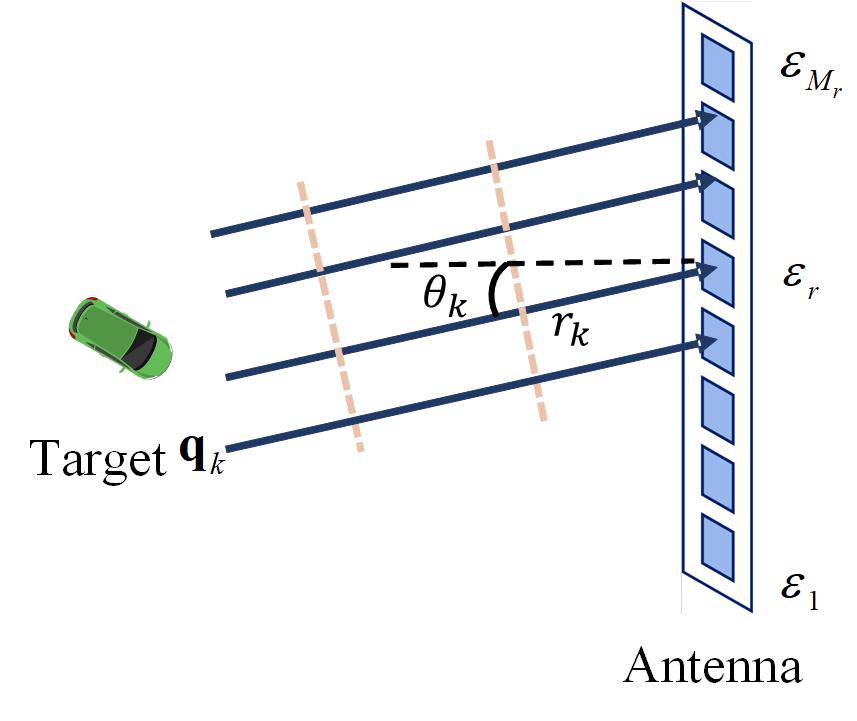}   
		\end{minipage}
		\label{upw}
	}\subfigure[Uniform spherical wave.] 
	{
		\begin{minipage}{4.5cm}
			\centering          
			\hspace{-6ex}
			\includegraphics[scale=0.37]{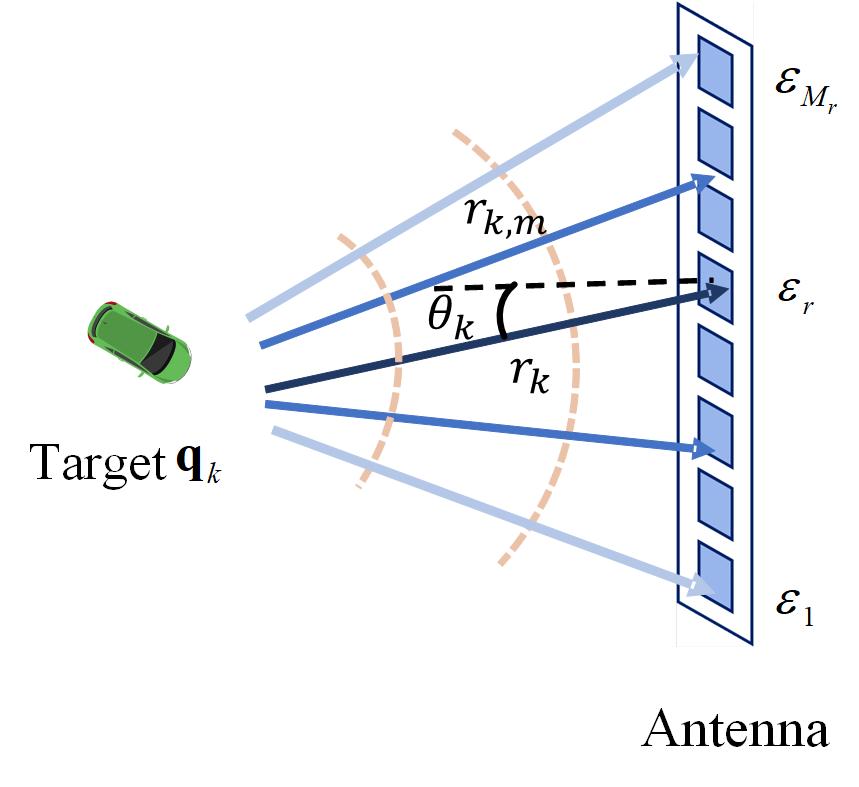}   
		\end{minipage}
		\label{nusw}
	}
	\caption{Illustration of far-field UPW versus near-field USW.} 
	\label{nusw-upw}  
	\vspace{-0.8cm}
\end{figure}

When the BS array aperture is small, targets are typically located in the far-field region, and the conventional UPW model accurately characterizes the variations of signal phases across the array, as shown in \mbox{Fig.~\ref{upw}}. However, for large-scale arrays, targets are more likely to be located in the near-field region, and the USW assumption should be used~\cite{b23,b39}, as shown in \mbox{Fig.~\ref{nusw}}. 

Let the location of target $k$ be represented as ${{\mathbf{q}}_k} = {[r_k\cos \theta_k, r_k\sin \theta_k]^T}, k=1,...,K$, where $r_k$ is the distance between target $k$ and a given reference antenna, and $\theta_k \in \left[ -\frac{\pi}{2},\frac{\pi}{2} \right]$ is the angle of target $k$ with respect to the normal vector of the array. Then, the receive steering vector $\mathbf{a}_{\mathrm{R}}(\mathbf{q}_k)$ can be expressed as
\begin{equation}
	\setlength\abovedisplayskip{2pt}
	\setlength\belowdisplayskip{2pt}
	\begin{aligned}
	\left[{\mathbf{a}}_{\mathrm{R}}({{\mathbf{q}}_k})\right]_m = {{e^{j\frac{{2\pi }}{\lambda }\left( {{r_{k,m}} - {r_k}} \right)}}}, m=1,...,M_r,
	\end{aligned}
	\label{steering_gen}
\end{equation}
where $r_{k,m}$ denotes the distance between the $k$-th target and the $m$-th receive antenna. For narrowband signals, half-wavelength inter-antenna spacing is typically assumed, i.e., $d=\frac{\lambda}{2}$. If we assume the $r$-th antenna is the reference located at the origin of the array's coordinate system, the positions of the $M_r$ antennas can be written as $[0,\varepsilon_md],m=1,\cdots,M_r$, where $|\varepsilon_md|$ is the distance between the $m$-th antenna and the reference, and $\varepsilon_r=0$. Thus, $r_{k,m}$ can be written as
\begin{equation}
	\setlength\abovedisplayskip{2pt}
	\setlength\belowdisplayskip{2pt}
	\begin{aligned}
	{r_{k,m}} = \sqrt {r_k^2 - 2{\varepsilon _m}d{r_k}\sin {\theta _k} + \varepsilon _m^2{d^2}}.
	\end{aligned}
	\label{steering_rkm}
\end{equation}
 
 The classical criterion for distinguishing the near- and far-field regions is the Rayleigh distance, given by ${r_\mathrm{Rayleigh}} = \frac{{2{D^2}}}{\lambda }$, where $D$ is the aperture of the antenna array~\cite{b39,b91,b92}. When $r_k>r_\mathrm{Rayleigh}$, the target is assumed to be located in the far-field, so that the UPW model is appropriate and $r_{k,m}$ can be simplified by performing a first-order Taylor series approximation of (\ref{steering_rkm}), as shown in Fig.~\ref{upw}. Thus, $r_{k,m}$ can be simplified as
\begin{equation}
	\setlength\abovedisplayskip{2pt}
	\setlength\belowdisplayskip{2pt}
	\begin{aligned}
	{r_{k,m}} \approx {r_k} - \varepsilon_m{d}\sin {\theta_k},
	\end{aligned}
	\label{rapp1}
\end{equation}
and the corresponding steering vector in (\ref{steering_gen}) becomes
\begin{equation}
	\setlength\abovedisplayskip{2pt}
	\setlength\belowdisplayskip{2pt}
	\begin{aligned}
	{{\mathbf{a}}_{{\mathrm{R}}}}({\theta _k}) = \left[ {{e^{-j\frac{{2\pi }}{\lambda }\varepsilon_m d\sin {\theta _k}}}} \right], m=1,...,M_r.
	\end{aligned}
	\label{steering_far}
\end{equation}
For near-field sensing with $r_k<r_\mathrm{Rayleigh}$ shown in Fig.~\ref{nusw}, substituting the more general expression in (\ref{steering_rkm}) into (\ref{steering_gen}) leads to
\begin{equation}
	\setlength\abovedisplayskip{2pt}
	\setlength\belowdisplayskip{2pt}
	\begin{aligned}
	\;{\bf{a}}_{\rm{R}}(r_k,\theta_k) = \left[ {{e^{j\frac{{2\pi }}{\lambda }\left( {\sqrt {r_k^2 - 2{\varepsilon _m}d{r_k}\sin {\theta _k} + \varepsilon _m^2{d^2}} - {r_k}} \right)}}} \right].
	\end{aligned}
	\label{steering_near}
\end{equation}

Comparing (\ref{steering_far}) and (\ref{steering_near}), we see that in the far-field, only the angle can be determined from one snapshot on different antenna elements, while both target angle and distance can be estimated in the near-field. The general expression in (\ref{steering_near}) can be simplified using a second-order Taylor approximation of (\ref{steering_rkm}) \cite{b39}, leading to
\begin{equation}
	\setlength\abovedisplayskip{2pt}
	\setlength\belowdisplayskip{2pt}
	\begin{aligned}
	{r_{k,m}} \approx {r_k} - \varepsilon_m{d}\sin {\theta_k} + \frac{{{\varepsilon_m^2}d^2}}{{2{r_k}}}{\cos ^2}{\theta_k},
	\end{aligned}
	\label{rapp}
\end{equation}
with corresponding array response vector
\begin{equation}
	\setlength\abovedisplayskip{2pt}
	\setlength\belowdisplayskip{2pt}
	\begin{aligned}
	\;{\bf{a}}_{\rm{R}}({r_k},{\theta_k}) = \left[ {{e^{j\left( {{\omega _k}{\varepsilon _m} + {\psi _k}\varepsilon _m^2} \right)}}} \right], m=1,...,M_r,
	\end{aligned}
	\label{steering_near_appro}
\end{equation}
where ${\omega _k} =  - \frac{{2\pi {d}\sin {\theta_k}}}{\lambda }$ and ${\psi _k} = \frac{{\pi d^2{{\cos }^2}{\theta_k}}}{{\lambda {r_k}}}$. Note that in this model $\omega_k$ only depends on $\theta_k$, while $\psi_k$ depends on both  $\theta_k$ and $r_k$. The quadratic phase model of (\ref{steering_near_appro}) is referred to as the Fresnel approximation for near-field sources.

In Sections \mbox{\ref{FF_Sensing}} and \mbox{\ref{near-field}}, we will present typical parameter estimation algorithms applicable to the four methods presented in \mbox{Fig.~\ref{PE}}, first for far-field and then for near-field scenarios. 

\subsection{Prototype Design and Hardware Requirements} \label{hardwares}

In this section, we first introduce a typical MIMO-OFDM ISAC prototype system.  Then, we elaborate on the fundamental hardware requirements and key parameters of ISAC prototype designs, including servers, development software, field programmable gate array (FPGA) and radio frequency (RF) development kits, and MIMO array design.

\subsubsection{Monostatic MIMO-OFDM ISAC Prototype} 
\begin{figure}[!t]
  \centering
  \includegraphics[width=1\linewidth]{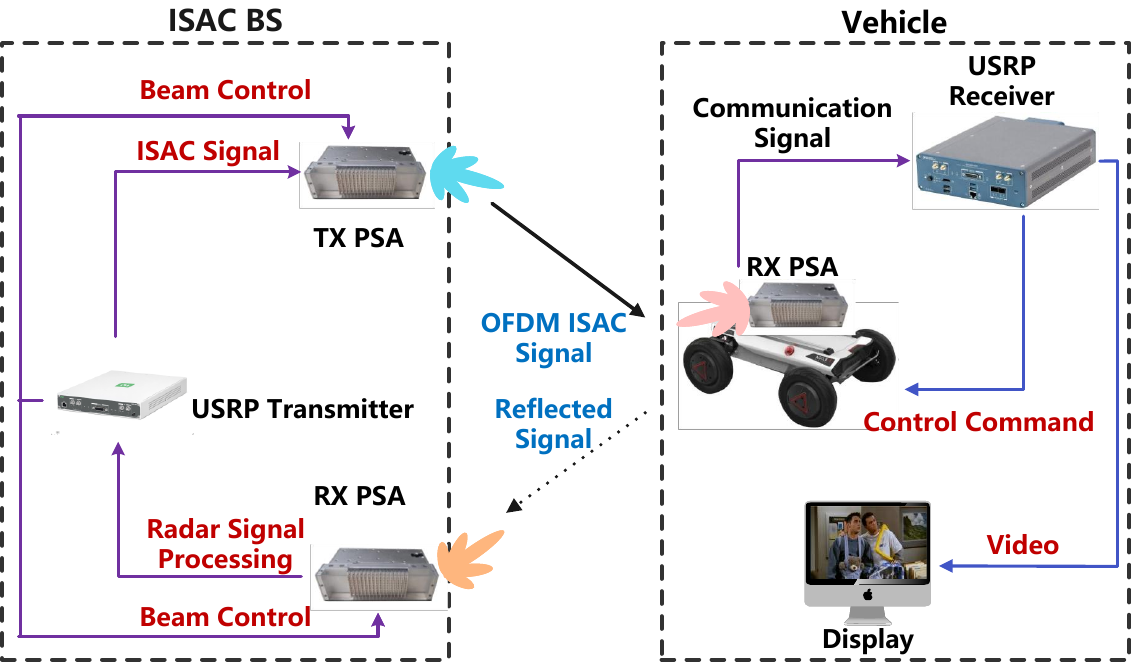}
  \caption{Monostatic MIMO-OFDM ISAC prototype system architecture~\cite{10233570}.}
  \label{prototype}
  \vspace{-0.6cm}
\end{figure}

The authors in \cite{10233570} describe a mmWave MIMO-OFDM monostatic ISAC prototype that they designed for internet of vehicles (IoV) applications. The ISAC BS provides downlink multimedia data to vehicles, while simultaneously utilizing the OFDM echoes to sense the position and velocity of the vehicles and surrounding obstacles to provide obstacle avoidance services. As shown in Fig.~\ref{prototype}, the ISAC BS is equipped with one universal software radio peripheral (USRP) and two mmWave phased arrays (PSAs) that function as ISAC-TX and Sen-RX, respectively. The vehicle is equipped with one USRP and a PSA, serving as Com-RX. At the ISAC BS, the USRP is responsible for generating OFDM ISAC signals, beam and vehicle control signaling, and sensing signal processing, while the two PSAs separately handle signal transmission and reception. On the vehicle side, the PSA receives communication signals, and the USRP performs communication signal processing, demodulates the video data stream and control commands, and forwards them to the vehicle systems. In the following, we discuss the fundamental hardware requirements and critical parameters for ISAC prototype development based on this system.

\begin{figure*}[!t]
  \centering
  \includegraphics[width=0.9\linewidth]{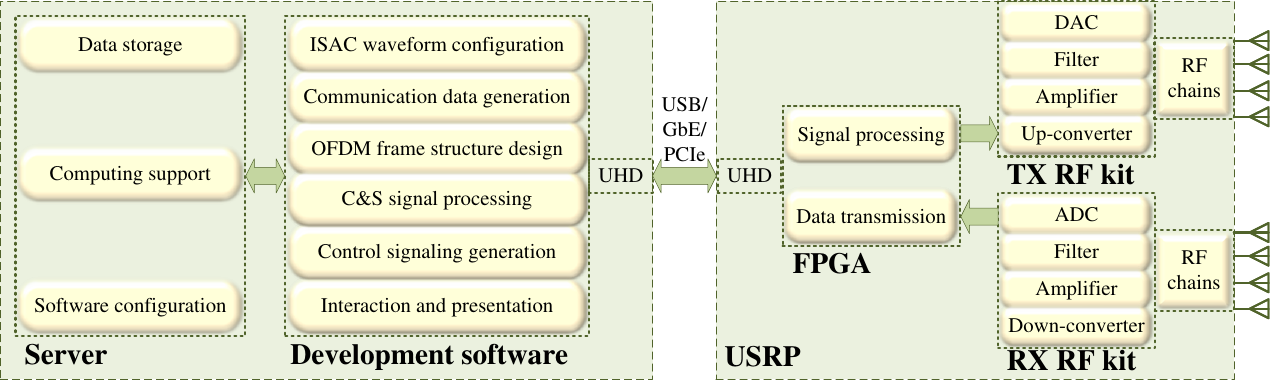}
  \caption{Hardware architecture of ISAC transceiver terminal.}
  \label{hardware}
  \vspace{-0.4cm}
\end{figure*}

\subsubsection{Hardware Requirements and Key Parameters} The basic architecture of the BS and UE prototypes is illustrated in Fig.~\ref{hardware}. Depending on their distinct roles in different ISAC modes, such as ISAC-TX, Com-RX, and Sen-RX, their transceiver arrays can operate either simultaneously or separately. For instance, as depicted in Fig.~\ref{prototype}, the transmit and receive arrays of the ISAC BS operate concurrently, while only the receive array is active at the UE. Next, we introduce the hardware components and their key parameters.

\textbf{Servers:} Servers are employed to fulfill data storage requirements, computing power support, software configuration and updates. The hardware configuration of the host, including the central processing unit (CPU), random access memory (RAM), and solid state drive (SSD), significantly impacts the operational efficiency, data processing, and storage capabilities of the experimental system. High-performance PCs generally feature a relatively large physical size, rendering them unsuitable for deployment on compact devices such as drones or mobile terminals. An effective solution is to deploy cloud servers for computing and storage, thereby enhancing computing power and storage resources. However, the latency induced by data upload and result retrieval may compromise the real-time performance of the system.

\textbf{Development software:} Development software comprises various programming tools designed to drive hardware operations, enabling core functionalities such as system parameter configuration, ISAC waveform parameter setup, C\&S signal processing, human-machine interaction, and result visualization. Specifically, the Verilog programming language facilitates the implementation of signal processing functionalities on FPGA chips, enabling high-efficiency and low-power processing. This approach enhances both the efficiency and real-time performance of the experimental system, albeit at the cost of high workload and implementation complexity.  
LabVIEW and GNU Radio are two graphical programming development environments that incorporate extensive libraries of functional modules tailored for communication and radar signal processing, effectively reducing the development challenges of experimental systems. LabVIEW is a commercial software platform developed by National Instruments (NI), while GNU Radio is an open-source software framework.  
In the prototype system illustrated in Fig.~\ref{prototype}, the authors implemented the core OFDM signal processing flow shown in Fig.~\ref{OFDMp} via the FPGAs, while more complex sensing signal processing was handled by LabVIEW running on the server. This strategy effectively balances system efficiency with development complexity.

 \textbf{FPGA and RF development kits:} The transceiver hardware primarily comprises FPGA chips for signal and data processing, along with various RF modules such as digital-to-analog converters (DAC), analog-to-digital converters (ADC), filters, amplifiers, and up/down converters. For prototype system development, commercial software-defined radio (SDR) platforms can be used, such as the USRP product series from NI. Several USRP hardware parameters are particularly critical for ISAC performance, including carrier frequency range, maximum bandwidth, and the number of TX/RX RF chains. Most current USRP products cover the sub-6 GHz frequency band, while experimental validation for higher frequency bands, such as mmWave and THz, requires additional high-frequency up/down converters. The system bandwidth supported by USRP restricts the communication rate and delay resolution of ISAC prototype systems. Finally, the number of USRP TX/RX RF chains is generally limited. For experimental systems requiring more RF chains, such as massive MIMO or XL-MIMO ISAC systems, multiple USRPs need to be jointly utilized with data aggregation and synchronization. Additionally, the server must be equipped with the USRP hardware driver (UHD) to facilitate data transmission and control signaling interaction with the USRP via interfaces such as USB, Ethernet, or PCIe.

\textbf{MIMO array design:} The antenna array is another critical hardware component in MIMO-OFDM ISAC systems, determining the near-field sensing region and spatial sensing resolution. Depending on the implementation methods used, MIMO arrays can be categorized as digital MIMO arrays, phased MIMO arrays, movable antenna arrays \cite{6DMA}, switched arrays, and ray antenna arrays (RAAs) \cite{raa1,raa2}. 
For a digital MIMO array, each antenna is connected to an individual RF chain, enabling precise digital beamforming at both transmitter and receiver. It supports diversity transmission at the transmitter and spatial sensing through array signal processing at the receiver. The theory described in this paper is based on the assumption that digital MIMO arrays are deployed. For large-scale or extremely large-scale arrays, the number of required RF chains increases correspondingly, leading to a sharp rise in hardware costs.
Phased MIMO arrays significantly reduce the number of required RF chains, down to a minimum of only one RF chain, but require a dedicated phase shifter for each antenna. Phased MIMO arrays usually rely on hybrid digital-analog or analog beamforming for signal transmission and reception, and beam scanning replaces the spatial signal processing shown in Fig.~\ref{PE}. The analog beamforming precision is limited by the phase shifters, resulting in degraded C\&S performance compared to digital MIMO arrays, albeit at a lower hardware cost.
Movable antenna arrays and switched arrays are designed based on the principles of synthetic aperture radar (SAR) \cite{SAR}. Movable antenna arrays rely on physical movement of antennas, while switched arrays use RF switches to achieve high-speed switching between array elements. Both approaches can achieve extremely large-scale aperture sensing at very low cost. However, because such architectures require time-division signal transmission and reception, they are only suitable for low-mobility scenarios and exhibit limited Doppler resolution and beamforming gain. 
The RAA is composed of multiple simple uniform linear arrays (ULAs) whose elements are connected to form a fixed beam at the array boresight without adaptive digital/analog beamforming, thus significantly reducing the hardware cost. Through careful design of the ULA orientations, this approach can achieve uniform angular resolution and enable the use of antenna elements with higher directionality, thus demonstrating considerable potential for C\&S applications. However, potential shadowing and mutual coupling effects among the ULAs may arise, necessitating specialized designs to mitigate these issues.
Furthermore, antennas and arrays in the sub-6G band are technologically mature and cost-effective, whereas higher-frequency antennas and arrays are considerably more expensive. Consequently, some research efforts are currently focused on developing low-cost and low-power high-frequency antenna arrays, such as holographic radar arrays \cite{Holographic0,Holographic1,NGAT} and reconfigurable antenna arrays \cite{Reconfigurable0,Reconfigurable1}.

\subsection{Lessons Learned}
The model depicted in Fig.~\ref{general model} provides a unified representation for the six commonly used ISAC modes illustrated in Fig.~\ref{ISAC_modes}. This unified model helps simplify the C\&S signal model derivation and signal processing workflow design, while also offering a foundational framework for ISAC system design and prototype validation. However, different ISAC modes have distinct channel characteristics. For instance, compared to UE-side sensing, an ISAC BS is more likely to be equipped with massive or XL-MIMO arrays. Consequently, targets in BS-side sensing are more likely located in the near-field region, necessitating the consideration of spherical wave propagation models and near-field spatial signal processing. In terms of coverage, monostatic sensing typically exhibits circular or spherical coverage patterns, whereas bistatic sensing generally follows Cassini oval curves or surfaces \cite{beamforming7,coverage1}. Finally, for the same target, the RCS may differ between monostatic and bistatic sensing due to variations in target scattering characteristics. Thus, single-mode ISAC may lead to detection blind spots or poor sensing performance, whereas multi-mode collaborative ISAC can enhance target detection accuracy and parameter estimation performance \cite{ding2025}.

For prototype development, beyond the fundamental hardware requirements discussed in Section~\ref{hardwares}, it is essential to design the ISAC system architecture as shown in Fig.~\ref{prototype} based on application scenarios and functional requirements. This involves configuring hardware parameters and time-frequency resources, designing signal processing workflows, and feedback mechanisms for upper-layer applications. The authors in \cite{ISAClayer} proposed a five-layer architecture to address the requirements and functions of ISAC systems at various levels, encompassing the terminal and signal layer, resource layer, function layer, application layer, and computation layer, and they conducted survey on existing ISAC prototype validation efforts. In general, significant progress has been made in the development of monostatic\cite{liu2019implementation,10233570,zhang2023integrated}, bistatic \cite{dd29}, and cell-free multistatic \cite{cfmimo} MIMO-OFDM ISAC prototypes, as well as in OFDM-based simultaneous localization and mapping (SLAM) \cite{yang2023multi}, SAR imaging \cite{8718390}, and ISAC-enabled CKM \cite{zhang2024prototyping} systems. However, most of these efforts focus on the far-field case, while near-field scenarios involving massive MIMO or XL-MIMO require further research into corresponding prototype designs and experimental validations.

\section{Far-field MIMO-OFDM ISAC}\label{FF_Sensing}
In this section, we elaborate on how to sense far-field targets based on the signal processing methods illustrated in Fig.~\ref{PE}, demonstrate the application of various standard radar algorithms under MIMO-OFDM ISAC, and analyze their complexity and performance.
\subsection{Far-field Sensing Signal Model}
For ease of exposition, we assume that the Sen-RX is equipped with a ULA with half-wavelength inter-antenna spacing. Setting the first element as the reference element, the position of the $m$-th element can be obtained as
\begin{equation}
	\setlength\abovedisplayskip{2pt}
	\setlength\belowdisplayskip{2pt}
	\begin{aligned}
        \varepsilon_{m}d=(m-1)d, m=1,\dots,M_r.
        \end{aligned}
	\label{29}
\end{equation}
Substituting \mbox{(\ref{29})} into \mbox{(\ref{steering_far})}, the steering vector $\mathbf{a}_\mathrm{R}(\mathbf{q}_k)=\mathbf{a}_\mathrm{R}(\theta_k) \in \mathbb{C}^{M_r\times 1}$ of the $k$-th target can be written as
\begin{equation}
	\setlength\abovedisplayskip{2pt}
	\setlength\belowdisplayskip{2pt}
	\begin{aligned}
		\mathbf{a}_\mathrm{R}\left(\theta_k\right)=\begin{bmatrix}1,e^{j2\pi \frac{d}{\lambda} \text{sin}\theta_k},\cdots,e^{j2\pi (M_r-1)\frac{d}{\lambda} \text{sin}\theta_k}\end{bmatrix}^H,
	\end{aligned}
	\label{doast}
\end{equation}
and the steering matrix $\mathbf{A}_{\theta} \in \mathbb{C}^{M_r\times K}$ is defined as
\begin{equation}
	\setlength\abovedisplayskip{2pt}
	\setlength\belowdisplayskip{2pt}
	\begin{aligned}
        \mathbf{A}_{\theta}=\begin{bmatrix}\mathbf{a}_\mathrm{R}\begin{pmatrix}\theta_1\end{pmatrix}, \mathbf{a}_\mathrm{R}\begin{pmatrix}\theta_2\end{pmatrix},\cdots,\mathbf{a}_\mathrm{R}\begin{pmatrix}\theta_K\end{pmatrix}\end{bmatrix}.
	\end{aligned}
	\label{doasm}
\end{equation}
Substituting \mbox{(\ref{doast})} into \mbox{(\ref{d161})} and \mbox{(\ref{16})}, the resulting sensing data tensor $\mathbf{Y}_s$ for far-field MIMO-OFDM ISAC can be expressed as
\begin{equation}
	\setlength\abovedisplayskip{2pt}
	\setlength\belowdisplayskip{2pt}
	\begin{aligned}
        \mathbf{Y}_s(:,n,p)=&b_{n,p}\sum_{k=1}^K\overline{\alpha}_{k}^{n,p} \mathbf{a}_\mathrm{R}(\theta_k) e^{-j2\pi n\Delta f\tau_k}e^{j2\pi pT_s\upsilon_{k}}\\
        &\qquad\qquad\qquad\qquad\qquad\qquad\quad+\bar{\mathbf{z}}^p[n].
        \end{aligned}
	\label{32}
\end{equation}

For far-field MIMO-OFDM ISAC, the parameters to be estimated for each sensing target are the AoA $\theta_k$, propagation delay $\tau_k$, and Doppler frequency $\upsilon_k$. In the following, we will discuss how various classic radar algorithms for parameter estimation can be applied to far-field MIMO-OFDM ISAC systems, such as the IDFT/DFT and subspace-based methods. IDFT/DFT-based methods like the periodogram \cite{dd32} can be easily implemented, but the resolution is limited by the array aperture, signal bandwidth, and CPI, respectively. On the other hand, subspace-based methods such as MUSIC \cite{music} and ESPRIT \cite{esprit} can achieve super resolution but they usually require high computational complexity. Additionally, CS and sparse recovery methods such as OMP \cite{OMP3} are also commonly used for parameter estimation. Compared to IDFT/DFT and subspace-based methods, CS achieves higher accuracy under low SNR and sparse channel conditions but incurs higher complexity, especially with large signal dimensions, making it difficult to implement in real time. Moreover, in scenarios with dense targets or clutter, CS methods are prone to locally optimal or suboptimal solutions, hence degrading estimation performance. 

The signals to/from different users can be separated using orthogonal time-frequency resources and then processed individually in OFDM ISAC systems. In the following, for ease of exposition, we consider the case of a single user with collocated time-frequency allocation, namely, the transmit beamforming vectors $\mathbf{w}_{n,p}$ in \mbox{(\ref{nISAC-TXsm})} are identical for all subcarriers and symbols $\overline{\alpha}_{k}^{n,p}=\overline{\alpha}_{k}={\alpha}_{k}\mathbf{a}_T^H(\mathbf{q}_k)\mathbf{w}e^{j2\pi\upsilon_{k}T_{cp}}, \forall n=0,1,\cdots,N-1, p=0,1,\cdots,P-1$. As a result, the signal in \mbox{(\ref{32})} reduces to
\begin{equation}
	\setlength\abovedisplayskip{2pt}
	\setlength\belowdisplayskip{2pt}
	\begin{aligned}
        \mathbf{Y}_s(:,n,p)=&b_{n,p}\sum_{k=1}^K\overline{\alpha}_{k} \mathbf{a}_\mathrm{R}(\theta_k) e^{-j2\pi n\Delta f\tau_k}e^{j2\pi pT_s\upsilon_{k}}\\
        &\qquad\qquad\qquad\qquad\qquad\qquad\quad+\bar{\mathbf{z}}^p[n].
        \end{aligned}
	\label{22}
\end{equation}

Based on the signal model in \mbox{(\ref{22})}, we detail below the various far-field MIMO-OFDM ISAC signal processing methods illustrated in \mbox{Fig.~\ref{PE}}.

\subsection{Decoupled Parameter Estimation}\label{far-doa}
Since the phase variations of the sensing signal in \mbox{(\ref{22})} in the angle, delay, and Doppler domains are completely decoupled, the three parameters $(\theta_k, \tau_k, \upsilon_k)$ can be estimated separately or sequentially using the parallel and sequential one-domain parameter estimation methods outlined in Fig.~\ref{PE}.

\subsubsection{Parallel One-domain Parameter Estimation}\label{parallel 1D}\ \\
 \indent \textbf{AoA estimation:} In far-field scenarios, the spatial domain processing can only provide AoA information for each target. In this case, the resulting OFDM sensing data in the subcarrier and symbol domains can be treated as snapshots. As such, the $M_r \times N \times P$ dimensional signal $\mathbf{Y}_s$ in \mbox{(\ref{22})} can be re-organized as an $M_r\times Q_{N,P}$ dimensional matrix $\mathbf{X}_{\theta}$, with $Q_{N,P}=NP$, where $\mathbf{X}_{\theta}(:,q_{n,p})=\mathbf{Y}_s(:,n,p),q_{n,p}=n+pN,  n\in[0,N-1],p\in[0,P-1]$. Thus, $\mathbf{X}_{\theta}$ can be equivalently expressed in matrix form as 
 \begin{equation}
	\setlength\abovedisplayskip{2pt}
	\setlength\belowdisplayskip{2pt}
	\begin{aligned}
	\mathbf{X}_{\theta}=\mathbf{A}_{\theta}\mathbf{S}_{\theta}+\mathbf{Z}_{\theta},
	\end{aligned}
	\label{23}
\end{equation}
 where $\mathbf{A}_{\theta} \in \mathbb{C}^{M_r \times K}$ is the steering matrix given in \mbox{(\ref{doasm})}, $\mathbf{Z}_{\theta}$ is the resulting noise, and $\mathbf{S}_{\theta} \in \mathbb{C}^{K \times Q_{N,P}}$ is given by
\begin{equation}
	\setlength\abovedisplayskip{2pt}
	\setlength\belowdisplayskip{2pt}
	\begin{aligned}
	\mathbf{S}_{\theta}(k,q_{n,p})=b_{n,p}\overline{\alpha}_{k}e^{-j2\pi n\Delta f\tau_k}e^{j2\pi pT_s\upsilon_{k}}.
	\end{aligned}
	\label{34}
\end{equation}
The AoA estimation problem corresponds to estimating the angles $\mathbb{S}_{\theta}=\{\theta_1,\cdots,\theta_K\}$ of the $K$ sensing targets embedded in steering matrix $\mathbf{A}_{\theta}$, given data matrix $\mathbf{X}_{\theta}$. Specific algorithms to address this problem will be presented in \mbox{Section~\ref{1d algorithm}}. 

\textbf{Delay estimation:} The random information-bearing symbols $b_{n,p}$ in \mbox{(\ref{22})} depend only on subcarrier index $n$ and symbol index $p$, not on antenna index $m$. They affect the manifold structure in the subcarrier and symbol domains, but not the spatial domain. Thus, unlike AoA estimation in \mbox{(\ref{23})}, to ensure the manifold structure in the subcarrier and symbol domains, the information-bearing symbols $b_{n,p}$ in \mbox{(\ref{22})} need to be removed. The transmitted data is known a priori at the Sen-RX in the monostatic case, or it can be demodulated and detected in the bistatic ISAC mode. The communication symbols can be removed by performing element-wise division on the received signal in \mbox{(\ref{22})}, i.e.,
\begin{equation}
  \begin{aligned} \label{d25}
	\setlength\abovedisplayskip{2pt}
	\setlength\belowdisplayskip{2pt}
        &\overline{\mathbf{Y}}_s(:,n,p)=\frac{\mathbf{Y}_s(:,n,p)}{b_{n,p}}\\
        &=\sum_{k=1}^K\overline{\alpha}_{k} \mathbf{a}_\mathrm{R}(\theta_k) e^{-j2\pi n\Delta f\tau_k}e^{j2\pi pT_s\upsilon_{k}} +\overline{\mathbf{Z}}(:,n,p), 
\end{aligned}  
\end{equation}
where $\overline{\mathbf{Z}}(:,n,p)=\frac{\bar{\mathbf{z}}^p[n]}{b_{n,p}}$ is the resulting noise.

For delay estimation, the data in the antenna and symbol domains can be treated as snapshots. As such, the $M_r \times N \times P$ dimensional data matrix after element-wise symbol division in \mbox{(\ref{d25})} can be restructured into an $N \times Q_{M_r,P}$ dimensional matrix $\mathbf{X}_{\tau}$, with $Q_{M_r,P}=M_rP$, where $\mathbf{X}_{\tau}(:,q_{m,p})=\overline{\mathbf{Y}}_s(m,:,p),q_{m,p}=m+pM_r, m\in[0,M_r-1],p\in[0,P-1]$. Then, matrix $\mathbf{X}_{\tau}$ can be expressed as
 \begin{equation}
	\setlength\abovedisplayskip{2pt}
	\setlength\belowdisplayskip{2pt}
	\begin{aligned}
	\mathbf{X}_{\tau}=\mathbf{A}_{\tau}\mathbf{S}_{\tau}+\mathbf{Z}_{\tau},
	\end{aligned}
	\label{d28}
\end{equation}
where $\mathbf{A}_{\tau} \in \mathbb{C}^{N\times K}$ is the steering matrix in the subcarrier domain
\begin{equation}
	\setlength\abovedisplayskip{2pt}
	\setlength\belowdisplayskip{2pt}
	\begin{aligned}
        \mathbf{A}_{\tau}=\left[\mathbf{a}_{\tau}(\tau_1),\cdots,\mathbf{a}_{\tau}(\tau_K)\right],
	\end{aligned}
	\label{Rsm}
\end{equation}
and $\mathbf{a}_{\tau}(\tau_k) \in \mathbb{C}^{N\times 1}$ is the steering vector in the delay domain
\begin{equation}
	\setlength\abovedisplayskip{2pt}
	\setlength\belowdisplayskip{2pt}
	\begin{aligned}
        \mathbf{a}_\tau\left(\tau_k\right)=\left[1,e^{j2\pi\Delta f\tau_k},\cdots,e^{j2\pi(N-1)\Delta f\tau_k}\right]^H.
	\end{aligned}
	\label{Rsv}
\end{equation}
The $(k,q_{m,p})$-th element of $\mathbf{S}_{\tau} \in \mathbb{C}^{K \times Q_{M_r,P}}$ is
\begin{equation}
	\setlength\abovedisplayskip{2pt}
	\setlength\belowdisplayskip{2pt}
	\begin{aligned}
        \mathbf{S}_{\tau}(k,q_{m,p})=\overline{\alpha}_{k}e^{-j2\pi m\frac{d}{\lambda} \sin (\theta_k)}e^{j2\pi pT_s\upsilon_{k}}.
	\end{aligned}
	\label{d3400}
\end{equation}
For the model in (\ref{d28}), the goal is to estimate the propagation delays $\mathbb{S}_\tau=\{\tau_1, \cdots, \tau_K\}$ of the $K$ sensing targets embedded in steering matrix $\mathbf{A}_{\tau}$, given data matrix $\mathbf{X}_{\tau}$ in \mbox{(\ref{d28})}.

\textbf{Doppler estimation:} For Doppler estimation, the data in the antenna and subcarrier domains can be treated as snapshots. The signal after element-wise symbol division in \mbox{(\ref{d25})} can be restructured into a $P \times Q_{M_r,N}$ dimensional  matrix $\mathbf{X}_{\upsilon}$, with $Q_{M_r,N}=M_rN$, where $\mathbf{X}_{\upsilon}(:,q_{m,n})={\overline{\mathbf{Y}}_s(m,n,:)},q_{m,n}=m+nM_r$. Then, matrix $\mathbf{X}_{\upsilon}$ can be written as
 \begin{equation}
	\begin{aligned}
	\mathbf{X}_{\upsilon}=\mathbf{A}_{\upsilon}\mathbf{S}_{\upsilon}+\mathbf{Z}_{\upsilon},
	\end{aligned}
	\label{d35}
\end{equation}
where $\mathbf{Z}_{\upsilon}$ is the corresponding noise, $\mathbf{A}_{\upsilon} \in \mathbb{C}^{P\times K}$ is the steering matrix in the Doppler domain
\begin{equation}
	\setlength\abovedisplayskip{2pt}
	\setlength\belowdisplayskip{2pt}
	\begin{aligned}
        \mathbf{A}_{\upsilon}=\left[\mathbf{a}_{\upsilon}(\upsilon_1),\cdots,\mathbf{a}_{\upsilon}(\upsilon_K)\right],
	\end{aligned}
	\label{Dsm}
\end{equation}
and $\mathbf{a}_{\upsilon}(\upsilon_k) \in \mathbb{C}^{P\times 1}$ is the steering vector in the symbol domain
\begin{equation}
	\setlength\abovedisplayskip{2pt}
	\setlength\belowdisplayskip{2pt}
	\begin{aligned}
        \mathbf{a}_{\upsilon}\left(\upsilon_k\right)=\left[1,e^{-j2\pi T_s\upsilon_k},\cdots,e^{-j2\pi(P-1)T_s\upsilon_k}\right]^H.
	\end{aligned}
	\label{Dsv}
\end{equation}
The $(k,q_{m,n})$-th element of $\mathbf{S}_{\upsilon} \in \mathbb{C}^{K \times Q_{M_r,N}}$ is given by
\begin{equation}
	\setlength\abovedisplayskip{2pt}
	\setlength\belowdisplayskip{2pt}
	\begin{aligned}
        \mathbf{S}_{\upsilon}(k,q_{m,p})=\overline{\alpha}_{k}e^{-j2\pi m\frac{d}{\lambda} \sin (\theta_k)}e^{-j2\pi n \Delta f \tau_{k}}.
	\end{aligned}
	\label{d36}
\end{equation}
For the model in \mbox{(\ref{d35})}, the goal is to estimate the Doppler frequencies $\mathbb{S}_{\upsilon}=\{\upsilon_1,\cdots,\upsilon_K\}$ of the $K$ sensing targets based on steering matrix $\mathbf{A}_{\upsilon}$, given data matrix $\mathbf{X}_{\upsilon}$ in \mbox{(\ref{d35})}.

After estimating the AoAs $\mathbb{S}_{\hat{\theta}}=\{\hat{\theta}_1,\cdots,\hat{\theta}_{K_{\hat{\theta}}}\}$, delays $\mathbb{S}_{\hat{\tau}}=\{\hat{\tau}_1,\cdots,\hat{\tau}_{K_{\hat{\tau}}}\}$, and Doppler frequencies $\mathbb{S}_{\hat{\upsilon}}=\{\hat{\upsilon}_1,\cdots,\hat{\upsilon}_{K_{\hat{\upsilon}}}\}$ separately via parallel one-domain parameter estimation, the individual estimation results have to be associated with the corresponding targets. In general, $K_{\hat{\theta}}, K_{\hat{\tau}}, 
K_{\hat{\upsilon}}\leq K$, since the targets are not necessarily distinguishable in all three domains. Two common parameter grouping methods are based on signal correlation and power detection, which first group the parameters estimated from the three domains one by one. 

There are a total of $K_{\hat{\theta}} K_{\hat{\tau}} K_{\hat{\upsilon}}$ possible parameter groups, indicated by $\left(\hat{\theta}_{k_{\hat{\theta}}},\hat{\tau}_{k_{\hat{\tau}}},\hat{\upsilon}_{k_{\hat{\upsilon}}}\right),\forall k_{\hat{\theta}}=1,\cdots,K_{\hat{\theta}},k_{\hat{\tau}}=1,\cdots,K_{\hat{\tau}},k_{\hat{\upsilon}}=1,\cdots,K_{\hat{\upsilon}}$. 
In the signal correlation method, the reconstructed signal $\hat{\mathbf{Y}}_{k_{\hat{\theta}},k_{\hat{\tau}},k_{\hat{\upsilon}}} \in \mathbb{C}^{M_r\times N \times P}$ for parameter group $\left(\hat{\theta}_{k_{\hat{\theta}}},\hat{\tau}_{k_{\hat{\tau}}},\hat{\upsilon}_{k_{\hat{\upsilon}}}\right)$ is expressed as
\begin{equation}
	\setlength\abovedisplayskip{2pt}
	\setlength\belowdisplayskip{2pt}
	\begin{aligned}
        \hat{\mathbf{Y}}(m,n,p)=e^{-j2\pi m\frac{d}{\lambda}\sin{\hat{\theta}_{k_{\hat{\theta}}}}} e^{-j2\pi n\Delta f\hat{\tau}_{k_{\hat{\tau}}}}e^{j2\pi pT_s\hat{\upsilon}_{k_{\hat{\upsilon}}}},
        \end{aligned}
	\label{rspp}
\end{equation}
where $m\in[0,M_r-1],n\in[0,N-1],p\in[0,P-1]$. Then, the correlation coefficient $\hat{C}$ between $\hat{\mathbf{Y}}$ and $\overline{\mathbf{Y}}_s$ in \mbox{(\ref{d25})} is calculated as
\begin{equation}
	\setlength\abovedisplayskip{2pt}
	\setlength\belowdisplayskip{2pt}
	\begin{aligned}
        \hat{C}=\left|\mathrm{mean}\left\{\hat{\mathbf{Y}}^{*} \odot \overline{\mathbf{Y}}_s\right\}\right|.
        \end{aligned}
	\label{correlation}
\end{equation}
The correct parameter groups will exhibit a strong correlation between the reconstructed signal and signal $\overline{\mathbf{Y}}_s$. Thus, the $K$ parameter pairs $\left(\hat{\theta}_{k_{\hat{\theta}}},\hat{\tau}_{k_{\hat{\tau}}},\hat{\upsilon}_{k_{\hat{\upsilon}}}\right)$ with the highest correlation coefficient $\hat{C}$ are assumed to correspond to the $K$ targets.

For power based detection, the power of $\overline{\mathbf{Y}}_s$ in \mbox{(\ref{d25})} in the angle-delay-Doppler domain at the location of each parameter group $\left(\hat{\theta}_{k_{\hat{\theta}}},\hat{\tau}_{k_{\hat{\tau}}},\hat{\upsilon}_{k_{\hat{\upsilon}}}\right)$ can be obtained as
\begin{equation}
	\setlength\abovedisplayskip{2pt}
	\setlength\belowdisplayskip{2pt}
	\begin{aligned}
\hat{P}= \frac{1}{M_rNP}\left|\mathbf{a}_{\tau}^H(\hat{\tau}_{k_{\hat{\tau}}})\left(\mathbf{a}_{\mathrm{R}}^H(\hat{\theta}_{k_{\hat{\theta}}})\overline{\mathbf{Y}}_s\right)\mathbf{a}_{\upsilon}^*(\hat{\upsilon}_{k_{\hat{\upsilon}}}) \right|^2,
        \end{aligned}
	\label{ppp}
\end{equation}
where $\mathbf{a}_{\mathrm{R}}(\hat{\theta}_{k_{\hat{\theta}}})$, $\mathbf{a}_{\tau}(\hat{\tau}_{k_{\hat{\tau}}})$, and $\mathbf{a}_{\upsilon}(\hat{\upsilon}_{k_{\hat{\upsilon}}})$ are the steering vectors in the spatial, delay, and Doppler domains in \mbox{(\ref{doast})}, \mbox{(\ref{Rsv})}, and \mbox{(\ref{Dsv})}, respectively. The $K$ parameter groups $\left(\hat{\theta}_{k_{\hat{\theta}}},\hat{\tau}_{k_{\hat{\tau}}},\hat{\upsilon}_{k_{\hat{\upsilon}}}\right)$ with the highest power $\hat{P}$ are taken to be the parameters of the $K$ sensing targets. 

\subsubsection{Sequential One-domain Parameter Estimation} \ \\
\indent The sequential one-domain parameter estimation method, as outlined in \mbox{Fig.~\ref{PE}}, sequentially estimates the target parameters over the three domains. The approach typically begins in the domain with the highest resolution so that more targets can be distinguished. Consider a case in which the array has many elements, while the bandwidth is narrow and the CPI is short. In this case, the angular resolution is highest, followed by the delay and Doppler resolutions, so it is best to first estimate the AoA of the sensing targets using $\mathbf{X}_{\theta}$ in \mbox{(\ref{23})}, and then use beamforming to extract the target reflections at different angles. Then, the delays are estimated for targets at each angle. Since there may be multiple targets at approximately the same angle, beamforming (matched filtering) can be applied in the delay domain to further distinguish targets with different delays. Finally, the corresponding Doppler frequency is estimated for targets at each delay. Thus, the target parameters in the three domains are directly grouped in this sequential method.

More specifically, assume that a total of $K_{\hat{\theta}}$ AoAs are estimated as $\hat{\theta}_{k_{\hat{\theta}}} , k_{\hat{\theta}}=1,\cdots,K_{\hat{\theta}}$, based on \mbox{(\ref{23})}, where $K_{\hat{\theta}} \leq K$ since not all targets are distinguishable in the spatial domain. Let $N_{k_{\hat{\theta}}} \geq 1$ denote the number of targets at each estimated angle and $K=\sum_{{k_{\hat{\theta}}}=1}^{K_{\hat{\theta}}} N_{k_{\hat{\theta}}}$. Before performing delay domain estimation, we first apply beamforming based on the estimated AoAs in order to exploit the beamforming gain and decouple the signals in the spatial domain. After applying beamforming to \mbox{(\ref{d25})}, the resulting signal $\mathbf{X}_{k_{\hat{\theta}}} \in \mathbb{C}^{N\times P}, k_{\hat{\theta}}=1,\cdots,K_{\hat{\theta}}$, is 
\begin{align} \label{d37}
	\setlength\abovedisplayskip{2pt}
	\setlength\belowdisplayskip{2pt}
        &{\mathbf{X}}_{k_{\hat{\theta}}}{(n,p)} =\mathbf{r}^H(\hat{\theta}_{k_{\hat{\theta}}})\overline{\mathbf{Y}}_s(:,n,p) \nonumber \\
        &=\sum_{k\in \Omega_{k_{\hat{\theta}}}}\overline{\alpha}_{k} \mathbf{r}^H(\hat{\theta}_{k_{\hat{\theta}}}) \mathbf{a}_\mathrm{R}(\theta_{k}) e^{-j2\pi n\Delta f\tau_{k}}e^{j2\pi pT_s\upsilon_{{k}}} \nonumber \\
        &+\sum_{k\notin \Omega_{k_{\hat{\theta}}}}\overline{\alpha}_{k} \mathbf{r}^H(\hat{\theta}_{k_{\hat{\theta}}}) \mathbf{a}_\mathrm{R}(\theta_k) e^{-j2\pi n\Delta f\tau_k}e^{j2\pi pT_s\upsilon_{k}} +\overline{z}\left(q_{n,p}\right) 
 \nonumber \\ 
        & \approx \sum_{i=1}^{N_{k_{\hat{\theta}}}}\tilde{\alpha}_{i,{k_{\hat{\theta}}}} e^{-j2\pi n\Delta f\tau_{i,{k_{\hat{\theta}}}}}e^{j2\pi pT_s\upsilon_{i,{k_{\hat{\theta}}}}} + \overline{z}\left(q_{n,p}\right), 
\end{align}
where $\Omega_{k_{\hat{\theta}}}=\left\{k|\theta_k\approx\hat{\theta}_{k_{\hat{\theta}}} \right\}$, $\tilde{\alpha}_{i,{k_{\hat{\theta}}}}= \overline{\alpha}_{i,{k_{\hat{\theta}}}}\mathbf{r}^H(\hat{\theta}_{k_{\hat{\theta}}}) \mathbf{a}_\mathrm{R}(\theta_{i,{k_{\hat{\theta}}}})$ is the resulting channel gain, $\tau_{i,{k_{\hat{\theta}}}}$ and $\upsilon_{i,{k_{\hat{\theta}}}}$ are the delay and Doppler of the $i$-th target at the estimated angle $\hat{\theta}_{k_{\hat{\theta}}}$, and $\mathbf{r}\Big(\hat{\theta}_{k_{\hat{\theta}}}\Big)$ is the beamforming vector for the estimated angle $\hat{\theta}_{k_{\hat{\theta}}}$. 

Classical beamforming methods that can be used in this approach include maximum-ratio combining (MRC), zero-forcing (ZF), and minimum mean-square error (MMSE) beamforming. The expressions for the corresponding beamforming vectors are given by
\begin{equation}
	\setlength\abovedisplayskip{2pt}
	\setlength\belowdisplayskip{2pt}
	\begin{aligned}  
        \mathbf{r}_{\mathrm{MRC}}\Big(\hat{\theta}_{{k_{\hat{\theta}}}}\Big)&\triangleq \frac{\mathbf{a}_{\mathrm{R}}\left(\hat{\theta}_{{k_{\hat{\theta}}}}\right)}{\left\|\mathbf{a}_{\mathrm{R}}\left(\hat{\theta}_{{k_{\hat{\theta}}}}\right)\right\|},\\
        \mathbf{r}_{\mathrm{ZF}}\Big(\hat{\theta}_{{k_{\hat{\theta}}}}\Big)&\triangleq \frac{\widehat{\mathbf{A}}_{{k_{\hat{\theta}}}}\mathbf{a}_{\mathrm{R}}\left(\hat{\theta}_{{k_{\hat{\theta}}}}\right)}{\left\|\widehat{\mathbf{A}}_{{k_{\hat{\theta}}}}\mathbf{a}_{\mathrm{R}}\left(\hat{\theta}_{{k_{\hat{\theta}}}}\right)\right\|}, \\
        \mathbf{r}_{\mathrm{MMSE}}\Big(\hat{\theta}_{{k_{\hat{\theta}}}}\Big)&\triangleq \frac{\mathbf{C}_{{k_{\hat{\theta}}}}^{-1}\mathbf{a}_{\mathrm{R}}\left(\hat{\theta}_{{k_{\hat{\theta}}}}\right)}{\left\|\mathbf{C}_{{k_{\hat{\theta}}}}^{-1}\mathbf{a}_{\mathrm{R}}\left(\hat{\theta}_{{k_{\hat{\theta}}}}\right)\right\|},
        \end{aligned}
	\label{d37-1}
\end{equation}
where $\widehat{\mathbf{A}}_{{k_{\hat{\theta}}}} \triangleq \mathbf{I}_{M_r}-\mathbf{A}_{k_{\hat{\theta}}}\left(\mathbf{A}_{k_{\hat{\theta}}}^H\mathbf{A}_{k_{\hat{\theta}}}\right)^{-1}\mathbf{A}_{k_{\hat{\theta}}}^H$, $\mathbf{A}_{{k_{\hat{\theta}}}} =\left[\mathbf{a}_{\mathrm{R}}\left(\hat{\theta}_1\right),\cdots,\mathbf{a}_{\mathrm{R}}\left(\hat{\theta}_{{k_{\hat{\theta}}}-1}\right),\mathbf{a}_{\mathrm{R}}\left(\hat{\theta}_{{k_{\hat{\theta}}}+1}\right),\cdots,\mathbf{a}_{\mathrm{R}}\left(\hat{\theta}_{K_{\hat{\theta}}}\right)\right]$, and $\mathbf{C}_{{k_{\hat{\theta}}}}=\sum_{k\neq {k_{\hat{\theta}}}}^{K_{\hat{\theta}}}\frac{|\overline{\alpha}_{k}|^2}{\sigma^2}\mathbf{a}_{\mathrm{R}}\left(\hat{\theta}_k\right)\mathbf{a}_{\mathrm{R}}^H\left(\hat{\theta}_k\right)+\mathbf{I}_{M_r}$. In the delay and Doppler domains, similar beamformers can be employed.

The approximation in \mbox{(\ref{d37})} holds if different AoA clusters are sufficiently separated such that the residual signals from other angles can be considered as noise. Then, $\mathbf{X}_{k_{\hat{\theta}}}$ in \mbox{(\ref{d37})} can be expressed in matrix form as
\begin{equation}
	\setlength\abovedisplayskip{2pt}
	\setlength\belowdisplayskip{2pt}
	\begin{aligned}
        \mathbf{X}_{k_{\hat{\theta}}}=\mathbf{A}_{\tau,{k_{\hat{\theta}}}}\mathrm{diag}(\tilde{\mathbf{\alpha}})\mathbf{A}_{\upsilon,{k_{\hat{\theta}}}}^T+\mathbf{Z}_{k_{\hat{\theta}}},
	\end{aligned}
	\label{d38}
\end{equation}
where $\tilde{\mathbf{\alpha}}=[\tilde{\alpha}_1,\cdots,\tilde{\alpha}_{N_{k_{\hat{\theta}}}}]^T$ is the resulting vector of complex-valued gains, $\mathbf{A}_{\tau,{k_{\hat{\theta}}}}=[\mathbf{a}_{\tau}\begin{pmatrix}\tau_1\end{pmatrix},\cdots,\mathbf{a}_{\tau}(\tau_{N_{k_{\hat{\theta}}}})]$ and $\mathbf{A}_{\upsilon,{k_{\hat{\theta}}}}=[\mathbf{a}_{\upsilon}(\upsilon_1),\cdots,\mathbf{a}_{\upsilon}(\upsilon_{N_{k_{\hat{\theta}}}})]$ are the steering matrices in the delay and Doppler domains for the targets located at angle $\hat{\theta}_{k_{\hat{\theta}}}$. Then, the matrices $\mathbf{X}_{k_{\hat{\theta}}}, {k_{\hat{\theta}}}=1,\cdots, K_{\hat{\theta}}$, can be exploited to estimate the delays $\mathbb{S}_{\tau,{k_{\hat{\theta}}}}=\left\{\tau_{i,k_{\hat{\theta}}}|i=1,\cdots,{N_{k_{\hat{\theta}}}}\right\}$ of the targets at different angles. Furthermore, for targets with different delays at each angle, further target discrimination can be achieved through beamforming in the delay domain. In particular, the signal for a target whose AoA and delay are estimated as 
$(\hat{\theta}_{k_{\hat{\theta}}},\hat{\tau}_{i,{k_{\hat{\theta}}}})$ can be reduced to $\mathbf{x}_{i,{k_{\hat{\theta}}}}^T=\mathbf{r}^H\left(\hat{\tau}_{i,{k_{\hat{\theta}}}}\right)\mathbf{X}_{k_{\hat{\theta}}}$, $i=1,\cdots,N_{k_{\hat{\theta}}}$. Then, the Doppler $\hat{\upsilon}_{i,{k_{\hat{\theta}}}}$ can be estimated from ${\mathbf{x}}_{{k_{\hat{\theta}}}}\in \mathbb{C}^{P\times 1}$ for the target.

\subsubsection{One-dimensional Parameter Estimation Algorithms}\label{1d algorithm} \ \\
\indent In the previous subsections, we have modeled the one-dimensional (1D) parameter estimation problems in \mbox{(\ref{23})}, \mbox{(\ref{d28})}, \mbox{(\ref{d35})}, \mbox{(\ref{d38})} for AoA, delay, and Doppler estimation. It is not difficult to see that all these problems share a similar structure, and thus the problems can be solved using essentially the same algorithms. In this section, we provide examples of 1D AoA estimation algorithms that can be easily modified for delay and Doppler estimation.

\textbf{Periodogram:} For any snapshot $q_{n,p}$ of signal $\mathbf{X}_{\theta}\in \mathbb{C}^{M_r \times Q_{N,P}}$ in \mbox{(\ref{23})}, the noise-free power spectrum ${f}(\theta)$ in the angular domain can be obtained as
\begin{align} \label{resolution}
	\setlength\abovedisplayskip{2pt}
	\setlength\belowdisplayskip{2pt} 
{f}(\theta )&= \frac{1}{M_r}\bigg|\mathbf{a}_{\mathrm{R}}^H(\theta)\mathbf{X}_{\theta}(:,q_{n,p})\bigg|^2 \nonumber \\
&= \frac{1}{M_r}\bigg|b_{n,p}\sum_{k=1}^K\overline{\alpha}_{k} e^{-j2\pi n\Delta f\tau_k}e^{j2\pi pT_s\upsilon_{k}} \nonumber \\
&\qquad\qquad\quad \times \sum_{m=0}^{M_r-1}e^{j2\pi m \frac{d}{\lambda}(\sin{\theta}-\sin{\theta_k})}\bigg|^2\\
&= \frac{1}{M_r}\bigg|b_{n,p}\sum_{k=1}^K\overline{\alpha}_{k} e^{-j2\pi n\Delta f\tau_k}e^{j2\pi pT_s\upsilon_{k}} \nonumber \\
&\qquad\qquad\quad \times{\frac{{\sin \left( {\pi\frac{{ d}}{\lambda } M_r(\sin \theta  - \sin {\theta _k})} \right)}}{{\sin \left( {\pi\frac{{ d}}{\lambda } (\sin \theta  - \sin {\theta _k})} \right)}}}\bigg|^2 \nonumber \\
 &\mathop  \approx \limits^{(a)} \frac{{{\left| {b_{n,p}} \right|}^2}}{M_r}\sum_{k=1}^K{{{\left| {{\overline{\alpha}}_k} \right|}^2}}{{{\left| {\frac{{\sin \left( {\pi\frac{{ d}}{\lambda } M_r(\sin \theta  - \sin {\theta _k})} \right)}}{{\sin \left( {\pi\frac{{ d}}{\lambda } (\sin \theta  - \sin {\theta _k})} \right)}}} \right|}^2}}, \nonumber
\end{align}
where $\mathbf{a}_{\mathrm{R}}(\theta)$ is the steering vector in the spatial domain in \mbox{(\ref{doast})}, $\theta \in \left[-90^\circ,90^\circ \right]$ is the observation angle, and $\theta_k$ is the $k$-th target angle. The approximation $(a)$ holds when the angular difference between the targets is larger than the angular resolution of the array. The angular resolution of an array is based on the Rayleigh limit \cite{b93}, which corresponds to half the width between the null points of a focused beam of the array. From \mbox{(\ref{resolution})}, the null points for a beam pointed at $\theta _k$ can be obtained by $\pi \frac{{ d}}{\lambda } M_r(\sin {\theta}  - \sin {\theta _k})=\pm \pi$, so the beamwidth is given by
\begin{equation}
    \begin{aligned}
       {{\mathrm{BW}}}_{\theta}=\sin {\theta}  - \sin {\theta _k} = \frac{{2\lambda }}{{M_r d}} \approx \frac{{2\lambda }}{{D}},
    \end{aligned}
\end{equation}
where $D=M_r d$ denotes the array aperture. The maximum resolution is achieved when $\theta_k \rightarrow 0^\circ$, which implies
\begin{equation}
    \begin{aligned}
        \Delta \theta = \arcsin\left(\frac{{\lambda }}{{M_rd}}\right) \approx \frac{{\lambda }}{D}.
    \end{aligned}
    \label{doares}
\end{equation}
Thus, the angular resolution of a ULA is inversely proportional to the effective array aperture; as the effective array aperture increases, the beamwidth becomes narrower, thus enhancing the ability of the array to resolve different targets. 

Similarly, for delay and Doppler estimation, the power spectrum in the delay and Doppler domains can be derived from signals $\mathbf{X}_\tau$ and $\mathbf{X}_\upsilon$ in \mbox{(\ref{d28})} and \mbox{(\ref{d35})}, respectively. By calculating the corresponding null-to-null bandwidth, the minimum resolvable differences in the target delay and Doppler frequency are given by
\begin{equation}
    \begin{aligned}
        \Delta \tau = \frac{{1}}{N\Delta f} = \frac{1}{{B}}, \  \Delta \upsilon = \frac{1}{PT_s}=\frac{1}{\mathrm{CPI}},
    \end{aligned}
    \label{distance-resolution}
\end{equation}
respectively. Thus, the delay and Doppler resolution are inversely proportional to system bandwidth $B$ and CPI, respectively. In contrast to angular resolution, which depends on the target angle, the delay and Doppler resolution are independent of the target parameters.

In practice, calculating the power spectrum in the angular domain (see \mbox{(\ref{resolution})}) involves an exhaustive search over $\theta\in [-90^\circ, 90^\circ]$, resulting in high computational complexity. To address this issue, the periodogram algorithm discretizes the phase $\varphi_{\theta}=\frac{d}{\lambda}\sin{\theta} \in \left[-\frac{d}{\lambda},\frac{d}{\lambda}\right]$ in the angular domain into $N_{\theta}^{\mathrm{IFFT}}$ points and employs the IFFT to accelerate the computation. When $N_{\theta}^{\mathrm{IFFT}}>M_r$, zero-padding is utilized to convert $\mathbf{X}_{\theta} \in \mathbb{C}^{M_r\times Q_{N,P}}$ in \mbox{(\ref{23})} to $\mathbf{X}_{\theta}^{\mathrm{ZP}} \in \mathbb{C}^{N_{\theta}^{\mathrm{IFFT}} \times Q_{N,P}}$. In particular, the periodogram $\mathbf{f}_{\theta}\in \mathbb{C}^{N_{\theta}^{\mathrm{IFFT}}\times 1}$ can be obtained by performing an IFFT on each column of matrix $\mathbf{X}_{\theta}^{\mathrm{ZP}}$ and then applying column-wise summation, given by  
\begin{equation}
	\setlength\abovedisplayskip{2pt}
	\setlength\belowdisplayskip{2pt}
	\begin{aligned}  \mathbf{f}_{\theta}&=\frac1{Q_{N,P}}\sum_{q=0}^{Q_{N,P}-1}\left|\mathrm{IFFT}_{N_{\theta}^{\mathrm{IFFT}}}\left\{\mathbf{X}_{\theta}^{\mathrm{ZP}}(:,q)\right\}\right|^2 \\
&=\frac1{Q_{N,P}}\sum_{q=0}^{Q_{N,P}-1}\frac1{M_r}\left|\mathbf{W}_{N_{\theta}^{\mathrm{IFFT}}}\mathbf{X}_{\theta}^{\mathrm{ZP}}(:,q)\right|^2,
        \end{aligned}
	\label{PFFT}
\end{equation}
where $|\cdot|$ denotes element-wise modulus, and $\mathbf{W}_N$ is the $N$-point IDFT matrix, expressed as
\begin{equation}
	\setlength\abovedisplayskip{2pt}
	\setlength\belowdisplayskip{2pt}
	\begin{aligned}  
\mathbf{W}_N=\begin{bmatrix}1&1&\cdots&1\\1&W^{-1}&\cdots&W^{-\left(N-1\right)}\\\vdots&\vdots&\ddots&\vdots\\1&W^{-\left(N-1\right)}&\cdots&W^{-\left(N-1\right)\left(N-1\right)}\end{bmatrix},
        \end{aligned}
	\label{IDFTM}
\end{equation}
with $ W=e^{-j\frac{2\pi}{N}}$. Additionally, since the default phase of the IFFT is $[0, \pi]$, the period of $\mathbf{f}_{\theta}$ must be shifted to $[-\pi/2, \pi/2]$. Then, the peak index $\hat{n}_{\theta}^k$ of the periodogram $\mathbf{f}_{\theta}\in \mathbb{C}^{N_{\theta}^{\mathrm{IFFT}}\times 1}$  indicates the phase $\frac{d}{\lambda}\sin{\hat{\theta}_k}=\frac{\hat{n}_{\theta}^k}{N_{\theta}^{\mathrm{IFFT}}}, \hat{n}_{\theta}^k \in \left[-\frac{N_{\theta}^{\mathrm{IFFT}}}{2},\frac{N_{\theta}^{\mathrm{IFFT}}}{2}\right]$, and the AoAs of the sensing targets can be estimated as
\begin{equation}
	\setlength\abovedisplayskip{2pt}
	\setlength\belowdisplayskip{2pt}
	\begin{aligned}  
 \hat{\theta}_k=\arcsin\biggl(\frac{\hat{n}_{\theta}^k\lambda}{N_{\theta}^\text{IFFT}d}\biggr), \forall k=1,\cdots,K.
        \end{aligned}
	\label{FFTDOA}
\end{equation}

\begin{figure}[!t]
  \centering
  \includegraphics[width=0.85\linewidth]{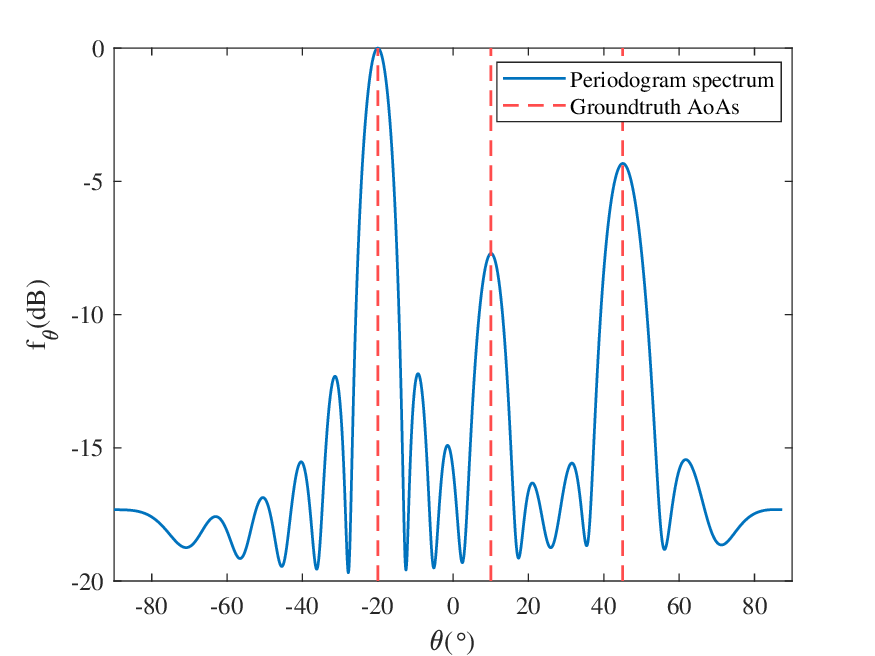}
  \caption{Periodogram spectrum for AoA estimation in MIMO-OFDM ISAC.}
  \label{Periodogram}
  \vspace{-0.6cm}
\end{figure}

\begin{table}[!t]
\renewcommand{\arraystretch}{1.2}
\centering
\caption{Simulation setup for monostatic MIMO-OFDM ISAC.}
\label{Ssetup}
\begin{tabular}{|c|c|}
\hline
Parameters                   & Values                                   \\ \hline
Number of antennas at the Sen-RX & $M_r=16$                                 \\ \hline
Number of subcarriers             & $N=128$                                  \\ \hline
Number of OFDM symbols            & $P=64$                                   \\ \hline
Subcarrier spacing           & $\Delta f=120\ \mathrm{kHz}$               \\ \hline
CP duration                  & $T_{\mathrm{cp}}=T/4=8.33\ \mu s$ \\ \hline
Carrier frequency            & $f_c=28\ \mathrm{GHz}$                     \\ \hline
\begin{tabular}[c]{@{}c@{}}$\mathrm{SNR}$ of the radar processing signal\\ at the Sen-RX in \mbox{(\ref{OSM})}\end{tabular} & $\mathrm{SNR=10\ dB}$                                 \\ \hline
Number of targets            & $K=3$                                    \\ \hline
AoA                          & $(-20,10,45)^{\circ}$                    \\ \hline
Distance                        & $(20,80,50)\ \mathrm{m}$                   \\ \hline
Delay                        & $(0.13,0.53,0.33)\ \mu s$                  \\ \hline
Radial velocity              & $(8,12,20)\ \mathrm{m/s}$                  \\ \hline
Doppler frequency            & $(1.49,2.24,3.73)\ \mathrm{kHz}$           \\ \hline
\end{tabular}
\vspace{-0.4cm}
\end{table}
Consider a monostatic ISAC system with parameters given in \mbox{Table \ref{Ssetup}}. The periodogram spectrum for AoA estimation after normalization and conversion to dB is shown in \mbox{Fig.~\ref{Periodogram}}. It is observed that higher spectral peaks occur at groundtruth AoAs, while lower sidelobes are present at other angles. The height of the spectral peaks corresponds to the strength of the target echoes, and targets with weaker echo signals may be obscured by the sidelobes of stronger targets. When the target angle approaches $0^{\circ}$, the width of the spectral peak becomes narrower, which enables the target to be more easily distinguished.

\begin{figure}[!t]
  \centering
  \includegraphics[width=0.8\linewidth]{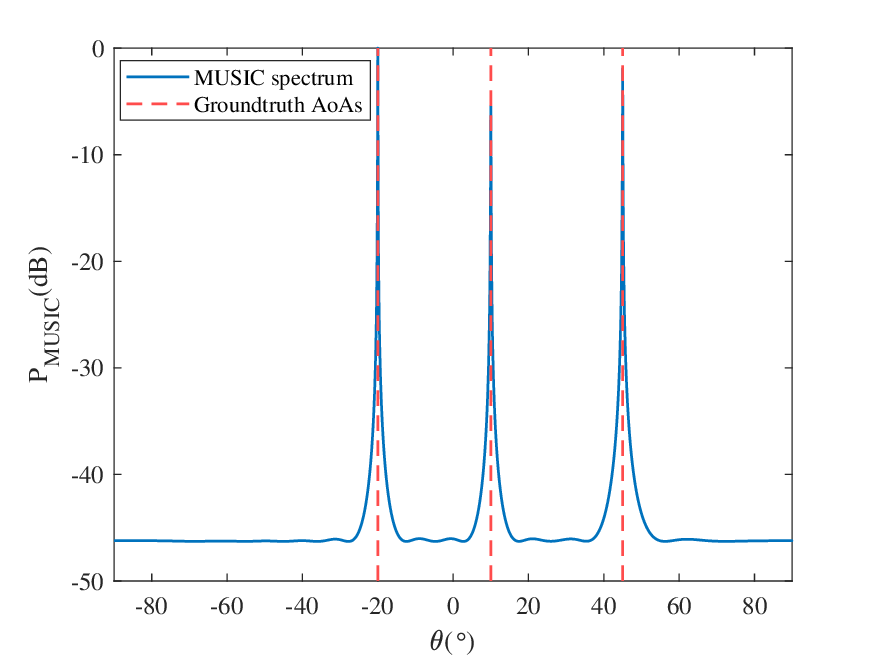}
  \caption{MUSIC spectrum for AoA estimation in MIMO-OFDM ISAC.}
  \label{SMUSIC}
  \vspace{-0.5cm}
\end{figure}

\textbf{MUSIC:} The MUSIC algorithm exploits the orthogonality between the noise and signal subspaces to estimate AoAs/delays/Doppler frequencies. These subspaces can be obtained via an eigenvalue decomposition (EVD) of the covariance matrix of $\mathbf{X}_{\theta}$ in \mbox{(\ref{23})}, which is expressed as
\begin{equation}
	\setlength\abovedisplayskip{2pt}
	\setlength\belowdisplayskip{2pt}
	\begin{aligned}  
\mathbf{R}_{\mathbf{X}}&=\frac{1}{NP}\mathbf{X}_{\theta}\mathbf{X}^H_{\theta}=\mathbf{A}_{\theta}\mathbf{R}_s\mathbf{A}_{\theta}^H + \sigma^2\mathbf{I}_{M_r} \\
&=\mathbf{E}_s\mathbf{\Sigma}_s\mathbf{E}_s^H+\mathbf{E}_n\mathbf{\Sigma}_n\mathbf{E}_n^H,
        \end{aligned}
	\label{EVD}
\end{equation}
where $\mathbf{R}_s=\frac{1}{NP}\mathbf{S}_{\theta}\mathbf{S}^H_{\theta}$, $\sigma^2$ is the noise power, $\mathbf{\Sigma}_s$ and $\mathbf{\Sigma}_n$ are diagonal matrices composed of the $K$ largest eigenvalues and the remaining $M_r-K$ eigenvalues, respectively, $\mathbf{E}_s$ and $\mathbf{E}_n$ denote the eigenvectors that span the signal and noise subspaces. The MUSIC spectrum can be written as
\begin{equation}
	\setlength\abovedisplayskip{2pt}
	\setlength\belowdisplayskip{2pt}
	\begin{aligned}  
        P_{\mathrm{MUSIC}}(\theta)=\frac{1}{\mathbf{a}_{\mathrm{R}}^H(\theta)\mathbf{E}_n\mathbf{E}_n^H\mathbf{a}_{\mathrm{R}}(\theta)},
        \end{aligned}
	\label{PMUSIC}
\end{equation}
 where $\mathbf{a}_R(\theta)$ is the steering vector in the spatial domain in \mbox{(\ref{doast})}, and the peaks of the MUSIC spectrum correspond to the AoAs of sensing targets.

 For the simulation setup in \mbox{Table \ref{Ssetup}}, the AoA spectrum of the MUSIC algorithm after normalization is shown in \mbox{Fig.~\ref{SMUSIC}}. Both the MUSIC and periodogram algorithms can accurately estimate the AoAs of all three targets, although the MUSIC and periodogram spectra differ in two aspects. First, the spectral peaks of MUSIC are sharper, implying higher angular resolution. Second, the MUSIC spectrum has essentially no sidelobes, allowing for better detection of weak targets. However, subspace methods such as MUSIC and ESPRIT (discussed next) rely on knowledge of the number of AoAs in order to work properly, which is a drawback compared with the non-parametric periodogram approach.

\textbf{ESPRIT:} The ESPRIT algorithm exploits the rotational invariance between two subarrays to estimate AoAs. For the signal model in \mbox{(\ref{23})}, we can take the first $M_r-1$ antennas of the ULA as subarray 1, and the last $M_r-1$ antennas as subarray 2, and the signals received by the two subarrays can be expressed as
\begin{equation}
	\setlength\abovedisplayskip{2pt}
	\setlength\belowdisplayskip{2pt}
	\begin{aligned}  
        \mathbf{X}_1&=\mathbf{X}_{\theta}(1:M_r-1,:)=\mathbf{A}_{1}\mathbf{S}_{\theta}+\mathbf{Z}_1,\\
        \mathbf{X}_2&=\mathbf{X}_{\theta}(2:M_r,:)=\mathbf{A}_{2}\mathbf{S}_{\theta}+\mathbf{Z}_2=\mathbf{A}_1\mathbf{\Phi}\mathbf{S}_{\theta}+\mathbf{Z}_2,
        \end{aligned}
	\label{ESPRITRS}
\end{equation}
where $\mathbf{A}_1=\mathbf{A}_{\theta}(1:M_r-1,:)$ and $\mathbf{A}_2=\mathbf{A}_{\theta}(2:M_r,:)$ denote the steering matrices of the two subarrays, respectively, $\mathbf{Z}_1$ and $\mathbf{Z}_2$ are the corresponding subarray noise matrices. The two subarray steering matrices are related by $\mathbf{A}_2=\mathbf{A}_1\mathbf{\Phi}$, where $\mathbf{\Phi}$ denotes a rotation operator determined by the target angles, expressed as
\begin{equation}
	\setlength\abovedisplayskip{2pt}
	\setlength\belowdisplayskip{2pt}
	\begin{aligned}  
        \boldsymbol{\Phi}=\mathrm{diag}\Bigg(e^{-j\frac{2\pi d\mathrm{sin}\theta_1}{\lambda}}, \cdots, e^{-j\frac{2\pi d\mathrm{sin}\theta_K}{\lambda}}\Bigg).
        \end{aligned}
	\label{RO}
\end{equation}
Thus, the target AoAs can be determined by finding the rotation operator.

The data from the two subarrays are concatenated into a new data matrix, i.e.,
\begin{equation}
	\setlength\abovedisplayskip{2pt}
	\setlength\belowdisplayskip{2pt}
	\begin{aligned}  
        \bar{\mathbf{X}}=\begin{bmatrix}\mathbf{X}_1\\\mathbf{X}_2\end{bmatrix}=\begin{bmatrix}\mathbf{A}_1\\\mathbf{A}_1\mathbf{\Phi}\end{bmatrix}\mathbf{S}_{\theta}+\begin{bmatrix}\mathbf{Z}_1\\\mathbf{Z}_2\end{bmatrix}.
        \end{aligned}
	\label{ESPRITRSMX}
\end{equation}
The signal subspace $\mathbf{E}_s$ and noise subspace $\mathbf{E}_n$ can be obtained by performing an EVD on the covariance matrix of the data matrix ${\mathbf{X}_{\theta}}$, given by
\begin{equation}
	\setlength\abovedisplayskip{2pt}
	\setlength\belowdisplayskip{2pt}
	\begin{aligned}  
        \mathbf{R}_{{\mathbf{X}_{\theta}}}=\frac{1}{NP}{\mathbf{X}_{\theta}}{\mathbf{X}_{\theta}}^H=\mathbf{E}_s\mathbf{\Lambda}_s\mathbf{E}_s^H+\mathbf{E}_n\mathbf{\Lambda}_n\mathbf{E}_n^H.
        \end{aligned}
	\label{ESPRITEVD}
\end{equation}
Due to the rotational invariance between the two subarrays, the signal subspace $\bar{\mathbf{E}}_s$ of the data matrix $\bar{\mathbf{X}}$ can be written as
\begin{equation}
	\setlength\abovedisplayskip{2pt}
	\setlength\belowdisplayskip{2pt}
	\begin{aligned}  
        \bar{\mathbf{E}}_s=\begin{bmatrix}\mathbf{E}_{s1}\\\mathbf{E}_{s2}\end{bmatrix}=\begin{bmatrix}\mathbf{A}_1\mathbf{T}\\\mathbf{A}_1\mathbf{\Phi}\mathbf{T}\end{bmatrix},
        \end{aligned}
	\label{Es}
\end{equation}
where $\mathbf{E}_{s1}=\mathbf{E}_{s}(1:M_r-1,:)$, $\mathbf{E}_{s2}=\mathbf{E}_{s}(2:M_r,:)$, and $\mathbf{T}$ denotes a nonsingular matrix. Therefore, we have
\begin{equation}
	\setlength\abovedisplayskip{2pt}
	\setlength\belowdisplayskip{2pt}
	\begin{aligned}  
        \mathbf{E}_{s2}=\mathbf{A}_1\mathbf{\Phi}\mathbf{T}=\mathbf{E}_{s1}\mathbf{T}^{-1}\mathbf{\Phi}\mathbf{T}.
        \end{aligned}
	\label{ESPRITphi}
\end{equation}
Let $\mathbf{\Psi}=\mathbf{T}^{-1}\mathbf{\Phi}\mathbf{T}$, and note that the eigenvalues of $\mathbf{\Psi}$ are given by the diagonal elements of $\mathbf{\Phi}$. A least squares estimate of $\mathbf{\Phi}$ can be obtained by
\begin{equation}
	\setlength\abovedisplayskip{2pt}
	\setlength\belowdisplayskip{2pt}
	\begin{aligned}  
        \hat{\mathbf{\Psi}}=\mathbf{E}_{s1}^{\dagger}\mathbf{E}_{s2}.
        \end{aligned}
	\label{ESPRITpsi}
\end{equation}
 The diagonal elements of $\mathbf{\Psi}$ can be obtained by computing the eigenvalues $\hat{\lambda}_k$ of the estimated $\hat{\mathbf{\Psi}}$, so the AoAs of the targets can be estimated as
\begin{equation}
	\setlength\abovedisplayskip{2pt}
	\setlength\belowdisplayskip{2pt}
	\begin{aligned}  
        \theta_k=-\arcsin\big(\lambda \mathrm{angle}\big(\hat{\lambda}_k\big)/2\pi d\big), k=1,2,\cdots,K.
        \end{aligned}
	\label{ESPRITdoa}
\end{equation}

\textbf{OMP:} The OMP algorithm is an efficient sparse recovery method based on iteratively selecting the basis vectors most correlated with the received signal to estimate target parameters, thereby avoiding complex optimization procedures. For AoA estimation, the observation angle $\theta \in [-90^\circ,90^\circ]$ is discretized into $N^s$ points, and the steering dictionary matrix is defined as
\begin{equation}
	\setlength\abovedisplayskip{2pt}
	\setlength\belowdisplayskip{2pt}
	\begin{aligned}  
        \mathbf{A}_{\mathrm{dic}}=\left[\mathbf{a}_\mathrm{R}(\theta_1),\cdots,\mathbf{a}_\mathrm{R}(\theta_{N_s}) \right]\in \mathbb{C}^{M_r\times N^s},
        \end{aligned}
	\label{doa_dic}
\end{equation}
where each spatial steering vector $\mathbf{a}_\mathrm{R}(\theta)$ serves as an atom. Then, the expression for $\mathbf{X}_{\theta}$ in (\ref{23}) is modified as
\begin{equation}
	\setlength\abovedisplayskip{2pt}
	\setlength\belowdisplayskip{2pt}
	\begin{aligned}  
\mathbf{X}_{\theta}=\mathbf{A}_{\theta}\mathbf{S}_{\theta}+\mathbf{Z}_{\theta}=\mathbf{A}_{\mathrm{dic}}\mathbf{S}_{\theta}^{\prime}+\mathbf{Z}_{\theta},
        \end{aligned}
	\label{doa_l0}
\end{equation}
where $\mathbf{S}_{\theta}^{\prime} \in \mathbb{C}^{N^s\times Q_{N,P}}$ is the sparse signal to be reconstructed with $K$ nonzero rows. Then, the sparse recovery problem can be formulated as
\begin{equation}
	\setlength\abovedisplayskip{2pt}
	\setlength\belowdisplayskip{2pt}
	\begin{aligned}  
&\min_{\mathbf{S}_\theta^{\prime}}\|\mathbf{S}_\theta^{\prime}\|_{2,0} \ \  \mathrm{s.t.} \  \|\mathbf{X}_\theta-\mathbf{A}_{\mathrm{dic}} \mathbf{S}_\theta^{\prime}\|_2<\epsilon\,
        \end{aligned}
	\label{doa_sr}
\end{equation}
where $\|\cdot\|_{2,0}$ is the $\ell_{2,0}$ mixed norm that denotes the number of nonzero rows, $\|\cdot\|_2$ is the Frobenius norm, $\epsilon$ is a small constant that represents the error tolerance governed by the noise level. Since $\ell_0$-norm minimization is NP-hard, greedy optimization algorithms are typically employed to approximate the original signal via a sparse representation. The OMP algorithm iteratively selects the most correlated atoms from the discretized dictionary, updates the residual via least squares minimization, and ultimately estimates the angles of $K$ targets.

In particular, OMP first initializes the residual as the signal $\mathbf{X}_\theta$ in (\ref{23}) and the support set as the empty set $\emptyset$, i.e.,
\begin{equation}
	\setlength\abovedisplayskip{2pt}
	\setlength\belowdisplayskip{2pt}
	\begin{aligned}  
    \mathbf{R}_0=\mathbf{X}_\theta, \mathbb{S}_{\hat{\theta}}=\emptyset.
        \end{aligned}
	\label{init}
\end{equation}
The algorithm then performs $K$ iterations, each selecting the atom most correlated with the current residual, expressed as
\begin{equation}
	\setlength\abovedisplayskip{2pt}
	\setlength\belowdisplayskip{2pt}
	\begin{aligned}  
    i_k=\mathrm{arg}\max_{i}\|\mathbf{A}_{\mathrm{dic}}(:,i)^H\mathbf{R}_{k-1}\|_2^2,i=1,2,\cdots,N^s,
        \end{aligned}
	\label{iter}
\end{equation}
and then the support set is updated as $\mathbb{S}_{\hat{\theta}}=\mathbb{S}_{\hat{\theta}} \cup \{i_k\}$. Here, $\mathbf{R}_k$ denotes the residual after the $k$-th iteration, where the sparse recovery signal is estimated via least squares as
\begin{equation}
	\setlength\abovedisplayskip{2pt}
	\setlength\belowdisplayskip{2pt}
	\begin{aligned}  
    \hat{\mathbf{S}}_{\theta,k}&=\mathrm{arg}\min_{\mathbf{S}}\|\mathbf{X}_\theta-\mathbf{A}_{\mathrm{dic}}(:,\mathbb{S}_{\hat{\theta}})\mathbf{S}\|_2 \\
    &=\mathbf{A}_{\mathrm{dic}}(:,\mathbb{S}_{\hat{\theta}})^{\dagger}\mathbf{X}_\theta, 
        \end{aligned}
	\label{ses}
\end{equation}
and the residual is obtained by
\begin{equation}
	\setlength\abovedisplayskip{2pt}
	\setlength\belowdisplayskip{2pt}
	\begin{aligned}  
    \mathbf{R}_k=\mathbf{X}_\theta-\mathbf{A}_{\mathrm{dic}}(:,\mathbb{S}_{\hat{\theta}})\hat{\mathbf{S}}_{\theta,k}. 
        \end{aligned}
	\label{residual}
\end{equation}
Finally, the AoAs are estimated as $\hat{\mathbf{\theta}}=\theta_{\mathrm{grid}}(\mathbb{S}_{\hat{\theta}})$, where $\theta_{\mathrm{grid}}$ denotes the discretized angular grid. 

\begin{figure}[!t]
  \centering
  \includegraphics[width=0.8\linewidth]{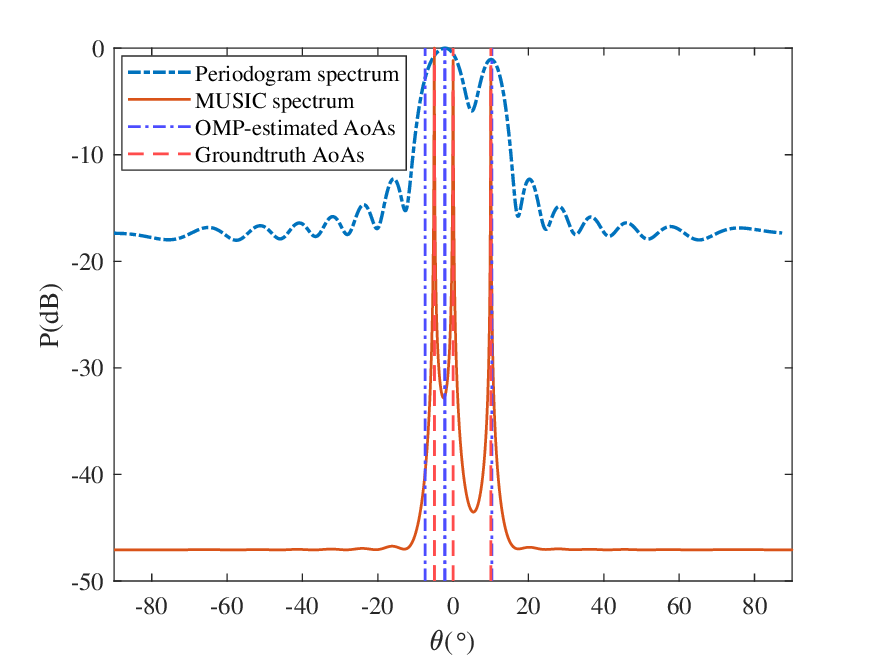}
  \caption{Comparison of periodogram, MUSIC, OMP for AoA estimation in MIMO-OFDM ISAC with closely spaced targets.}
  \label{periodogramMUSIC}
  \vspace{-0.6cm}
\end{figure}

We consider another scenario with more closely spaced targets located at $(-5, 0, 10)^{\circ}$, while the other parameters in \mbox{Table \ref{Ssetup}} remain unchanged. We show a comparison in \mbox{Fig.~\ref{periodogramMUSIC}} between the periodogram, MUSIC, and OMP. It is observed that when the angle difference between targets is smaller than the resolution limit of the array ($7.1^\circ$ for a 16-element ULA), their spectral peaks blend together in the periodogram, making them indistinguishable. However, the OMP and MUSIC algorithms are still able to resolve all three targets. Although the OMP algorithm can also achieve super-resolution, unlike MUSIC its estimation accuracy significantly deteriorates due to high correlation between the steering vectors of the closely spaced targets. In general, the OMP algorithm exhibits degraded estimation performance in scenarios where targets are densely distributed. For the OMP algorithm, the parameters of each target are estimated separately, preventing weak targets from being masked by strong ones.

\begin{figure}[!t]
	\centering
	\setlength{\abovecaptionskip}{0cm}
	\includegraphics[width=0.85\linewidth]{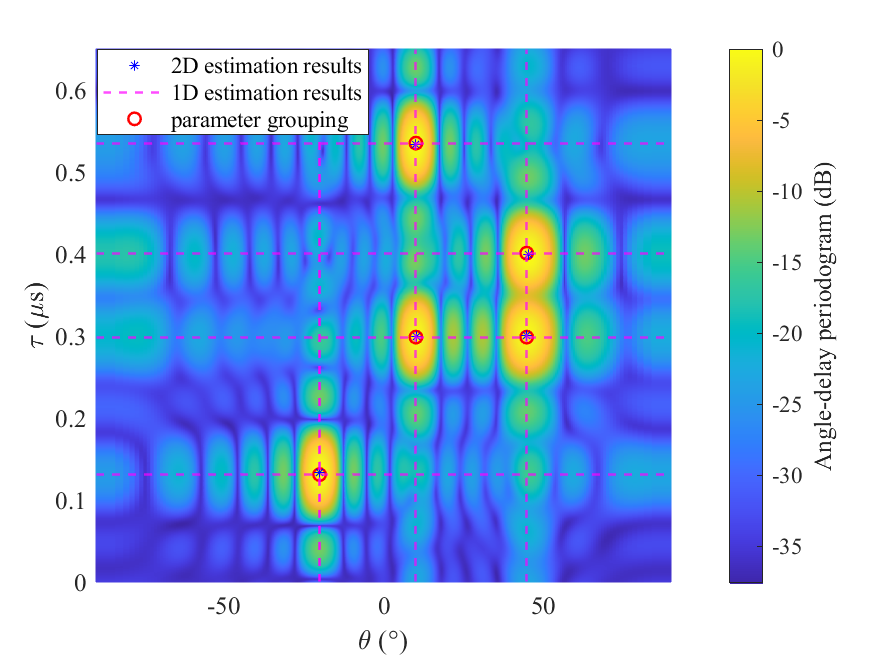}
	\caption{Parallel one-domain parameter estimation based on the periodogram in MIMO-OFDM ISAC.}
	\label{P1D}
	\vspace{-0.6cm}
\end{figure}

\begin{figure}[!t]
  \centering
  \setlength{\abovecaptionskip}{0cm}
  \includegraphics[width=0.85\linewidth]{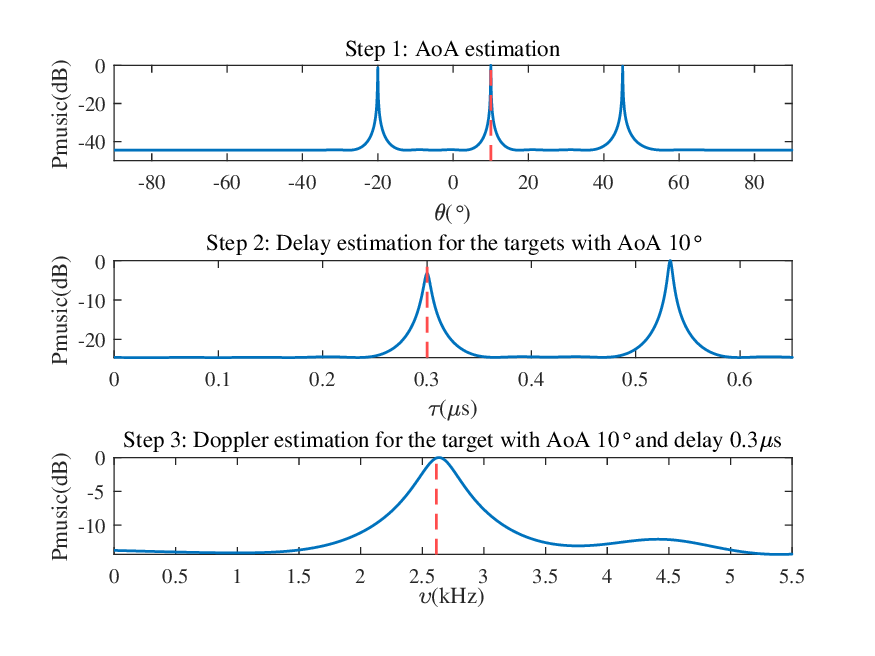}
  \caption{Sequential one-domain parameter estimation based on MUSIC for estimating the target with parameters $(10^\circ,0.3\mu s,2.61\mathrm{kHz})$ in MIMO-OFDM ISAC.}
  \label{S1D}
  \vspace{-0.6cm}
\end{figure}

In \mbox{Section~\ref{far-doa}}, we introduced two decoupled parameter estimation methods and their corresponding 1D estimation algorithms. In the following, we provide an example comparing parallel and sequential one-domain estimation. For the simulation setup in \mbox{Table \ref{Ssetup}}, we consider $K=5$ targets with the AoA, distance, and radial velocity parameters ($-20^\circ$,$20\ \mathrm{m}$,$8\ \mathrm{m/s}$), ($10^\circ$,$45\ \mathrm{m}$,$14\ \mathrm{m/s}$), ($10^\circ$,$80\ \mathrm{m}$,$20\ \mathrm{m/s}$), $\left(45^\circ,60\ \mathrm{m},6\ \mathrm{m/s}\right)$, $\left(45^\circ,45\ \mathrm{m},12\ \mathrm{m/s}\right)$. The corresponding delays and Doppler frequency shifts of the five sensing targets are $\left(0.13\ \mu s,1.49\ \mathrm{kHz}\right)$, $\left(0.3\ \mu s,2.61\ \mathrm{kHz}\right)$, $\left(0.53\ \mu s,3.73\ \mathrm{kHz}\right)$, $\left(0.4\ \mu s,1.12\ \mathrm{kHz}\right)$, $\left(0.3\ \mu s,2.24\ \mathrm{kHz}\right)$, respectively. The estimation and grouping results for the angles and delays of the five targets for parallel one-domain estimation are illustrated in Fig.~\ref{P1D}, and the results for sequential one-domain estimation in the three domains of the target with parameters $\left(10^\circ,0.3\ \mu s,2.61\ \mathrm{kHz}\right)$ are shown in \mbox{Fig.~\ref{S1D}}, where ZF-based beamforming is exploited to separate the target signals with different AoAs and delays. Using parallel estimation, the 1D periodogram can only resolve three and four targets in the angular and delay domains, respectively. However, after parameter grouping and power detection based on (\ref{ppp}), the sensing performance closely follows that of the 2D periodogram. The sequential estimation involves three steps in sensing a target. For step 1, only three AoAs can be estimated for the five sensing targets. After spatial domain beamforming towards $10^\circ$, two delays are estimated in step 2. Finally, after delay domain beamforming for the target with delay $0.3\ \mu s$, the Doppler frequency $2.61\ \mathrm{kHz}$ can be estimated in step 3.

The sequential domain method first performs AoA estimation once, followed by spatial domain beamforming and delay estimation $K_{\hat{\theta}}$ times, and finally delay domain beamforming and Doppler estimation $K$ times. Compared to sequential one-domain estimation, parallel one-domain estimation requires performing AoA, delay, and Doppler estimation only once, thus reducing complexity. However, the parallel method requires the extra step of parameter grouping, whereas the sequential method associates the parameters automatically. If the sensing targets cannot be distinguished in a single domain, the parallel method may result in poor estimation performance. In contrast, the sequential method can use beamforming to extract the signal of each target as long as they can be distinguished based on a single parameter, thus enabling accurate estimation of the other two target parameters. The sequential method can also effectively prevent weaker targets from being obscured by first separating out the signals of the stronger targets \cite{b97}. Therefore, the parallel method is preferred if the AoA, delay, and Doppler resolutions are all high; otherwise, the sequential method is preferable. Moreover, for sequential one-domain parameter estimation, since the failure to distinguish different targets before performing beamforming will degrade estimation performance, the target parameter with the highest resolution should be estimated first, followed by the parameter with the next highest resolution, and so on.

\subsection{Joint Two-domain Parameter Estimation}\label{dd_OFDM}
\subsubsection{Joint Two-domain Parameter Estimation Model} \ \\
\indent In this subsection, we present the joint two-domain parameter estimation methods outlined in \mbox{Fig.~\ref{PE}}. Compared to the decoupled parameter estimation discussed in \mbox{Section~\ref{far-doa}}, joint two-domain parameter estimation can achieve higher accuracy at the cost of higher computational complexity. The two-dimensional (2D) joint target parameter estimation can be divided into three cases, corresponding to joint delay-Doppler estimation, joint AoA-delay estimation, and joint AoA-Doppler estimation. Since the three cases have similar mathematical forms, we will focus only on the joint delay-Doppler estimation, while the other two cases can be handled using similar methods.

As illustrated in \mbox{Fig.~\ref{PE}}, before performing joint delay-Doppler estimation, we first estimate the AoAs of the sensing targets based on \mbox{(\ref{23})}, and then perform beamforming on \mbox{(\ref{d25})} to decouple signals from different angles. However, joint two-domain parameter estimation aims to jointly estimate the delay and Doppler for the targets at each angle, rather than sequentially estimating these two parameters. Assume that $K_{\hat{\theta}}$ AoAs are estimated, with $N_{k_{\hat{\theta}}}$ targets at angle $\hat{\theta}_{k_{\hat{\theta}}}$. After spatial domain beamforming, the extracted signals $\mathbf{X}_{k_{\hat{\theta}}}$ in \mbox{(\ref{d38})} for each estimated angle are exploited to simultaneously estimate the target delay $\tau_k$ and Doppler frequency shift $\upsilon_k$.

\subsubsection{2D Joint Parameter Estimation Algorithms} \ \\
\indent \textbf{2D-Periodogram:} Based on the signal in \mbox{(\ref{d38})}, the noise-free 2D power spectrum $f(\tau,\upsilon)$ in the delay-Doppler domain can be obtained in \mbox{(\ref{2DP})} as shown at the top of the next page,
\begin{figure*}[hbt]
\centering
 \begin{equation}
	\setlength\abovedisplayskip{2pt}
	\setlength\belowdisplayskip{2pt}
	\begin{aligned}  
f(\tau,\upsilon)&= \frac{1}{NP}\bigg|\mathbf{a}_{\tau}^H(\tau)\mathbf{X}_{k_{\hat{\theta}}}\mathbf{a}_{\upsilon}^*(\upsilon) \bigg|^2 \\
&= \frac{1}{NP}\Bigg|\sum_{i=1}^{N_{k_{\hat{\theta}}}}\tilde{\alpha}_{i,{k_{\hat{\theta}}}} \sum_{n=0}^{N-1} e^{j2\pi n\Delta f \left(\tau-\tau_{i,{k_{\hat{\theta}}}}\right)} \sum_{p=0}^{P-1}e^{j2\pi pT_s \left(\upsilon_{i,{k_{\hat{\theta}}}}-\upsilon \right)}\Bigg|^2\\
 &\mathop  \approx \limits^{(a)} \sum_{i=1}^{N_{k_{\hat{\theta}}}}\frac{{{\left| {{{\tilde{\alpha} }_{i,{k_{\hat{\theta}}}}}} \right|}^2}}{NP}{{\left| {\frac{{\sin \left( \pi N\Delta f \left(\tau-\tau_{i,{k_{\hat{\theta}}}}\right) \right)}}{{\sin \left( \pi \Delta f \left(\tau-\tau_{i,{k_{\hat{\theta}}}}\right) \right)}}} \right|}^2}  {{\left| {\frac{{\sin \left( \pi P T_s \left(\upsilon-\upsilon_{i,{k_{\hat{\theta}}}} \right) \right)}}{{\sin \left( \pi T_s \left(\upsilon-\upsilon_{i,{k_{\hat{\theta}}}} \right) \right)}}} \right|}^2}.
        \end{aligned}
	\label{2DP}
\end{equation}   
\vspace{-0.5cm}
\end{figure*}
where $\mathbf{a}_{\tau}(\tau)$ and $\mathbf{a}_{\upsilon}(\upsilon)$ are the steering vectors in the delay and Doppler domains in \mbox{(\ref{Rsv})} and \mbox{(\ref{Dsv})}, $\tau \in \left[0,\frac{1}{\Delta f} \right]$ and $\upsilon \in \left[-\frac{1}{2T_s},\frac{1}{2T_s}\right]$ are the observation delay and Doppler frequency, respectively. The approximation $(a)$ in \mbox{(\ref{2DP})} holds if the delay or Doppler difference between sensing targets exceeds the corresponding resolution. Eq. \mbox{(\ref{2DP})} reveals that the power spectrum in the delay-Doppler domain is the product of the powers in the two domains, and the two components are mutually independent. Therefore, as long as the targets can be distinguished in either domain, the power spectrum can be used to distinguish them.

Similar to the 1D-periodogram, to reduce the computational complexity of searching $f(\tau,\upsilon)$ over the delay $\tau$ and Doppler $\upsilon$, an $N_{\tau}^{\mathrm{IFFT}}$-point IFFT and $N_{\upsilon}^{\mathrm{FFT}}$-point FFT can be performed in the subcarrier and symbol domains of $\mathbf{X}_{k_{\hat{\theta}}}$ in \mbox{(\ref{d38})} for fast computation. When ${N_{\tau}^\mathrm{IFFT}}>N$ or ${N_{\upsilon}^\mathrm{FFT}}>P$, zero-padding is used to convert $\mathbf{X}_{k_{\hat{\theta}}} \in \mathbb{C}^{N\times P}$ to $\mathbf{X}_{k_{\hat{\theta}}}^{\mathrm{ZP}} \in \mathbb{C}^{{N_{\tau}^\mathrm{IFFT}}\times {N_{\upsilon}^\mathrm{FFT}}}$. Then, the resulting 2D-periodogram $\mathbf{F}\in \mathbb{C}^{{N_{\tau}^\mathrm{IFFT}}\times {N_{\upsilon}^\mathrm{FFT}}}$ can be expressed as
\begin{equation}
	\setlength\abovedisplayskip{2pt}
	\setlength\belowdisplayskip{2pt}
	\begin{aligned}  
        \mathbf{F}&=\left|\mathrm{FFT}_{N_{\upsilon}^\mathrm{FFT}}^2\left\{\mathrm{IFFT}_{N_{\tau}^\mathrm{IFFT}}^1\left\{\mathbf{X}_{k_{\hat{\theta}}}^{\mathrm{ZP}}\right\}\right\}\right|^2 \\
        &= \frac1{NP}\left|\mathbf{W}_{N_{\tau}^\mathrm{IFFT}} \mathbf{X}_{k_{\hat{\theta}}}^{\mathrm{ZP}} \mathbf{W}^{*}_{N_{\upsilon}^\mathrm{FFT}}\right|^2,
        \end{aligned}
	\label{2DFFT}
\end{equation}
where $\mathbf{W}^{*}$ is the conjugate of IDFT matrix $\mathbf{W}$. Negative Doppler frequency estimation necessitates a spectral shift in the Doppler domain of $\mathbf{F}$ to center the zero frequency component within the spectrum. Then, the peak index $\big(\hat{n}_{\tau}^{i,{k_{\hat{\theta}}}},\hat{n}_{\upsilon}^{i,{k_{\hat{\theta}}}}\big)$ of $\mathbf{F}$ indicates the phases $\Delta f \hat{\tau}_{i,{k_{\hat{\theta}}}}=\frac{\hat{n}^{i,{k_{\hat{\theta}}}}_{\tau}}{{N_{\tau}^\mathrm{IFFT}}}$ and $T_s \hat{\upsilon}_{i,{k_{\hat{\theta}}}}=\frac{\hat{n}^{i,{k_{\hat{\theta}}}}_{\upsilon}}{{N_{\upsilon}^\mathrm{FFT}}}$ in the delay and Doppler domains, respectively. Thus, the delay $\hat{\tau}_{i,{k_{\hat{\theta}}}}$ and Doppler frequency $\hat{\upsilon}_{i,{k_{\hat{\theta}}}}$, $\forall i=1,\cdots,{N_{k_{\hat{\theta}}}}$, of the $N_{k_{\hat{\theta}}}$ sensing targets at angle $\hat{\theta}_{k_{\hat{\theta}}}$ can be estimated as
\begin{equation}
	\setlength\abovedisplayskip{2pt}
	\setlength\belowdisplayskip{2pt}
	\begin{aligned}  
        \hat{\tau}_{i,{k_{\hat{\theta}}}}=\frac{\hat{n}_{\tau}^{i,{k_{\hat{\theta}}}}}{\Delta f N_{\tau}^{\mathrm{IFFT}}},\hat{\upsilon}_{i,{k_{\hat{\theta}}}}=\frac{\hat{n}_{\upsilon}^{i,{k_{\hat{\theta}}}}}{T_sN_{\upsilon}^{\mathrm{FFT}}}.
        \end{aligned}
	\label{FFTtauD}
\end{equation}

\begin{figure}[!t]
	\centering
	\subfigbottomskip=-4pt %
	\subfigcapskip=-2pt %
	\subfigure[2D-Periodogram spectrum.]{
		\begin{minipage}[!t]{0.85\linewidth}
			\includegraphics[width=1\linewidth]{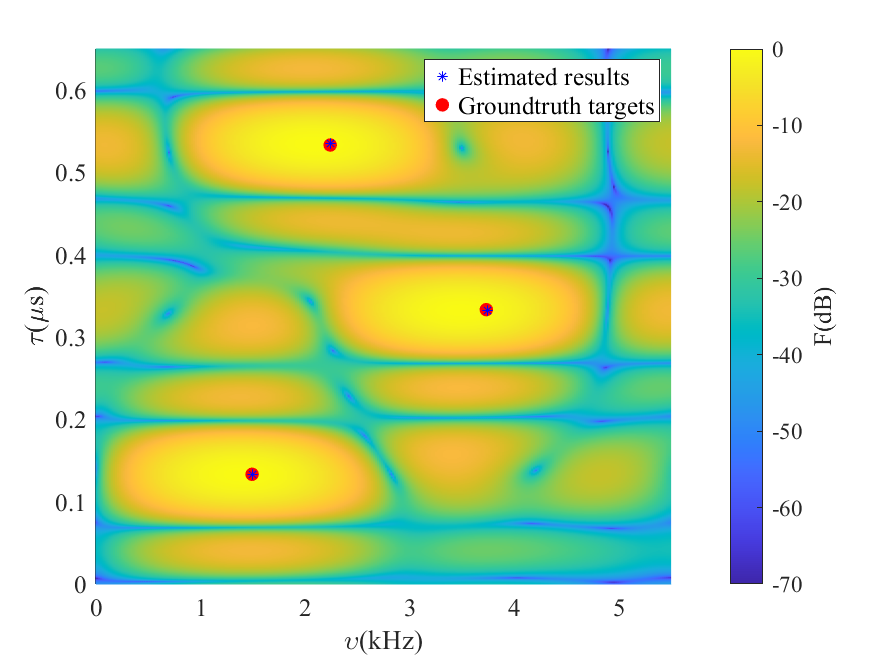}
			\label{2DFFTS}
		\end{minipage}%
	}%
	
	\subfigure[2D-MUSIC spectrum.]{
		\begin{minipage}[!t]{0.85\linewidth}
			\includegraphics[width=1\linewidth]{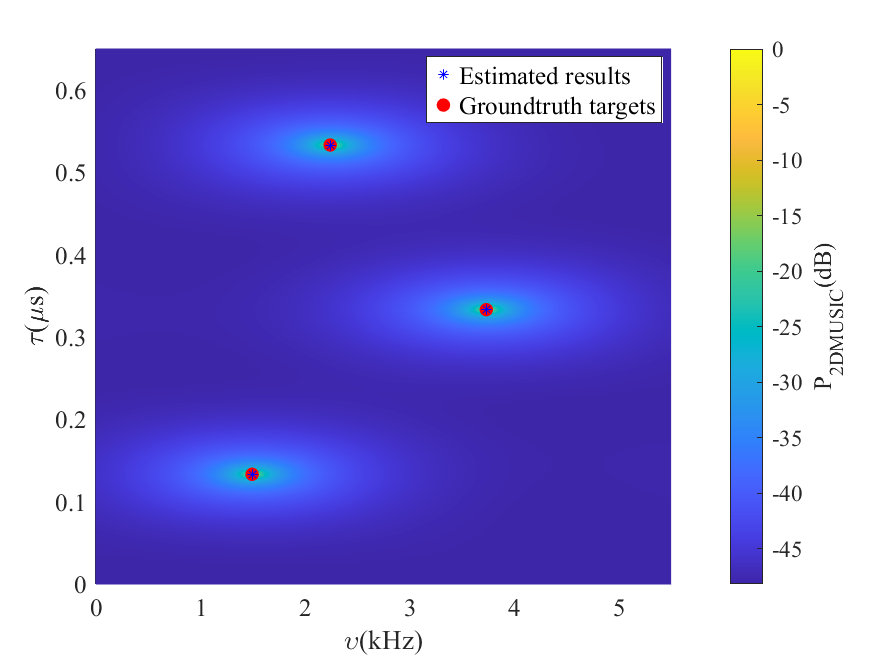}
			\label{2DMUSICS}
		\end{minipage}%
	}%
	\caption{A comparison of 2D-periodogram and 2D-MUSIC for joint delay-Doppler estimation in MIMO-OFDM ISAC.}\label{2DPAM}
	\vspace{-0.5cm}
\end{figure}

\textbf{2D-MUSIC:} To apply super-resolution algorithms to the joint delay-Doppler estimation problem, the data matrix $\mathbf{X}_{k_{\hat{\theta}}} \in \mathbb{C}^{N \times P}$ in (\ref{d38}) is vectorized as $\mathbf{x}_{k_{\hat{\theta}}} \in \mathbb{C}^{NP \times 1}$, leading to
\begin{equation}
	\setlength\abovedisplayskip{2pt}
	\setlength\belowdisplayskip{2pt}
	\begin{aligned}  
        \mathbf{x}_{k_{\hat{\theta}}}&=\operatorname{vec}\left(\mathbf{X}_{k_{\hat{\theta}}}\right)\\
        &=\sum_{i=1}^{N_{k_{\hat{\theta}}}} \tilde{\alpha}_{i,{k_{\hat{\theta}}}} \mathbf{a}_{\upsilon}(\upsilon_{i,{k_{\hat{\theta}}}})\otimes\mathbf{a}_{\tau}(\tau_{i,{k_{\hat{\theta}}}})+\operatorname{vec}\left(\mathbf{Z}_{{k_{\hat{\theta}}}}\right).
        \end{aligned}
	\label{vec}
\end{equation}
Based on this observation vector, the delay and Doppler can be jointly estimated based on the corresponding steering vector $\mathbf{a}\left(\tau,\upsilon\right)=\mathbf{a}_{\upsilon}\left(\upsilon\right)\otimes\mathbf{a}_{\tau}\left(\tau \right)\in\mathbb{C}^{NP\times 1}$.
However, since there is only one snapshot in \mbox{(\ref{vec})} for joint delay-Doppler estimation, it is not possible to estimate the covariance matrix for $\mathbf{x}_{{k_{\hat{\theta}}}}$. Thus, modified ``spatial smoothing" preprocessing (MSSP) \cite{dd42} is performed to achieve decoherence. By performing MSSP in the subcarrier and symbol domains of $\mathbf{X}_{k_{\hat{\theta}}}\in \mathbb{C}^{N \times P}$ in (\ref{d38}) with a $N_{\mathrm{sub}}\times P_{\mathrm{sub}}$ window, and reformulating each submatrix into a vector, $N_{\mathrm{snap}}=\begin{pmatrix}N-N_{\mathrm{sub}}+1\end{pmatrix}\times \begin{pmatrix}P-P_{\mathrm{sub}}+1\end{pmatrix}$ snapshots can thus be obtained. Mathematically, the submatrix of each smoothing window is
\begin{equation}
	\setlength\abovedisplayskip{2pt}
	\setlength\belowdisplayskip{2pt}
	\begin{aligned}  
        {\mathbf{X}}_{{k_{\hat{\theta}}}}^{n_s,p_s} = {{\mathbf{X}}_{k_{\hat{\theta}}}}\left( {n_s:n_s + {N_{{\mathrm{sub}}}} - 1,p_s:p_s + {P_{{\mathrm{sub}}}} - 1} \right), \\
        n_s = 1, \cdots ,N - {N_{{\rm{sub}}}} + 1, p_s = 1, \cdots ,P - {P_{{\rm{sub}}}} + 1.
        \end{aligned}
	\label{Swsm}
\end{equation}
By reconstructing each submatrix into a column vector with $\mathbf{x}_{q_{n_s,p_s}}^S=\mathrm{vec}\left({\mathbf{X}}_{k_{\hat{\theta}}}^{n_s,p_s}\right) \in\mathbb{C}^{N_{\mathrm{sub}} P_{\mathrm{sub}}\times 1}, q_{{n_s,p_s}}=n_s+p_s(N- {N_{{\rm{sub}}}} + 1)$, $N_{\mathrm{snap}}$ vectors can be obtained. Then, a new observation matrix can be created by combining these vectors as columns of the matrix 
\begin{equation}
	\setlength\abovedisplayskip{2pt}
	\setlength\belowdisplayskip{2pt}
	\begin{aligned}  
        \mathbf{X}_{k_{\hat{\theta}}}^S=\left[\mathbf{x}_1^S,\mathbf{x}_2^S,\cdots,\mathbf{x}_{N_{\mathrm{snap}}}^S\right] \in\mathbb{C}^{N_{\mathrm{sub}} P_{\mathrm{sub}}\times N_{\mathrm{snap}}}.
        \end{aligned}
	\label{smoothing}
\end{equation}
Similar to (\ref{EVD}) and (\ref{PMUSIC}), the 2D-MUSIC spectrum for joint delay and Doppler estimation can be obtained by
\begin{equation}
	\setlength\abovedisplayskip{2pt}
	\setlength\belowdisplayskip{2pt}
	\begin{aligned}  
        P_{\mathrm{MUSIC}}\left(\tau,\upsilon\right)=\frac1{\mathbf{a}_S^H\left(\tau,\upsilon\right)\mathbf{E}_n\mathbf{E}_n^H\mathbf{a}_S\left(\tau,\upsilon\right)},
        \end{aligned}
	\label{2DMUISC}
\end{equation}
 where $\mathbf{E}_n$ denotes the noise subspace, $\mathbf{a}_{S}\left(\tau,\upsilon\right)=\mathbf{a}_{\upsilon}\left(\upsilon,P_{\mathrm{sub}}\right)\otimes\mathbf{a}_{\tau}\left(\tau,N_{\mathrm{sub}}\right)\in\mathbb{C}^{N_{\mathrm{sub}}P_{\mathrm{sub}}\times1}$ is the corresponding steering vector, $\mathbf{a}_{\tau}\left(\tau,N_{\mathrm{sub}}\right)$ and $\mathbf{a}_{\upsilon}\left(\upsilon,P_{\mathrm{sub}}\right)$ are the steering vectors in the delay and Doppler domains in \mbox{(\ref{Rsv})} and \mbox{(\ref{Dsv})}, respectively. The peaks of $P_{\mathrm{MUSIC}}\left(\tau,\upsilon\right)$ correspond to the delays and Doppler frequency shifts of the sensing targets.

For the simulation setup in \mbox{Table \ref{Ssetup}}, consider the challenging scenario where the AoAs of the three targets are all $10^\circ$, while the other parameters remain unchanged. A comparison of the 2D-periodogram and 2D-MUSIC methods for joint delay-Doppler estimation in this case is shown in \mbox{Figs.~\ref{2DFFTS}} and \mbox{\ref{2DMUSICS}}, where MSSP is perfomed for the 2D-MUSIC algorithm with a window size of $N_{\mathrm{sub}}\times P_{\mathrm{sub}}=\frac{N}{2}\times \frac{P}{2}=64\times 32$. Similar to the 1D-periodogram and 1D-MUSIC, the spectrum of 2D-MUSIC is smoother and exhibits sharper peaks compared to that of the 2D-periodogram. Consequently, 2D-MUSIC offers higher delay and Doppler resolution and can prevent weak targets from being obscured by the sidelobes of strong targets, thereby reducing inter-target interference and achieving better sensing accuracy. However, for 2D-MUSIC, one must correctly choose the dimension of the signal subspace in order to include the impact of the weak targets, and this is a difficult task in general.

\subsection{Joint Three-domain Parameter Estimation}

In this subsection, we present the joint three-domain parameter estimation method outlined in \mbox{Fig.~\ref{PE}}. Joint AoA-delay-Doppler estimation simultaneously determines the angles, delays, and Doppler frequency shifts of the sensing targets using the signal $\overline{\mathbf{Y}}_s$ after element-wise symbol division in \mbox{(\ref{d25})}.

\textbf{3D-Periodogram:} To reduce the computational complexity of calculating $f(\theta,\tau,\upsilon)$ over the observation angle $\theta$, delay $\tau$, and Doppler $\upsilon$, an $N_{\theta}^{\mathrm{IFFT}}$-point IFFT, $N_{\tau}^{\mathrm{IFFT}}$-point IFFT, and $N_{\upsilon}^{\mathrm{FFT}}$-point FFT are performed in the antenna, subcarrier, and symbol domains of $\overline{\mathbf{Y}}_s$ in \mbox{(\ref{d25})}. When $N_{\theta}^{\mathrm{IFFT}}>M_r$, ${N_{\tau}^\mathrm{IFFT}}>N$ or ${N_{\upsilon}^\mathrm{FFT}}>P$, zero-padding is utilized to convert $\overline{\mathbf{Y}}_s \in \mathbb{C}^{M_r\times N\times P}$ into $\overline{\mathbf{Y}}_s^{\mathrm{ZP}} \in \mathbb{C}^{N_{\theta}^{\mathrm{IFFT}}\times {N_{\tau}^\mathrm{IFFT}}\times {N_{\upsilon}^\mathrm{FFT}}}$. Then, the resulting 3D-periodogram $\mathbf{F}_{\mathrm{3D}}\in \mathbb{C}^{N_{\theta}^{\mathrm{IFFT}}\times {N_{\tau}^\mathrm{IFFT}}\times {N_{\upsilon}^\mathrm{FFT}}}$ can be expressed as
\begin{equation}
	\setlength\abovedisplayskip{2pt}
	\setlength\belowdisplayskip{2pt}
	\begin{aligned}
\mathbf{F}_{\mathrm{3D}}=\left|\mathrm{FFT}_{N_{\upsilon}^\mathrm{FFT}}^3\left\{\mathrm{IFFT}_{N_{\tau}^\mathrm{IFFT}}^2\left\{ \mathrm{IFFT}_{N_{\theta}^\mathrm{IFFT}}^1\left\{ \overline{\mathbf{Y}}_s^{\mathrm{ZP}} \right\}\right\}\right\}\right|^2.
        \end{aligned}
	\label{d68}
\end{equation}
Similar to the 1D and 2D cases, a spectral shift in the angle and Doppler domains of $\mathbf{F}_{\mathrm{3D}}$ to center the zero frequency component within the spectrum is required. Then, from the peak indices $\big(\hat{n}^k_{\theta},\hat{n}^k_{\tau},\hat{n}^k_{\upsilon}\big)$ of $\mathbf{F}_{\mathrm{3D}}$, the angle $\hat{\theta}_k$, delay $\hat{\tau}_k$, and Doppler frequency $\hat{\upsilon}_k$, $\forall k=1,\cdots,K$, of the $K$ sensing targets can be estimated as
\begin{equation}
	\setlength\abovedisplayskip{2pt}
	\setlength\belowdisplayskip{2pt}
	\begin{aligned}  
        \hat{\theta}_k=\arcsin\biggl(\frac{\hat{n}_{\theta}^k\lambda}{N_{\theta}^\text{IFFT}d}\biggr), \hat{\tau}_k=\frac{\hat{n}_{\tau}^k}{\Delta f N_{\tau}^{\mathrm{IFFT}}},\hat{\upsilon}_{k}=\frac{\hat{n}_{\upsilon}^k}{T_sN_{\upsilon}^{\mathrm{FFT}}}.
        \end{aligned}
	\label{3DFFTtauD}
\end{equation}

\textbf{3D-MUSIC:} Similar to 2D-MUSIC, MSSP is essential for decoherence. With a smoothing window size of $M_{\mathrm{sub}} \times N_{\mathrm{sub}}\times P_{\mathrm{sub}}$ over $\overline{\mathbf{Y}}_s$ in \mbox{(\ref{d25})}, $N_{\mathrm{snap}}=\begin{pmatrix}M-M_{\mathrm{sub}}+1\end{pmatrix}\times\begin{pmatrix}N-N_{\mathrm{sub}}+1\end{pmatrix}\times \begin{pmatrix}P-P_{\mathrm{sub}}+1\end{pmatrix}$ snapshots can be obtained. Then, the resulting observation matrix  $\tilde{\mathbf{Y}}^S\in\mathbb{C}^{M_\mathrm{sub} N_{\mathrm{sub}} P_{\mathrm{sub}}\times N_{\mathrm{snap}}}$ after MSSP is
\begin{equation}
	\setlength\abovedisplayskip{2pt}
	\setlength\belowdisplayskip{2pt}
	\begin{aligned}  
        \tilde{\mathbf{Y}}^S=\left[\overline{\mathbf{y}}_1^S,\overline{\mathbf{y}}_2^S,\cdots,\overline{\mathbf{y}}_{N_{\mathrm{snap}}}^S\right].
        \end{aligned}
	\label{3d-smoothing}
\end{equation}
Thus, the 3D-MUSIC spectrum for joint AoA-delay-Doppler estimation can be obtained as
\begin{equation}
	\setlength\abovedisplayskip{2pt}
	\setlength\belowdisplayskip{2pt}
	\begin{aligned}  
        P_{\mathrm{MUSIC}}\left(\theta,\tau,\upsilon\right)=\frac1{\mathbf{a}_S^H\left(\theta,\tau,\upsilon\right)\mathbf{E}_n\mathbf{E}_n^H\mathbf{a}_S\left(\theta,\tau,\upsilon\right)},
        \end{aligned}
	\label{3DMUISC}
\end{equation}
where $\mathbf{a}_{S}\left(\theta, \tau, \upsilon\right)=\mathbf{a}_{\upsilon}\left(\upsilon,P_{\mathrm{sub}}\right)\otimes\mathbf{a}_{\tau}\left(\tau,N_{\mathrm{sub}}\right)\otimes\mathbf{a}_{\mathrm{R}}\left(\theta,M_{\mathrm{sub}}\right)\in\mathbb{C}^{M_{\mathrm{sub}} N_{\mathrm{sub}} P_{\mathrm{sub}}\times1}$ is the corresponding steering vector.
The peaks of $P_{\mathrm{MUSIC}}\left(\theta,\tau,\upsilon\right)$ correspond to the AoA, delay, and Doppler frequency shift of the sensing targets, provided the signal subspace dimension is correctly chosen.

\begin{table*}[!t]
\renewcommand{\arraystretch}{1.4}
\centering
\caption{Complexity of 1D parameter estimation algorithms for AoA estimation in MIMO-OFDM ISAC.}
\label{cx1D}
\begin{tabular}{|c|c|}
\hline
Algorithm & Theoretical complexity \\ \hline
    Periodogram \cite{dd32}       &    $Q{N_{\theta}^{{\text{IFFT}}}}{\log _2}{N_{\theta}^{{\text{IFFT}}}}$                    \\ \hline
    MUSIC \cite{music}       &      $Q{M_r^2} + {M_r^3} + \left( {2M_r\left( {M_r - K} \right) + M_r} \right) {N_{\theta}^{\text{s}}}$                  \\ \hline
    PM-MUSIC~\cite{dd39,dd40}       &  \begin{tabular}[c]{@{}c@{}}$Q{M_r^2} + M_rK\left( {M_r - K} \right) + \left( {2M_r\left( {M_r - K} \right) + M_r} \right)  {N_{\theta}^{\text{s}}}$\end{tabular}         \\ \hline
    FFT-MUSIC \cite{dd41}       &      $Q{M_r^2} + {M_r^3} + \left( {M_r - K} \right){N_{\theta}^{{\text{IFFT}}}}{\log _2}{N_{\theta}^{{\text{IFFT}}}}$                       \\ \hline
    LS-ESPRIT \cite{esprit}      &         $Q{M_r^2} + {M_r^3} + {K^2}\left( {M_r - 1} \right) + {K^3} $                   \\ \hline
    TLS-ESPRIT \cite{esprit}      &          $Q{M_r^2} + {M_r^3} + {\left( {2K} \right)^2}\left( {M_r - 1} \right) + 10{K^3} $                  \\ \hline
    PM-ESPRIT~\cite{dd39,dd40}       &      \begin{tabular}[c]{@{}c@{}}$ Q{M_r^2} + M_rK\left( {M_r - K} \right) + {Q^2}\left( {M_r - 1} \right) + {K^3}$ \end{tabular}   \\ \hline
    OMP \cite{OMP1,OMP2,OMP3}      &             $ KMN_{\theta}^sQ+MK^3+MK^2Q$               \\ \hline
\end{tabular}
\vspace{0.2cm}
\end{table*}
\begin{table*}[!t]
\renewcommand{\arraystretch}{1.4}
\centering
\caption{The complexity of 1D algorithms along different antenna number for AoA estimation in MIMO-OFDM ISAC.}
\label{cx1DA}
\begin{tabular}{|c|cccccccccc|}
\hline
\multirow{4}{*}{Algorithms} & \multicolumn{10}{c|}{Basic setup and complexity}                                                                                                                                                                                                                                                                   \\ \cline{2-11} 
                            & \multicolumn{10}{c|}{$N=128, P=64, Q=NP=8192, K=3, N^s=N_{\theta}^{\mathrm{IFFT}}=1800$}                                                                                                                                                                                                                           \\ \cline{2-11} 
                            & \multicolumn{2}{c|}{$M_r=16$} & \multicolumn{2}{c|}{$M_r=64$}   & \multicolumn{2}{c|}{$M_r=256$}    & \multicolumn{2}{c|}{$M_r=512$}  & \multicolumn{2}{c|}{$M_r=1024$} \\ \cline{2-11} 
                            & \multicolumn{1}{c|}{Theoretical} & \multicolumn{1}{c|}{Time(s)} & \multicolumn{1}{c|}{Theoretical} & \multicolumn{1}{c|}{Time(s)} & \multicolumn{1}{c|}{Theoretical} & \multicolumn{1}{c|}{Time(s)} & \multicolumn{1}{c|}{Theoretical} & \multicolumn{1}{c|}{Time(s)} & \multicolumn{1}{c|}{Theoretical} & Time(s) \\ \hline
Periodogram                 & \multicolumn{1}{c|}{1.596e8}     & \multicolumn{1}{c|}{0.0658}  & \multicolumn{1}{c|}{1.596e8}     & \multicolumn{1}{c|}{0.0671}    & \multicolumn{1}{c|}{1.596e8}     & \multicolumn{1}{c|}{0.0785}  & \multicolumn{1}{c|}{1.596e8}     & \multicolumn{1}{c|}{0.0928} & \multicolumn{1}{c|}{1.596e8}     & \multicolumn{1}{c|}{0.1020}  \\ \hline
MUSIC                       & \multicolumn{1}{c|}{2.879e6}     & \multicolumn{1}{c|}{0.0020}  & \multicolumn{1}{c|}{4.799e7}     & \multicolumn{1}{c|}{0.0061}    & \multicolumn{1}{c|}{7.874e8}     & \multicolumn{1}{c|}{0.0287}  & \multicolumn{1}{c|}{3.221e9}     & \multicolumn{1}{c|}{0.0946} & \multicolumn{1}{c|}{1.343e10}     & \multicolumn{1}{c|}{0.3772} \\ \hline
PM-MUSIC                    & \multicolumn{1}{c|}{2.876e6}     & \multicolumn{1}{c|}{0.0018}  & \multicolumn{1}{c|}{4.774e7}     & \multicolumn{1}{c|}{0.0056}    & \multicolumn{1}{c|}{7.708e8}     & \multicolumn{1}{c|}{0.0256}  & \multicolumn{1}{c|}{3.088e9}     & \multicolumn{1}{c|}{0.0791}  & \multicolumn{1}{c|}{1.236e10}     & \multicolumn{1}{c|}{0.2790} \\ \hline
ROOT-MUSIC                  & \multicolumn{1}{c|}{/}           & \multicolumn{1}{c|}{0.0014}  & \multicolumn{1}{c|}{/}           & \multicolumn{1}{c|}{0.0110}    & \multicolumn{1}{c|}{/}           & \multicolumn{1}{c|}{0.3239}  & \multicolumn{1}{c|}{/}           & \multicolumn{1}{c|}{1.1291} & \multicolumn{1}{c|}{/}           & \multicolumn{1}{c|}{4.7785} \\ \hline
FFT-MUSIC                   & \multicolumn{1}{c|}{2.354e6}     & \multicolumn{1}{c|}{0.0015}  & \multicolumn{1}{c|}{3.500e7}     & \multicolumn{1}{c|}{0.0049}   & \multicolumn{1}{c|}{5.586e8}     & \multicolumn{1}{c|}{0.0230}  & \multicolumn{1}{c|}{2.292e9}     & \multicolumn{1}{c|}{0.0704}  & \multicolumn{1}{c|}{9.684e9}     & \multicolumn{1}{c|}{0.3018}  \\ \hline
LS-ESPRIT                   & \multicolumn{1}{c|}{2.101e6}     & \multicolumn{1}{c|}{0.0011}  & \multicolumn{1}{c|}{3.382e7}     & \multicolumn{1}{c|}{0.0038}   & \multicolumn{1}{c|}{5.537e8}     & \multicolumn{1}{c|}{0.0239}  & \multicolumn{1}{c|}{2.282e9}     & \multicolumn{1}{c|}{0.0889} & \multicolumn{1}{c|}{9.664e9}     & \multicolumn{1}{c|}{0.4400}   \\ \hline
TLS-ESPRIT                  & \multicolumn{1}{c|}{2.102e6}     & \multicolumn{1}{c|}{0.0011}  & \multicolumn{1}{c|}{3.382e7}     & \multicolumn{1}{c|}{0.0038}   & \multicolumn{1}{c|}{5.537e8}     & \multicolumn{1}{c|}{0.0247}  & \multicolumn{1}{c|}{2.282e9}     & \multicolumn{1}{c|}{0.0903}  & \multicolumn{1}{c|}{9.664e9}     & \multicolumn{1}{c|}{0.4380} \\ \hline
PM-ESPRIT                   & \multicolumn{1}{c|}{2.098e6}     & \multicolumn{1}{c|}{0.0010}  & \multicolumn{1}{c|}{3.357e7}     & \multicolumn{1}{c|}{0.0033}   & \multicolumn{1}{c|}{5.371e8}     & \multicolumn{1}{c|}{0.0159}  & \multicolumn{1}{c|}{2.148e9}     & \multicolumn{1}{c|}{0.0502}  & \multicolumn{1}{c|}{8.593e9}     & \multicolumn{1}{c|}{0.1900}  \\ \hline
OMP                   & \multicolumn{1}{c|}{7.094e8}     & \multicolumn{1}{c|}{0.2543}  & \multicolumn{1}{c|}{2.837e9}     & \multicolumn{1}{c|}{0.3004}    & \multicolumn{1}{c|}{1.135e10}     & \multicolumn{1}{c|}{0.5505}  & \multicolumn{1}{c|}{2.270e10}     & \multicolumn{1}{c|}{0.9933} & \multicolumn{1}{c|}{4.540e10}     & \multicolumn{1}{c|}{1.9099} \\ \hline
\end{tabular}
\end{table*}

The ESPRIT and OMP algorithms can also be applied to the 2D and 3D parameter estimation problems. Similar to the MUSIC algorithm, spatial smoothing is also required for ESPRIT to achieve decorrelation. The conventional 2D and 3D ESPRIT algorithms require parameter grouping \cite{700985} to match the estimation results of the different domains, whereas some modified ESPRIT algorithms~\cite{dd35,dd36} can achieve automatic grouping. Furthermore, extensive efforts have been devoted to reducing the complexity of super-resolution methods. For example, the ROOT-MUSIC~\cite{dd37,dd38} algorithm replaces the spectrum search in MUSIC with polynomial rooting, the propagator method (PM)~\cite{dd39,dd40} uses an alternative approach to achieve the eigenvalue decomposition, and the FFT-MUSIC \cite{dd41} algorithm accelerates the MUSIC spectrum peak search using FFT operations. In the following, we analyze the complexity and estimation accuracy of the aforementioned algorithms.

\subsection{Complexity and Performance Analysis}
\subsubsection{Complexity Analysis}  \ \\
\indent We examine the theoretical computational complexity of the aforementioned algorithms in terms of their required number of complex-valued multiplications. For the analysis, we assume the complexity associated with multiplying $M\times Q$ and $Q\times N$ matrices is $\mathcal{O}(QMN)$, while the EVD of an $M\times M$ matrix requires $\mathcal{O}(M^3)$ multiplies. The complexity of various 1D parameter estimation algorithms for AoA estimation is shown in \mbox{Table~\ref{cx1D}}, and these results also apply to delay and Doppler estimation. In these expressions, $Q$ denotes the number of snapshots, $N^{\mathrm{FFT}}/N^{\mathrm{IFFT}}$ and $N^{\mathrm{s}}$ denote the number of FFT/IFFT points and spectrum search points respectively, which are determined by the estimation interval $\eta$, and $N^{\mathrm{FFT}}=N^{\mathrm{IFFT}}=N^{\mathrm{s}}=\frac{\Delta_{\mathrm{max}}}{\eta}$ where $\Delta_{\mathrm{max}}$ represents the maximum estimation range. The estimation interval $\eta$ affects estimation accuracy, and it should be adjustable in order to achieve specific accuracy requirements. It is difficult to express the theoretical complexity of ROOT-MUSIC due to the difficulty of quantifying the complexity of the polynomial rooting process. For a typical example, we set the number of OFDM subcarriers and symbols as $N=128$ and $P=64$, respectively, the number of targets as $K=3$, the estimation interval as $\eta=0.1 ^{\circ}$, and the estimation range as $[-90^{\circ},90^{\circ}]$. Then, the theoretical complexity and code runtime of the various 1D estimation algorithms for different numbers of Sen-RX antennas $M_r$ is shown in \mbox{Table~\ref{cx1DA}}, where the code runtime represents the average runtime of implementing the algorithms over 100 iterations. For reference, all algorithms are implemented using MATLAB, and the specific computer specifications are 12-th Generation Intel(R) Core(TM) i7-12700 CPU, 32GB RAM, and 1TB SSD. Given the limitation inherent in the random access machine models for computational complexity \cite{ram1, ram2}  and the practical discrepancies in MATLAB operations such as accelerated processing for large-size matrix multiplications, in the following we analyze the complexity of each algorithm based on theoretical complexity, with the time complexity provided only for reference.

\begin{table*}[!t]
\renewcommand{\arraystretch}{1.4}
\centering
\caption{Complexity of 2D algorithms for joint delay-Doppler estimation and 3D algorithms for joint AoA-delay-Doppler estimation in MIMO-OFDM ISAC.}
\label{cx2D}
\begin{tabular}{|c|c|}
\hline
Algorithms      & Theoretical complexity \\ \hline
2D-Periodogram & $M_r(NN_\upsilon^\mathrm{FFT}\mathrm{log}_2N_\upsilon^\mathrm{FFT}+N_\upsilon^\mathrm{FFT}N_\tau^\mathrm{IFFT}\mathrm{log}_2N_\tau^\mathrm{IFFT})$          \\ \hline
2D-MUSIC       & $M_rN^2P^2+{(NP)}^3+(2NP\left(NP-K\right)+NP) {N_{\tau}^{\text{s}}} {N_{\upsilon}^{\text{s}}}$          \\ \hline
2D-ESPRIT      & $2M_rN^2P^2+{2(NP)}^3+K^2N\left(P-1\right)+K^2P\left(N-1\right)+2K^3$          \\ \hline
3D-Periodogram & $NPN_\theta^{\mathrm{IFFT}}\log_2N_\theta^{\mathrm{IFFT}}+NN_\theta^{\mathrm{IFFT}}N_\upsilon^{\mathrm{FFT}}\log_2N_\upsilon^{\mathrm{FFT}}+N_\theta^{\mathrm{IFFT}}N_\upsilon^{\mathrm{FFT}}N_\tau^{\mathrm{IFFT}}\log_2N_\tau^{\mathrm{IFFT}}$          \\ \hline
3D-MUSIC       & ${(M_rNP)}^2+{(M_rNP)}^3+(2M_rNP\left(M_rNP-K\right)+M_rNP){N_{\theta}^{\text{s}}}{N_{\tau}^{\text{s}}}{N_{\upsilon}^{\text{s}}}$          \\ \hline
\end{tabular}
\vspace{-0.2cm}
\end{table*}

\begin{figure*}[!t]
\centering
\subfigcapskip=4pt 
	\subfigure[AoA estimation RMSE.]{
		\begin{minipage}[hbt]{0.33\linewidth}
		\centering
		\includegraphics[width=0.95\linewidth]{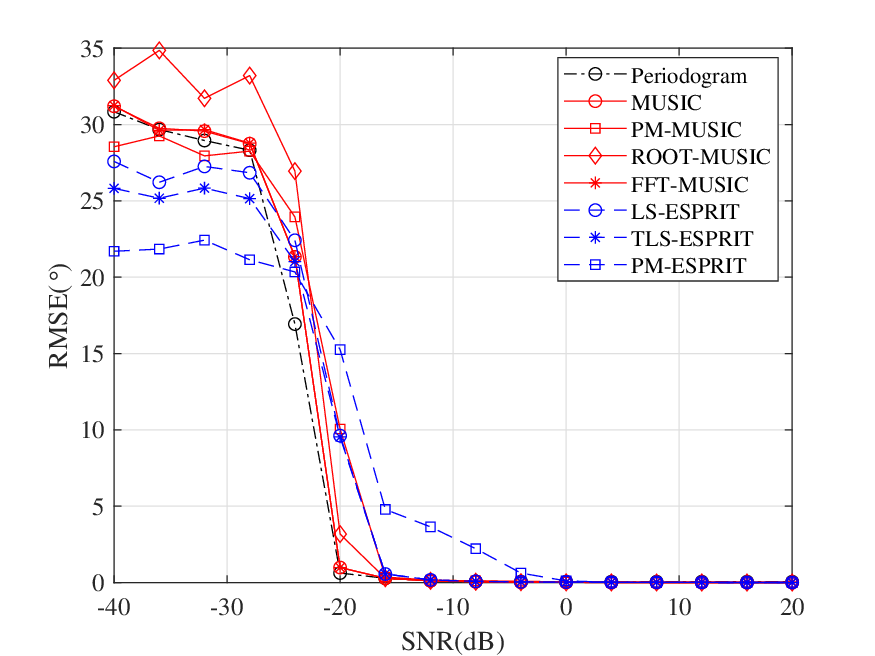}
		\label{RMSE_AoA}
		\end{minipage}%
	}\hspace{-3mm}%
	\subfigure[Delay estimation RMSE.]{
		\begin{minipage}[hbt]{0.33\linewidth}
		\centering
		\includegraphics[width=0.95\linewidth]{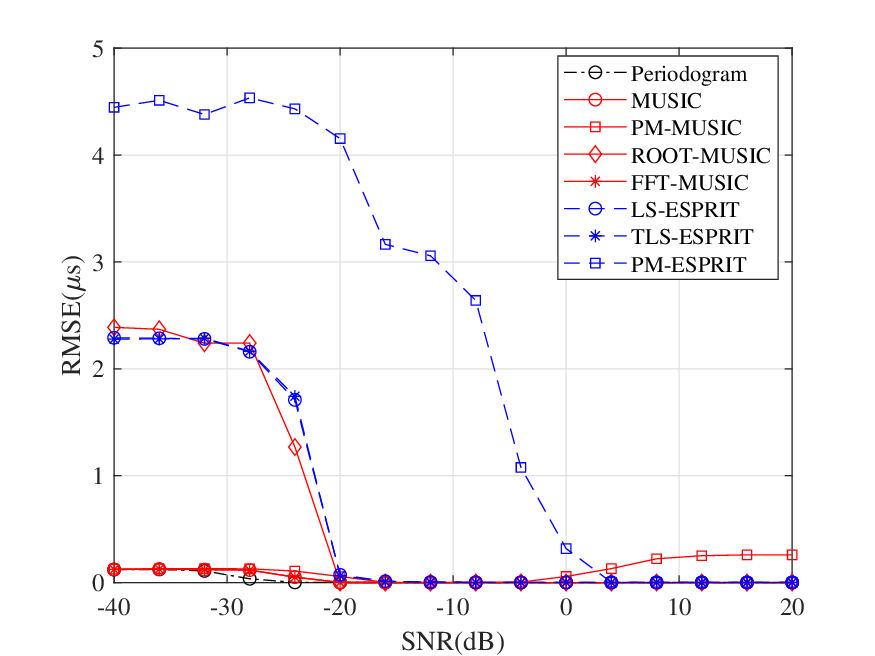}
		\label{RMSE_R}
		\end{minipage}%
	}\hspace{-3mm}%
	\subfigure[Doppler estimation RMSE.]{
		\begin{minipage}[hbt]{0.33\linewidth}
		\centering
		\includegraphics[width=0.95\linewidth]{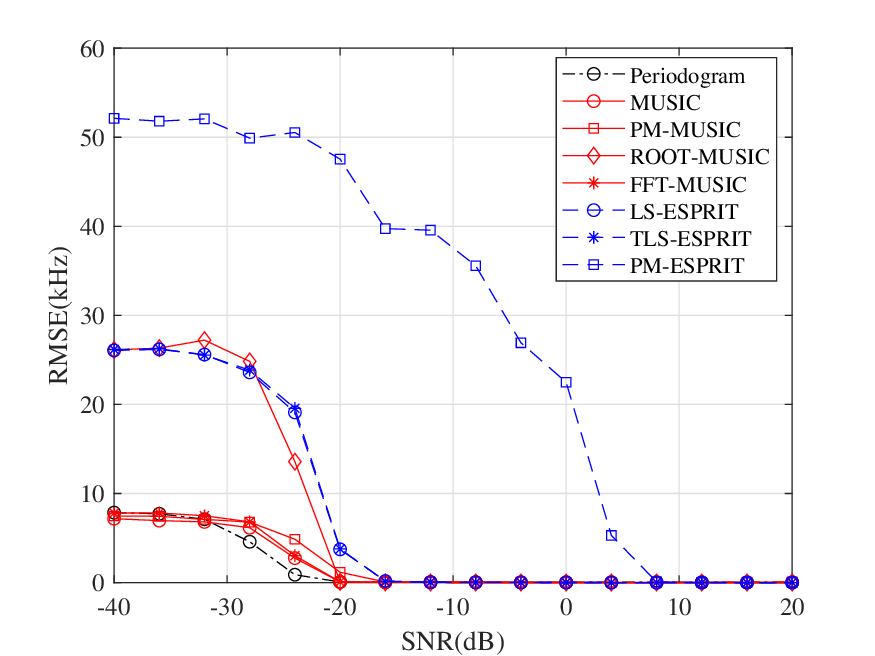}
		\label{RMSE_D}
		\end{minipage}
	}\hspace{-3mm}%
\centering
\caption{Estimation accuracy of various 1D algorithms in the three individual domains for MIMO-OFDM ISAC.}
\label{RMSE3}
\vspace{-0.3cm}
\end{figure*}

As shown in \mbox{Table~\ref{cx1DA}}, when the number of antennas is small and the number of subcarriers and OFDM symbols is large, numerous snapshots are available for AoA estimation, and the periodogram algorithm even exhibits higher complexity than MUSIC and ESPRIT. However, the situation changes as the number of antennas increases. The periodogram becomes more computationally efficient due to the fast computation of the FFT, followed by ESPRIT and MUSIC, while OMP has the highest complexity. The periodogram and OMP can estimate the desired parameters using only partial snapshots or even a single snapshot to reduce complexity, at the cost of increased error. Among the variants of MUSIC, FFT-MUSIC has the lowest complexity, followed by ROOT-MUSIC, with PM-MUSIC offering the least reduction in complexity. Total LS (TLS)-ESPRIT \cite{esprit} is a variation of LS-ESPRIT that offers slightly improved performance, but with a corresponding slight increase in complexity. Unlike MUSIC, the PM algorithm has the potential to significantly reduce the complexity of ESPRIT. Overall, the periodogram, FFT-MUSIC, and PM-ESPRIT algorithms offer the lowest complexity.

The complexity of 2D algorithms for joint delay-Doppler estimation and 3D algorithms for joint AoA-delay-Doppler estimation is presented in Table~\ref{cx2D}. Since the complexity analysis for the 2D/3D algorithms follows a similar approach to that of 1D algorithms, it will not be elaborated on further here. Note that 2D/3D algorithms not only have a high computational complexity but also involve calculations with large-dimensional matrices, imposing significant computational demands. To achieve high-accuracy parameter estimation while mitigating these challenges, coarse-fine grid search \cite{bar2009efficient} and signal decimation \cite{OMP3,dai2024low} techniques can significantly reduce both computational complexity and hardware requirements.

\subsubsection{Estimation Accuracy and Resolution Analysis} \ \\
\indent To study the performance of the 1D parameter estimation algorithms discussed above, we consider a scenario similar to that outlined in Table \ref{Ssetup}, but with the spacing between the parameters increased to guarantee that all the algorithms resolve the targets. We conducted 300 Monte Carlo experiments to estimate the target parameters in three domains via these 1D parameter estimation algorithms for the SNR in \mbox{(\ref{d161})} varying from -40 dB to 20 dB. The root mean-squared error (RMSE) for each algorithm in the three domains is shown in \mbox{Fig.~\ref{RMSE3}}.

We observe a threshold effect at approximately -20 dB where the algorithm performance improves dramatically for the above setup. The periodogram shows the best performance, followed by MUSIC, with ESPRIT providing the lowest estimation accuracy. The performance of PM and ESPRIT in the spatial domain surpasses that in the delay and Doppler domains since ICI affects the latter, while signals in spatial domain remain approximately orthogonal. For delay estimation, PM-MUSIC fails due to signal coherence at high SNR, although this can be addressed by spatial smoothing. FFT-MUSIC and MUSIC have similar estimation accuracy, while ROOT-MUSIC performs poorly at low SNR but approaches MUSIC at high SNR. Finally, TLS-ESPRIT has slightly higher accuracy than LS-ESPRIT at low SNR. PM-ESPRIT exhibits superior performance exclusively at high SNR, particularly for delay and Doppler estimation.

\begin{figure}[!t]
\centering
\subfigcapskip=4pt 
\subfigure[Probability of successfully distinguishing two AoAs for $\mathrm{SNR}=-10\mathrm{dB}$.]{
\begin{minipage}[hbt]{0.9\linewidth}
\centering
\includegraphics[width=0.9\linewidth]{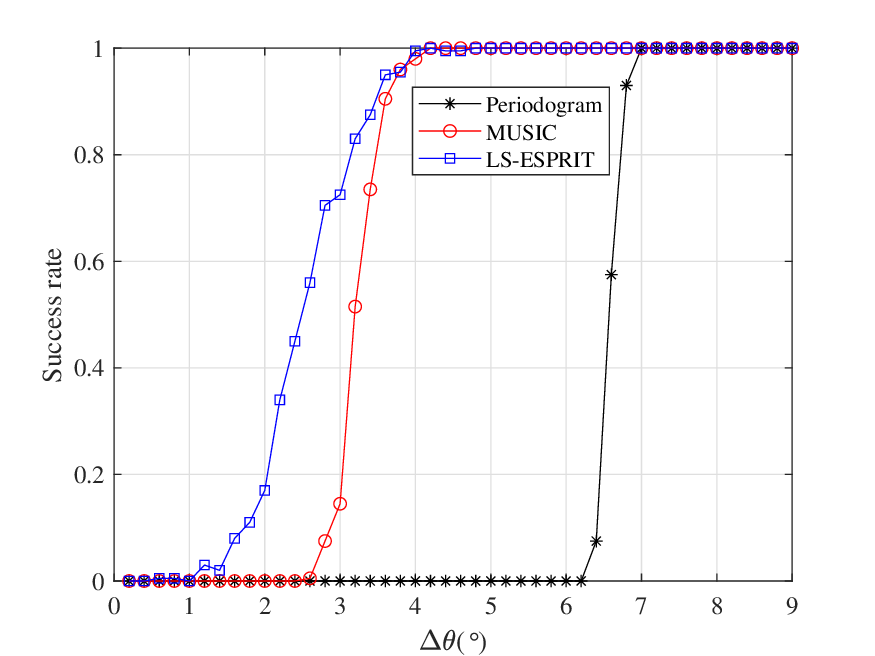}
\label{res0}
\end{minipage}%
}

\subfigure[Probability of successfully distinguishing two AoAs for $\mathrm{SNR}=10\mathrm{dB}$.]{
\begin{minipage}[hbt]{0.9\linewidth}
\centering
\includegraphics[width=0.9\linewidth]{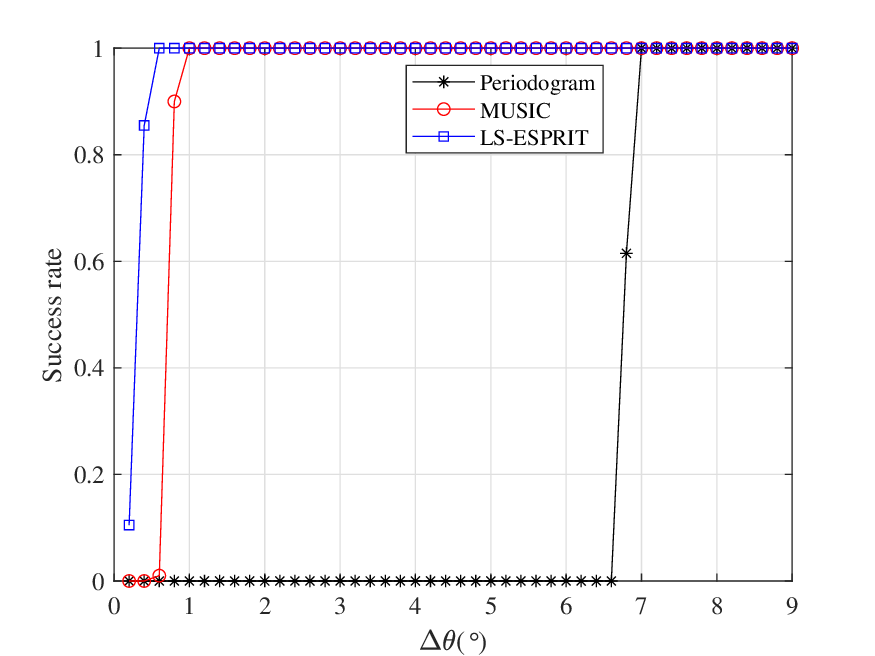}
\label{res10}
\end{minipage}%
}%
\centering
\caption{Probability of successfully resolving two targets using the periodogram, MUSIC, and ESPRIT algorithms for $\mathrm{SNR}=-10\mathrm{dB}$ and $\mathrm{SNR}=10\mathrm{dB}$ with the angle difference varying from $0.2^{\circ}$ to $9^{\circ}$ in MIMO-OFDM ISAC.}
\label{RES}
\vspace{-0.6cm}
\end{figure}

Earlier, we obtained the theoretical resolution of the periodogram algorithm for AoA, delay, and Doppler estimation in (\ref{doares}) and (\ref{distance-resolution}). However, deriving the theoretical resolution for super-resolution algorithms such as MUSIC and ESPRIT is challenging. Therefore, we design a simulation experiment to study this issue. Using the estimation error of two closely positioned targets as a constraint, we statistically determine the probability of successfully resolving the two targets for each of the considered algorithms. Taking the angular resolution as an example, the Sen-RX is chosen to be a ULA with $M_r=16$ half-wavelength-spaced antennas. The ISAC-TX transmits an OFDM signal with $N=128$, the CPI is $P=64$ symbols, subcarrier spacing $\Delta f=120\ \mathrm{kHz}$, and CP duration $T_{\mathrm{cp}}=\frac{1}{4}T$. Since the theoretical resolution of the periodogram is $7.1^{\circ}$ under these conditions, we assume one target located at $0^\circ$ and the other at different angles between $0.2^\circ$ and $9^\circ$. An algorithm is considered to have resolved the two targets successfully if the RMSE is within $0.5^{\circ}$ for $\mathrm{SNR}=-10\ \mathrm{dB}$, and within $0.3^{\circ}$ for $\mathrm{SNR}=10\ \mathrm{dB}$. As the angle of the second target changes, the probabilities of successfully resolving the two targets using the periodogram, MUSIC, and ESPRIT algorithms for $\mathrm{SNR}=-10\mathrm{dB}$ and $\mathrm{SNR}=10\mathrm{dB}$ are shown in \mbox{Figs.~\ref{res0}} and \ref{res10}, respectively, based on $200$ Monte Carlo experiments. 

The results in Figs.~\ref{res0} and \ref{res10} show the angular resolution of the periodogram algorithm is $7^{\circ}$ for both $\mathrm{SNR}=-10\mathrm{dB}$ and $\mathrm{SNR}=10\mathrm{dB}$, consistent with the theoretical resolution in (\ref{doares}). The resolution of the MUSIC and ESPRIT algorithms depends on the SNR. For lower SNR, MUSIC and ESPRIT exhibit similar resolution, approximately $4^{\circ}$. For higher SNR, ESPRIT achieves superior resolution around $0.6^{\circ}$, whereas MUSIC provides resolution around $1^{\circ}$. Overall, as expected, super-resolution algorithms significantly enhance parameter estimation resolution compared to the periodogram, particularly at high SNR levels.

\subsection{Lessons Learned}
In far-field scenarios, the full decoupling of the three-domain signal processing and the linear array-manifold-like structure in each domain provide significant flexibility for target sensing. Based on the four methods depicted in Fig.~\ref{PE} and different radar sensing algorithms, various trade-offs between complexity, resolution, and accuracy can be achieved. In particular, these methods can accomplish target sensing under very low SNR conditions, which exacerbates the disparity between C\&S coverage. Specifically, communication typically relies on the LoS and multipaths between the BS and users, whereas sensing depends on the two-hop path from the ISAC-TX to the target and then to the Sen-RX. Consequently, the communication SNR is generally higher than that for sensing. However, there are also significant differences in the SNR requirements for C\&S signal processing. This discrepancy in both signal SNR and SNR requirements between C\&S leads to differences in their coverage ranges, which warrants further in-depth investigation.

For spatial sensing, super-resolution sensing algorithms are not directly applicable to ISAC systems equipped with analog or hybrid beamforming-based MIMO arrays. Nevertheless, beamforming optimization holds promise for facilitating narrower beams than the Rayleigh limit, thereby enhancing spatial sensing resolution. For target detection, non-super-resolution algorithms can apply constant false alarm rate (CFAR) \cite{CFAR} detection to target sensing periodograms to accurately estimate the number of targets. In contrast, subspace-based super-resolution algorithms divide the signal and noise subspaces based on the magnitude of the eigenvalues of the covariance matrix, which may lead to higher missed detection or false alarm rates. Target detection methods for super-resolution sensing algorithms still require further exploration. Finally, implementing super-resolution sensing algorithms on FPGA chips for efficient computation remains challenging and requires more research effort.

\section{Near-field MIMO-OFDM ISAC}\label{near-field}
In this section, we extend the results in Section \ref{FF_Sensing} to the case where the targets are located in the near-field region of the Sen-RX. While the main signal processing strategies outlined in Fig.~\ref{PE} are still valid in the near-field, the spatial domain estimation must be extended from AoA estimation only in the far-field to joint angle-range estimation in the near-field by exploiting the characteristics of spherical wavefronts. Therefore, this section focuses mainly on spatial domain estimation. 

\begin{figure}[t!]
	\centering
	\includegraphics[width=0.9\linewidth]{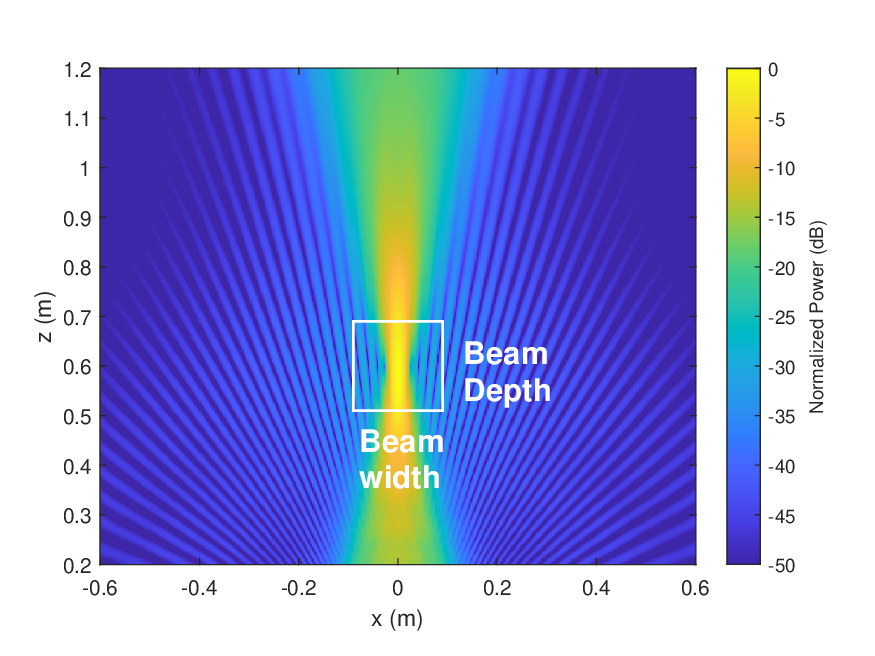}
	\caption{Illustration of beam characteristics in near-field ISAC.}
	\label{BD}
	\vspace{-0.5cm}
\end{figure}

In the near-field region, propagation is described by spherical wavefronts, which produce a nonlinear variation in phase from element to element of the array. By leveraging the distance information embedded in the spherical wavefront, near-field beamforming can focus energy precisely at a designated location. Therefore, the energy can be concentrated in both the angle and distance dimensions, producing a narrow beam width (BW) and beam depth (BD), respectively \cite{BWBD}, as shown in Fig.~\ref{BD}. In wideband scenarios, near-field propagation causes the beams at different frequencies to point to different physical locations, an effect referred to as near-field beam splitting. While this effect may result in significant loss of array gain since it is difficult to achieve beam alignment across a wide frequency band, one can control the angular coverage of the beams at different frequencies in a way that enables rapid CSI acquisition \cite{WBT}.

\subsection{Near-field Sensing Signal Model}
\begin{figure}[t!]
	\centering
	\includegraphics[width=0.9\linewidth]{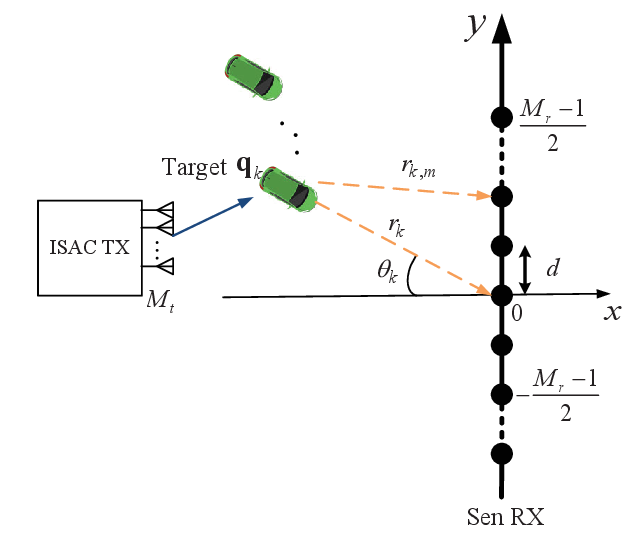}
	\caption{Near-field sensing for MIMO-OFDM ISAC.}
	\label{near-up}
\vspace{-0.6cm}
\end{figure}
As shown in Fig.~\ref{near-up}, we assume that the Sen-RX is equipped with an extremely large-scale ULA, which is placed along the $y$-axis with $\varepsilon_r=0,$ and $ \varepsilon_{m}=m-(M_r+1)/2$ in (\ref{steering_near}). The corresponding steering matrix is denoted as $\mathbf{A}_{\rm{R}}(r,\theta)=[\mathbf{a}_\mathrm{R}(r_1,\theta_1),\cdots,\mathbf{a}_\mathrm{R}(r_K,\theta_K)]$. Based on the model in (\ref{23}), the $M_r \times Q_{N,P}$ received signal is given~by
\begin{equation}
	\setlength\abovedisplayskip{2pt}
	\setlength\belowdisplayskip{2pt}
	\begin{aligned}
		\mathbf{X}=\mathbf{A}_{\rm{R}}(r,\theta)\mathbf{S}+\mathbf{Z},
	\end{aligned}
	\label{near_receive}
\end{equation}
where $\mathbf{S} = [\mathbf{s}_1,\mathbf{s}_2,\cdots,\mathbf{s}_K]^T \in \mathbb{C}^{K \times Q_{N,P}}$ denotes the equivalent reflected target signals described in (\ref{34}). Note that compared with the far-field model in (\ref{23}), the array response vector for near-field sensing is a function of both the target angle and distance, hence allowing simultaneous estimation of both parameters independent of delay-based distance estimation. This will not only improve the distance estimation accuracy (especially if the distance estimation is performed jointly in the AoA and delay domains), but it can also alleviate the synchronization errors that typically arise in traditional delay-based distance estimation techniques. In order to estimate the angle and distance parameters based on spatial-domain processing, several methods are introduced below that either ignore the near-field model and directly apply far-field algorithms or extend far-field algorithms to the near-field case.

\subsection{Directly Applying Far-field Algorithms to Near-field}
\begin{figure}[t!]
	\centering
	\includegraphics[width=0.8\linewidth]{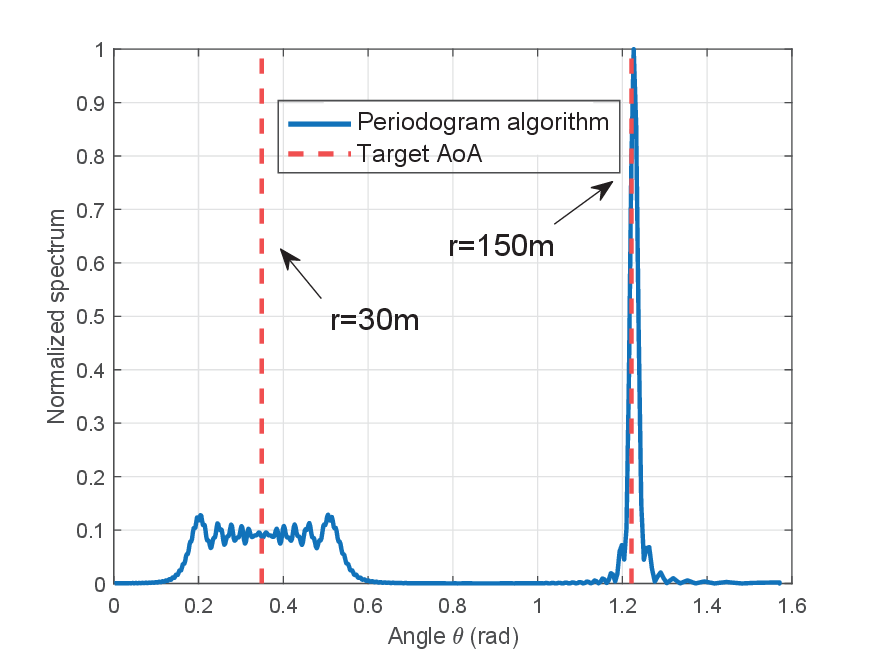}
	\caption{Directly applying far-field algorithms in near-field scenario for AoA estimation.}
	\label{nmfs}
    \vspace{-0.6cm}
\end{figure}

In this section, we illustrate the performance of directly applying the far-field algorithms discussed in Section \ref{far-doa} to near-field scenarios while ignoring the spherical wave propagation. The result for the standard Periodogram is shown in Fig.~\ref{nmfs}, where $K=2$ targets are considered. We see that only the target with $r=150m$ has an accurate AoA estimate based on (\ref{FFTDOA}). For the nearer target at $r=30m$, multiple peaks exist around the ground-truth AoA, and it is due to the mismatch between the near-field channel and the assumed far-field model. This is known as the so-called energy spread effect, which illustrates the necessity of developing dedicated near-field target localization algorithms.

\subsection{Extending Far-field Algorithms to Near-field} 
Below we discuss how to extend algorithms developed for far-field parameter estimation to the near-field case.
\subsubsection{Near-field Beam Focusing}
For near-field models, the distance and angle parameters are closely coupled in the near-field array response vector. In addition, the non-linear phase variation from antenna to antenna in near-field models also means that the IFFT is no longer directly applicable. Therefore, in this section we introduce 2D beamforming in both the angle and distance domains. Based on (\ref{near_receive}), the spatial domain sample covariance matrix of $\bf X$ is given by
\begin{equation}
	\begin{aligned}
		{{\bf{R}}_X} = \frac{1}{{Q_{NP}}}{{\bf{X}}}{{{\bf{X}}}^H},
	\end{aligned}
	\label{near-rx}
\end{equation}
and the 2D periodogram spectrum can be written as
\begin{equation}
	\begin{aligned}
		{P_\mathrm{2D - BF}}\left( {r,\theta } \right) = {{{\bf{a}}_{\rm{R}}^H}}(r,\theta ){{\bf{R}}_X}{\bf{a}}_{\rm{R}}(r,\theta ).
	\end{aligned}
	\label{bfspec}
\end{equation}
The peaks of ${P_\mathrm{2D - BF}}$ in the 2D angle-distance plane can be used to estimate the target angle and distance parameters.
\subsubsection{Near-field MUSIC}
Similar to (\ref{PMUSIC}), by performing the EVD of the covariance matrix ${\bf R}_X$ and obtaining the noise subspace ${\bf E}_n$, the 2D MUSIC spectrum can be expressed as
\begin{equation}
	\begin{aligned}
		{P_\mathrm{2D - MUSIC}}\left( {r, \theta} \right) = \frac{1}{{{{\bf{a}}_{\rm{R}}^H}(r,\theta ){{\bf{E}}_n}{\bf{E}}_n^H{\bf{a}}_{\rm{R}}(r,\theta )}}.
	\end{aligned}
	\label{2DMUSIC}
\end{equation}
\subsubsection{Near-field ESPRIT and Root-MUSIC}
As shown in Section \ref{far-doa}, the far-field ESPRIT algorithm exploits the rotational invariance between identical subarrays, requiring a linear phase variation with the parameters between the subarrays. Therefore, conventional ESPRIT is not directly applicable in near-field sensing. However, for symmetric antenna arrays, a generalized version of ESPRIT can be performed, which will be introduced in Section \ref{symme}. Root-MUSIC also relies on the assumption of steering vectors with a linear phase shift between adjacent elements, and thus is also not directly applicable in the near-field case.
 
\subsubsection{Simulations and discussions}
\begin{figure}[!t]
\centering
\subfigcapskip=-4pt 
\subfigure[Near-field beam focusing spectrum.]{
\begin{minipage}[t!]{0.9\linewidth}
\centering
\includegraphics[width=1\linewidth]{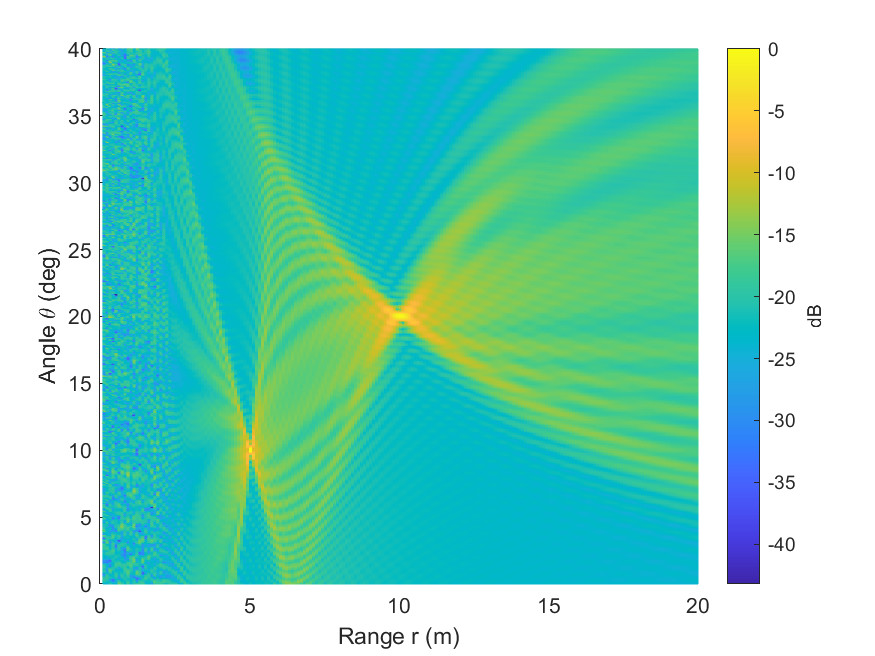}
\label{2dbf}
\end{minipage}%
}

\subfigure[Near-field 2D-MUSIC spectrum.]{
\begin{minipage}[t!]{0.9\linewidth}
\centering
\includegraphics[width=1\linewidth]{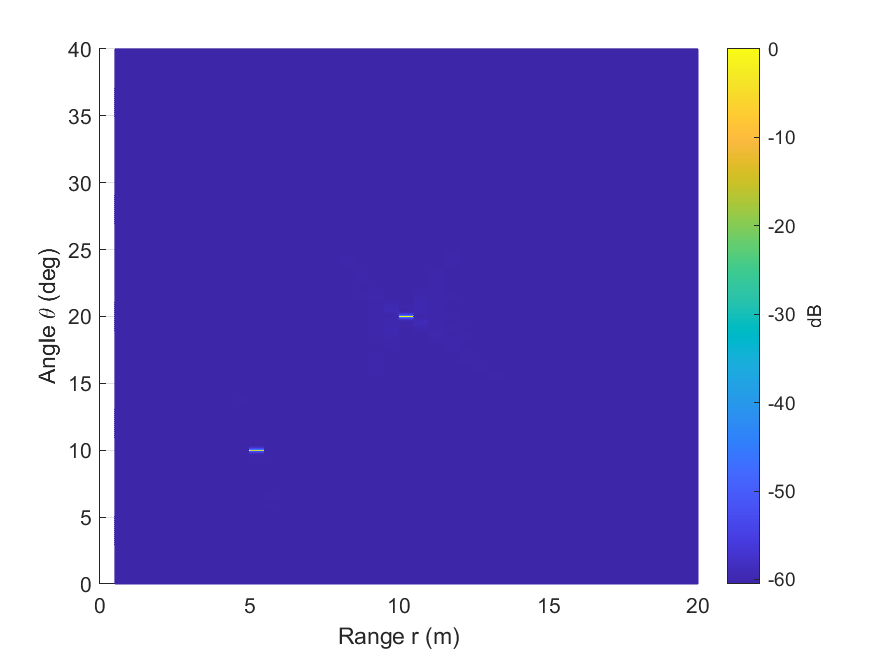}
\label{2dmu}
\end{minipage}%
}%
\centering
\caption{Near-field angle-distance estimation.}
\vspace{-0.6cm}
\end{figure}

Figs.~\ref{2dbf} and~\ref{2dmu} show the 2D spectra for near-field beam focusing and MUSIC respectively. The number of antennas is $M_r=256$, and the locations of the two targets are $(r_1,\theta_1)=(5 \mathrm{m},10^{\circ})$ and $(r_2,\theta_2)=(10 \mathrm{m},20^{\circ})$, respectively. It is observed that both beamforming and MUSIC can localize the closer target with higher accuracy than the distant target, primarily due to the limited accuracy with respect to distance.

We see that several issues arise that complicate parameter estimation in near-field scenarios. Besides the fact that some algorithms can no longer be directly applied, we see that 2D rather than 1D searches are necessary to jointly estimate the AoAs and ranges, which adds complexity. The accuracy of the distance estimate is considerably less reliable than the AoA estimate, especially as the distance increases to the point where the far-field model holds \cite{near-resolution}. This is due to the decreasing sensitivity of the model with distance in such scenarios, and is a factor that must be taken into account when jointly estimating the distance by including the delay domain.

To address these challenges, one feasible approach is to decouple estimation of the angle and distance parameters to reduce the complexity of implementing the parameter optimization. The low-complexity near-field sensing algorithms presented next are based on this idea.

\subsection{Low-complexity Near-Field Estimation Algorithms}
Recently, several algorithms have been proposed to improve the accuracy and resolution of near-field sensing with low computational complexity by efficiently decoupling the distance and angle parameters. Many such methods exploit the approximate Fresnel array response model in (\ref{steering_near_appro}) based on a second-order Taylor approximation, where the parameters to be estimated are taken to be ${\omega _k} =  - \frac{{2\pi {d}\sin {\theta_k}}}{\lambda }$ and ${\psi _k} = \frac{{\pi d^2{{\cos }^2}{\theta_k}}}{{\lambda {r_k}}}$ instead of $\theta_k$ and $r_k$.
\subsubsection{Second-order cumulant (SoC)} 
Note that in (\ref{steering_near_appro}), $\omega_{k}$ only depends on $\theta_k$, while $\psi_{k}$ depends on both $\theta_k$ and $r_k$. Therefore, a dedicated SoC can be constructed to transform the multi-dimensional search into two single-parameter estimation problems. Based on (\ref{near_receive}) and (\ref{steering_near_appro}), the noise-free correlation sequence between the $(am+b)$th and $m$th antenna elements can be written as~\cite{b66}
\begin{align} \label{corre}
	\setlength\abovedisplayskip{2pt}
	\setlength\belowdisplayskip{2pt}
		&r(am + b,m)= \frac{1}{{{Q_{N,P}}}}{\bf{X}}(am \!+ \!b \!+ \!J \!+ \!1,:){{\bf{X}}^H}(m \!+\! J\! +\! 1,:) \nonumber \\
		&= \frac{1}{{{Q_{N,P}}}}{{\bf{A}}_{R,am + b + J + 1}}{\bf{S}}{{\bf{S}}^H}{{\bf{A}}^H_{R,m + J + 1}}\\
		&\mathop  \approx \limits^{(a)} \sum\limits_{k = 1}^K {{p_k}} {e^{j[(a - 1)m + b]{\omega _k} + j\left[ {\left( {{a^2} - 1} \right){m^2} + 2abm + {b^2}} \right]{\psi _k}}}, \nonumber
\end{align}
where ${{\bf{A}}_{R,m}}$ is the $m$th row of ${{\bf{A}}_{R}}$. Approximation $(a)$ holds by assuming ${\bf{s}}_p^H{{\bf{s}}_q} \approx 0,p \ne q$. $J=\frac{M_r-1}{2}$ and $p_{k}=\frac{1}{Q_{N,P}}{\bf{S}}(k,:){{\bf{S}}^H}(k,:)$ denotes the equivalent transmitted signal. Furthermore, by properly choosing $a$ and $b$, $\psi_k$ can be eliminated from the resulting expression, and estimation of the distance and angle parameters can be decoupled. For instance, substituting $(a,b)=(-1,0)$ and $(a,b)=(1,1)$ into (\ref{corre}), we obtain the following correlation sequences
\begin{align} \label{equivalent soc}
		r_1( - m,m) &= \sum\limits_{k = 1}^K {p_{k} {e^{ - 2j{\omega _k}m}}}, m=0,1,...,J,\\
		r_2(m + 1,m) &= \sum\limits_{k = 1}^K {p_{k} {e^{j\left( {{\omega _k} + {\psi _k}} \right)}}{e^{ 2j{\psi _k}m}}},m=0,1,...,J-1. \nonumber
\end{align}
Stacking the $r_1$ and $r_2$ terms into vectors, we obtain the signal model
\begin{equation}
	\begin{aligned}
		{{\bf{r}}_1} = {{\bf{A}}_1}{{{\bf{p}}}_1} \in {\mathbb{C}^{(J+1) \times 1}},\ 
		{{\bf{r}}_2} = {{\bf{A}}_2}{{{\bf{p}}}_2} \in {\mathbb{C}^{J \times 1}},
	\end{aligned}
\end{equation}
where the $k$-th column of the $(J+1)\times K$ matrix ${{\bf{A}}_1}$ is ${{\bf{a}}_1}_k = {[1,{e^{ - 2j{\omega _k}}},...,{e^{ - j({M_r} - 1){\omega _k}}}]^T}$, and ${{\bf{p}}_1} = {[{p_1},...,{p_K}]^T}.$ Similarly, the $k$th column of ${\bf{A}}_2$ and ${\bf{p}}_2$ can be expressed as ${{\bf{a}}_2}_k = {[1,{e^{2j{\psi _k}}},...,{e^{ - j({M_r} - 3){\psi _k}}}]^T}$ and ${{\bf{p}}_2} = {[{p_1}{e^{j({\omega _1} + {\psi _1})}},...,{p_K}{e^{j({\omega _K} + {\psi _K})}}]^T}$, respectively. 

It can be observed that ${\bf{r}}_1$ and ${\bf{r}}_2$ have a form similar to the far-field model. Thus, $\omega_k$ and $\psi_k$ can be obtained using conventional AoA estimation algorithms. In the following, we refer to this approach as ``SoC-AoA". However, the spatial resolution of this approach is reduced to $BW=\frac{4\lambda}{D}$ since the equivalent antenna aperture is half of the physical aperture. Based on SoC, a weighted linear prediction algorithm has been proposed for near-field localization~\cite{b66} and higher-order cumulants (HoC) have been proposed for higher accuracy and wider application~\cite{b99,b68}.

\subsubsection{Reduced dimension (RD) and reduced rank (RR) algorithms} Another method to decouple $\omega_k$ and $\psi_k$ is by decomposing ${\bf a}_{\rm{R}}(r,\theta)$ in (\ref{steering_near_appro}) as~\cite{b65}
\begin{equation}
	\setlength\abovedisplayskip{2pt}
	\setlength\belowdisplayskip{2pt}
	\begin{aligned}
		&{\bf{a}}_{\rm{R}}\left( {\omega ,\psi } \right) =\\
		&\underbrace {\left[ {\begin{array}{*{20}{c}}
					{{e^{j( - J)\omega }}}&{}&{}&{}\\
					{}&{{e^{j( - J + 1)\omega }}}&{}&{}\\
					{}&{}& \ddots &{}\\
					{}&{}&{}&1\\
					{}&{}& {\mathinner{\mkern2mu\raise1pt\hbox{.}\mkern2mu
							\raise4pt\hbox{.}\mkern2mu\raise7pt\hbox{.}\mkern1mu}} &{}\\
					{}&{{e^{j(J - 1)\omega }}}&{}&{}\\
					{{e^{jJ\omega }}}&{}&{}&{}
			\end{array}} \right]}_{{\bf{\Gamma }}\left( \omega \right)}
		\cdot \underbrace {\left[ {\begin{array}{*{20}{l}}
					{{e^{j{{( - J)}^2}\psi }}}\\
					{{e^{j{{( - J + 1)}^2}\psi }}}\\
					\vdots \\
					{{e^{j{{( - 1)}^2}\psi }}}\\
					1
			\end{array}} \right]}_{{\bf{\xi }}\left( \psi  \right)\;},
	\end{aligned}
	\label{rdrr}
\end{equation}
where $J=\frac{M_r-1}{2}$, ${\bf{\Gamma }}\left( \omega \right)\in \mathbb{C}^{(2J+1)\times(J+1)}$ only depends on the angle, while
${\bf{\xi }}\left( \psi  \right)\;\in \mathbb{C}^{(J+1)\times 1}$ is related to both angle and distance. Note that (\ref{rdrr}) allows the 2D search for 2D-MUSIC (\ref{2DMUSIC}) to be converted into two 1D searches, thus reducing the complexity. Specifically, the MUSIC spectrum can be constructed as 
\begin{equation}
	\begin{aligned}
		{P_\mathrm{2D - MUSIC}}\left( {\omega ,\psi} \right)\; &= \frac{1}{{{{\bf{\xi }}^H}(\psi ){{\bf{\Gamma }}^H}(\omega ){{\bf{E}}_n}{\bf{E}}_n^H{\bf{\Gamma }}(\omega ){\bf{\xi }}(\psi )}}\\
		&= \frac{1}{{{{\bf{\xi }}^H}(\psi ){\bf{Q}}(\omega ){\bf{\xi }}(\psi )}}.
		\label{rd_rr}
	\end{aligned}
\end{equation}
In the following, we present angle and distance estimation algorithms for the RD and RR cases, respectively.

\textbf{Reduced-rank algorithm:} 
Since the rank of ${\bf E}_n\in\mathbb{C}^{M_r\times(M_r-K)}$ will be no smaller than $M_r-K$, ${\bf Q}(\omega)={{\bf{\Gamma }}^H}(\omega ){{\bf{E}}_n}{\bf{E}}_n^H{\bf{\Gamma }}(\omega )\in  \mathbb{C}^{(J+1)\times (J+1)}$ can be obtained by performing an EVD on ${\bf R}_X$, and ${\bf {Q}}(x)$ generally has full rank if $M_r-K\ge (J+1)$, i.e., $K\le J$. On the other hand, when $\omega 
\in \Omega=\{\omega_1,...,\omega_K\}$, we have ${{\bf{\xi }}^H}(\psi ){\bf{Q}}({\omega _k}){\bf{\xi }}(\psi ) \to 0$ based on the MUSIC algorithm, which is true only if ${\bf Q}(\omega)$ is rank deficient since generally ${\bf{\xi }}(\psi )\ne 0$~\cite{b94}. Based on these observations, the value of $\theta_k$ can be obtained from the $K$ highest peaks of the following 1D spectral function 
\begin{equation}
	\begin{aligned}
		{\hat \omega _k} = \mathop {\arg \max }\limits_\omega  \frac{1}{{\det \left( {{\bf{Q}}\left( \omega  \right)} \right)}},
	\end{aligned}
\end{equation}
and the corresponding $r_k$ is obtained by searching for peaks in the distance domain using
\begin{equation}
	\begin{aligned}
		{\hat r_k} = \mathop {\arg \max }\limits_r {P_\mathrm{2D - MUSIC}}\left( {{{\hat \theta }_k},r} \right).
	\end{aligned}
\end{equation}

\textbf{Reduced-dimension algorithm:} 
The spectrum search in (\ref{rd_rr}) can also be viewed as the following optimization problem
\begin{equation}
	\begin{aligned}
		\mathop {\min }\limits_{\psi ,\omega }~ {{\bf{\xi }}^H}(\psi ){\bf{Q}}(\omega ){\bf{\xi }}(\psi )\quad
		\rm s.t.~{\bf{e}}_1^H{\bf{\xi }}(\psi ) = 1,
	\end{aligned}
	\label{rd_opti}
\end{equation}
where ${{\bf{e}}_1} = {[0,0,...,1]^T} \in {\mathcal{C}^{(J + 1) \times 1}}$. Solving (\ref{rd_opti}), ${\bf{\xi }}(\psi )$ can be expressed as ${\bf{\xi }}(\psi ) = \frac{{{{\bf{Q}}^\dagger }(\omega ){{\bf{e}}_1}}}{{{\bf{e}}_1^H{{\bf{Q}}^\dagger }(\omega ){{\bf{e}}_1}}}$. Therefore, (\ref{rd_rr}) can be transformed into a 1D search in terms of the angle parameter~\cite{b65}
\begin{equation}
	\begin{aligned}
		{\hat \omega _k} = \mathop {\arg \min }\limits_\omega  \frac{1}{{{\bf{e}}_1^H{{\bf{Q}}^{ \dagger}}\left( \omega  \right){{\bf{e}}_1}}}.
	\end{aligned}
\end{equation}
Furthermore, $\psi_k$ can be obtained by constructing the LS problem 
\begin{equation}
	\begin{aligned}
		\mathop {\min }\limits_{{\bf{c}}_k} \left\| {{\bf{P}}{{\bf{c}}_k} - {{\widehat {\bf{g}}}_k}} \right\|^2,
		\label{ls}
	\end{aligned}
\end{equation}
where ${\bf{c}}_k=[c_{k,0},\psi_k]^T$, ${\widehat {\bf{g}}_k} = {\rm{angle}}\left( {\frac{{{{\bf{Q}}^{ \dagger}}\left( {\hat \omega } \right){{\bf{e}}_1}}}{{{\bf{e}}_1^H{{\bf{Q}}^{ \dagger}}\left( {\hat \omega } \right){{\bf{e}}_1}}}} \right)$, and ${\bf{P}} = {\left( \begin{array}{l}
		1,...,1\\
		{J^2},...,0
	\end{array} \right)^T}.$ After finding $\psi_k$ using (\ref{ls}), the angle and distance parameters can be estimated as
\begin{equation}
	\begin{aligned}
		\;{\hat \theta _k} =  - \arcsin \left( {\frac{{{{\hat \omega }_k}\lambda }}{{2\pi d}}} \right),\ 
		{\hat r_k} = \frac{{\pi {d^2}{{\cos }^2}{{\hat \theta }_k}}}{{\lambda {{\hat \psi }_k}}}.
	\end{aligned}
\end{equation}
\subsubsection{FFT-enhanced low-complexity and super-resolution algorithm}
In order to further reduce the complexity, the authors in \cite{FFTsuper} proposed an efficient algorithm to narrow the search region for 2D angle-distance estimation in (\ref{2DMUSIC}). The main idea of this algorithm is to first eliminate those regions that do not contain any targets. Afterwards, 2D MUSIC is performed within a region of reduced size, effectively balancing computational complexity and super-resolution. This algorithm involves three steps:

\textbf{FFT-enhanced angle cluster determination:} 
Applying an $S$-point FFT and IFFT to the columns and rows of sample covariance matrix ${\bf R}_X$ in (\ref{near-rx}), respectively, the output can be used to eliminate angular regions where no targets exist. The resulting spectrum can be expressed as ${{\bf{p}}} = \text{diag}({\bf{W}}{{\bf{R}}_X}{{\bf{W}}^{ - 1}}),$
where $\bf W$ is the DFT matrix, and $\text{diag(\bf A)}$ extracts the diagonal
elements of $\bf A$ to form a vector. The $n$th angle cluster is represented by the minimum and maximum angle value within this cluster, denoted as ${\bm{\alpha} }_n = [{\underline{\alpha}_n},{{\overline \alpha }_n}]$ for $n=1,\cdots,L$, where $L$ is the number of clusters.

\textbf{Distance cluster determination:}
For each angle cluster ${\bm{\alpha} }_n = [{\underline{\alpha}_n},{{\overline \alpha }_n}]$, 1D beamforming is performed over the distance domain for $\underline{\alpha_n}$ and ${{\overline \alpha }_n}$, 
\begin{equation}
	\begin{aligned}
		{P_{low}}(r) = {\bf{a}}_{\rm R}^H(r,{\underline{\alpha}_n}){{\bf{R}}_X}{{\bf{a}}_{\rm R}}(r,{\underline{\alpha}_n}),\\
		{P_{up}}(r) = {\bf{a}}_{\rm R}^H(r,{{\overline \alpha }_n}){{\bf{R}}_X}{{\bf{a}}_{\rm R}}(r,{{\overline \alpha }_n}).
	\end{aligned}
\end{equation}
By setting a threshold $\Gamma$, the distance cluster ${{{\bm{\underline \beta }}}_n}$ can be determined by the condition $P_{low}(r)<\Gamma$, and ${{{\bm{\bar \beta }}}_n}$ can be obtained similarly. Therefore, the distance cluster is expressed as ${\bm{\beta}}_n = {{{\bm{\underline \beta }}}_n} \cup {{{\bm{\bar \beta }}}_n}.$

\textbf{Super-resolution 2D sensing:}
Finally, for each cluster $n$, 2D MUSIC is performed for $\theta\in {\bm \alpha}_n$ and $r\in {\bm {\beta}}_n$. By clustering the angle and distance dimensions, the number of grid points for the 2D search and the overall complexity is significantly reduced.

\subsubsection{Symmetry-based near-field estimation algorithms}\label{symme}

Another strategy to reduce the computational complexity of the 2D peak search is by exploiting the symmetry of the array. Here, we present two approaches to decouple the problems of angle and distance estimation by leveraging the symmetric structure of the antenna array for multiple targets.

\textbf{Modified MUSIC~\cite{6111312}:} In the absence of noise, the anti-diagonal elements of the covariance ${\bf{R}}_X$ in (\ref{near-rx}) are given by
\begin{align}
		&{\bf{R}}_X\left[n,M_r+1-n\right]=\mathbb{E}\{{{\bf{X}}}{{{\bf{X}}}^H}\} \nonumber \\
		&=\mathbb{E}\left\{ \left(\sum\limits_{k = 1}^K {{{e^{j\left( n{\omega _k} + {n^2}{\psi _k} \right)}}}{\mathbf{s}_k^T}}\right)\right.
		\left.  \left(\sum\limits_{k = 1}^K {{{e^{-j\left(-n{\omega _k} + {n^2}{\psi _k} \right)}}}{{\mathbf{s}^*_k}}}\right)\right\} \nonumber \\
		&=\sum\limits_{k = 1}^K p_k{{{e^{j2 n{{\omega _k} } }}}}, n=1,2,\cdots,M,
\end{align}
where $p_k$ denotes the equivalent reflected signal power of target $k$, i.e., $\mathbb{E}\{{\mathbf{s}^H_k}{\mathbf{s}_k}\} =  p_k$. 
From the anti-diagonal samples of the covariance, we can reconstruct a new vector that depends only on the angle parameters, given by
\begin{equation}
	[\tilde{\mathbf{y}}]_m = \sum_{k=1}^K p_k e^{j 2m{\omega _k}}.
\end{equation}
To perform 1D MUSIC, a full-rank set of observations is required, so spatial smoothing techniques are employed to split $ \tilde{\mathbf{y}} $ into $ L $ subvectors, each comprising $ (M+1-L) $ entries. 
The $\ell$-th split signal can be expressed as

\begin{equation}
	\tilde{\mathbf{y}}_{\ell} = \left[[\tilde{\mathbf{y}}]_{\ell}, [\tilde{\mathbf{y}}]_{\ell +1},\cdots,[\tilde{\mathbf{y}}]_{\ell+M-L}\right]^H.
\end{equation}
Then, $ \tilde{\mathbf{y}}_{\ell} $ can be rewritten as $ \tilde{\mathbf{y}}_{\ell} = \mathbf{A}\mathbf{p}_\ell $,
where { $ \mathbf{A} = [\mathbf{a}(\alpha_1),\mathbf{a}(\alpha_2),\cdots, \mathbf{a}(\alpha_K)] $} with $ [\mathbf{a}(\alpha_k)]_i = e^{j 2i \omega_k}, i = 1,2,\cdots,M+1-L$.
Moreover, the $ k $-th entry of $ \mathbf{p}_\ell $ is  $ [\mathbf{p}_\ell]_k = p_k e^{\ 2(\ell-1) \omega_k}$.
The matrix $ \mathbf{A} $ has the form of a far-field steering matrix with an array of dimension $ (M+1-L) $.
Thus, classical 1D MUSIC can be applied to search for the target angles.
The covariance matrix of $ \tilde{\mathbf{y}}_{\ell} $ is given by
\begin{equation}
	\tilde{\mathbf{R}} = \frac{1}{L} \sum_{\ell=1}^L \tilde{\mathbf{y}}_\ell \tilde{\mathbf{y}}_\ell^{H}=\frac{1}{L} \mathbf{A} \mathbf{R}_p \mathbf{A}^{H},
\end{equation} 
where $ \mathbf{R}_p = \sum_{\ell=1}^L \mathbf{p}_\ell \mathbf{p}_\ell^{H} $. To ensure that $ \mathbf{R}_p $ is full-rank, we impose the constraint $ K < L $. After obtaining the target angles, we can perform a 1D search in the distance domain.

\begin{table*}[t!]
	\centering
	\caption{Performance comparison for near-field sensing algorithms}
	\renewcommand\arraystretch{1.5}
	\begin{tabular}{|c|c|c|c|c|c|}
		\hline
		Algorithm                          & \makecell[c]{Antenna\\ spacing}         & \makecell[c]{Estimation \\resolution} & \makecell[c]{Parameter \\pairing\\ required?} & Complexity ($M_r=2J+1$)                        \\ \hline
		2D Beamforming \cite{b96}                 & $\le \frac{\lambda}{2}$  & Low                & No          & $M_r^2NP+n_gn_lM_r^2$                            \\ \hline
		2D MUSIC \cite{b95}                      & $\le \frac{\lambda}{2}$  & Highest             & No          & $M_r^3+M_r^2NP+n_gn_l(M_r-K)(M_r+1)$               \\ \hline
		Second order cumulant   \cite{b61}        & $\le \frac{\lambda}{4}$ & High                & Yes      &   $J^2NP+Slog_2S+Kn_lM^2$                                           \\ \hline
		Reduced rank algorithm  \cite{b94}               & $\le \frac{\lambda}{4}$ & High                & No          & $M_r^3+M_r^2NP+(M_r-K)[n_g(J+1)(M_r+J+1)+n_lK(M_r+1)]$ \\ \hline
		Reduced dimension algorithm \cite{b65}           & $\le \frac{\lambda}{4}$ & High                & Yes      & $M_r^3+M_r^2NP+n_g[(M_r-K)(J+1)(M+J+1)+(J+1)^3]$ \\ \hline
		FFT-enhanced near-field sensing \cite{FFTsuper}    & $\le \frac{\lambda}{2}$  & High      & No          &   $M_r^3+M_r^2NP+2M_rSlog_2(S)+2Ln_lM_r^2+Ln_g'n_l'(M_r-K)(M_r+1)$      \\ \hline
		Modified MUSIC~\cite{6111312}     & $\le \frac{\lambda}{4}$  & High      & No          &   $M_r^3+M_r^2NP+(J+1)^3+n_g(J+1)^2+n_lM_r^2K$      \\ \hline
		Generalized ESPRIT~\cite{4202630}     & $\le \frac{\lambda}{4}$  & High      & No          &   $M_r^3+M_r^2NP+n_gK^4(KJ+J^2)+n_lM_r^2K$      \\ \hline
	\end{tabular}
	\label{near-perfor}
    \vspace{-0.2cm}
\end{table*}	
\textbf{Generalized ESPRIT~\cite{4202630}:} For generalized ESPRIT, the antenna array is partitioned into two subarrays, with the first subarray comprising the first $J$ antennas arranged in ascending order, and the second subarray consisting of the last $J$ antennas arranged in descending order. The steering vectors of the two subarrays can be expressed as
\begin{equation}\begin{aligned}
		\mathbf{A}_{1}=&[\mathbf{a}_1(r_1,\theta_1),\ldots,\mathbf{a}_1(r_K,\theta_K)]\in\mathbb{C}^{J\times K},\\
		\mathbf{A}_{2}=&[\mathbf{a}_2(r_1,\theta_1),\ldots,\mathbf{a}_2(r_K,\theta_K)]\in\mathbb{C}^{J\times K}.
\end{aligned}\end{equation}
For the $k$-th target, we have
\begin{equation}
\begin{aligned}
	\mathbf{a}_{1}\!\!&=\!\!\left[e^{j\left(-N\omega_{k}+N^{2}\psi_{k}\right)},\cdots,e^{j\left((-N+J-1)\omega_{k}+(N-J+1)^{2}\psi_{k}\right)}\right]^{\mathrm{T}},\\
\mathbf{a}_{2}\!\!&=\!\!\left[e^{j\left(N\omega_{k}+N^{2}\psi_{k}\right)},\cdots,e^{j\left((N-J+1)\omega_{k}+(N-J+1)^{2}\psi_{k}\right)}\right]^{\mathrm{T}}.
\end{aligned}
\end{equation}
The array response matrix $\mathbf{A}$ can thus be divided as
\begin{equation}\mathbf{A}=\begin{bmatrix}\mathbf{A}_1\\\\\text{Last }(M-J)\text{rows}\end{bmatrix}=\begin{bmatrix}\text{First }(M-J)\text{rows}\\\\\mathbf{J}\mathbf{A}_2\end{bmatrix},\end{equation}
where $\mathbf{J}$ is used to reverse the order of the matrix rows, and $\mathbf{J}^2 = \mathbf{I}$. The array response vector of the second subarray can be represented as
\begin{equation}\mathbf{A}_2=\left[\mathbf{D}(\theta_1)\mathbf{a}_1(r_1,\theta_1),\ldots,\mathbf{D}(\theta_K)\mathbf{a}_1(r_K,\theta_K)\right],\end{equation}
where 
\begin{equation}\mathbf{D}(\theta_k)=\mathrm{diag}\left[e^{-j2N\omega_k},\ldots,e^{-j2(N-L+1)\omega_k}\right].\end{equation}
Similar to (\ref{Es}), the rotation invariance between the two subarrays is manifest in terms of the signal subspace eigenvectors as follows:
\begin{equation}
	\mathbf{E}_{s1} = \mathbf{A}_{1}\mathbf{T},~
	\mathbf{E}_{s2} = \mathbf{J}\mathbf{A}_{2}\mathbf{T}.		
\end{equation}
Then, we define the new steering matrix 
\begin{equation}\begin{aligned}
		\mathbf{A}_3(\theta)&=\left[\mathbf{D}(\theta)\mathbf{a}_1(r_1,\theta_1),\ldots,\mathbf{D}(\theta)\mathbf{a}_1(r_K,\theta_K)\right]\\
		&=\mathbf{D}(\theta)\mathbf{A}_1.
\end{aligned}\end{equation}
When $\theta$ aligns with the $i$-th target, the $i$-th column of $\mathbf{A}_3$ is  identical to the  $i$-th column of $\mathbf{A}_1$,
and $\mathbf{F}(\theta)=\mathbf{J}\mathbf{E}_{s2}-\mathbf{D}(\theta)\mathbf{E}_{s1}$ becomes rank deficient, which means that for any arbitrary full column rank matrix $\mathbf{W}\in \mathbb{C}^{J\times K}$, $\mathbf{W}^H \mathbf{F}(\theta_k)$ is singular and its determinant is zero. For instance, $\mathbf{W}$ can be a semi-unitary matrix satisfying $\mathbf{W}^H \mathbf{W} = \mathbf{I}_K$. 
A simple example is 
\begin{equation}
\mathbf{W} =
\begin{bmatrix}
\mathbf{I}_K \\
\mathbf{0}_{(J-K)\times K}
\end{bmatrix}.
\end{equation}
The generalized ESPRIT algorithm exploits this observation by defining the following spectrum function whose peaks will correspond to the target angles
\begin{equation}f(\theta)=\frac{1}{\det\left(\mathbf{W}^H\mathbf{F}(\theta)\right)}.\end{equation}
Once the target angles are estimated, a 1D search can be conducted to estimate the distance parameters.

\subsection{Performance Analysis and Comparison} 

\begin{figure}[t!]
	\centering
	\includegraphics[width=0.9\linewidth]{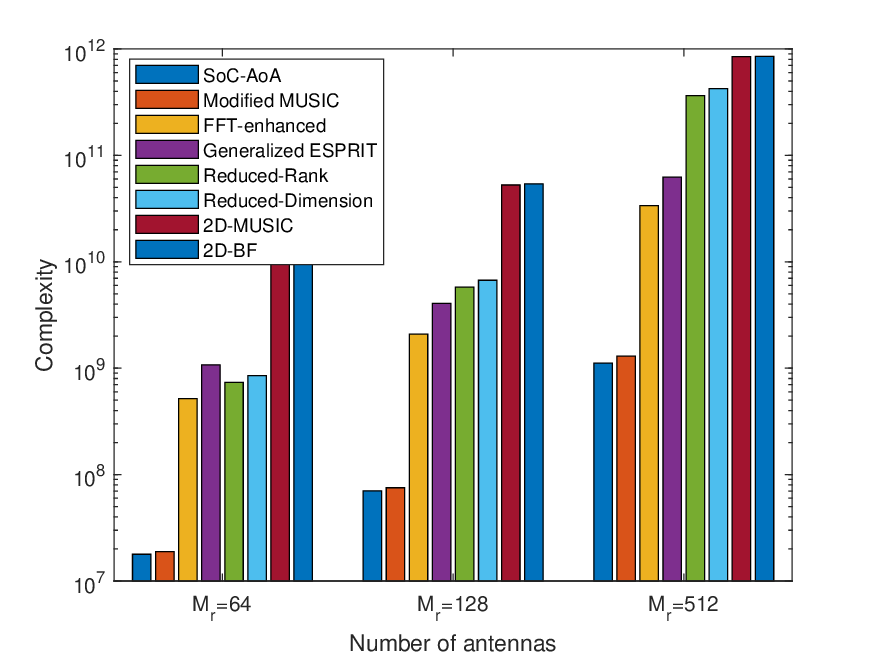}
	\caption{Theoretical complexity of algorithms for near-field MIMO-OFDM ISAC. ($K=4,N=256, P=10$)}
	\label{complexity_bar}
    \vspace{-0.5cm}
\end{figure}
Table \ref{near-perfor} shows the properties of the near-field sensing algorithms discussed above, including their complexity and resolution capability.
Fig.~\ref{complexity_bar} gives a numerical representation of the theoretical complexity of each algorithm based on the values in Table \ref{near-perfor} for receive array sizes $M_r=64, 128$ and $512$. The search grids for the angle and distance parameters are chosen to have $n_g=\frac{180}{0.05}=3600$ and $n_l=\frac{90}{0.1}=900$ cells, respectively. It is observed that for all $M_r$, SoC and modified MUSIC have the lowest complexity, while 2D MUSIC and 2D BF have the highest. However, as $M_r$ increases, the complexity gap between 2D MUSIC/2D BF and other algorithms becomes smaller, because the complexity of the EVD becomes more dominant for large $M_r$.

\begin{figure}[!t]
	\centering
	\includegraphics[width=0.8\linewidth]{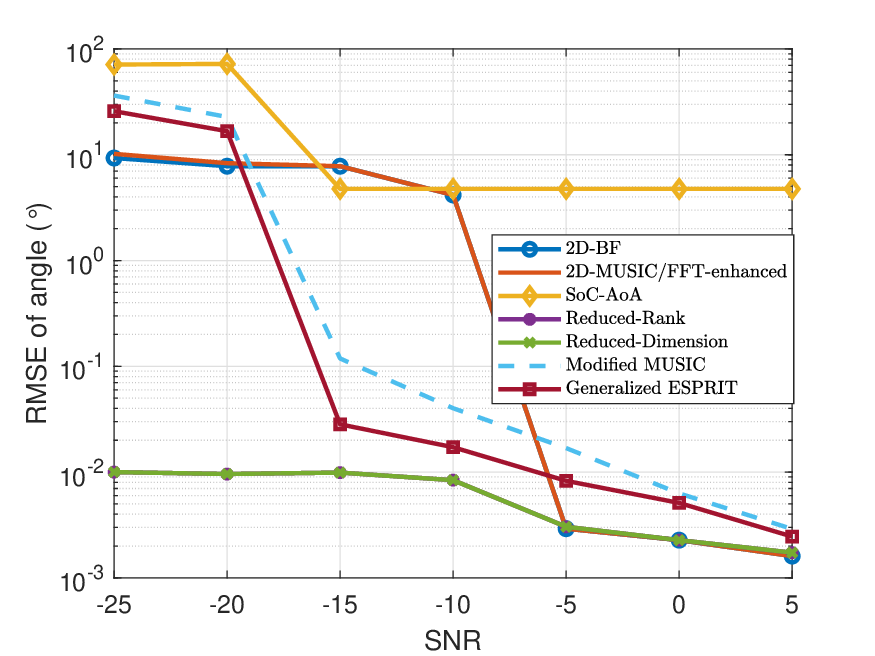}
	\caption{RMSE of angle parameter for near-field MIMO-OFDM ISAC.}
	\label{aoa_angle}
    \vspace{-0.2cm}
\end{figure}

\begin{figure}[!t]
	\centering
	\includegraphics[width=0.8\linewidth]{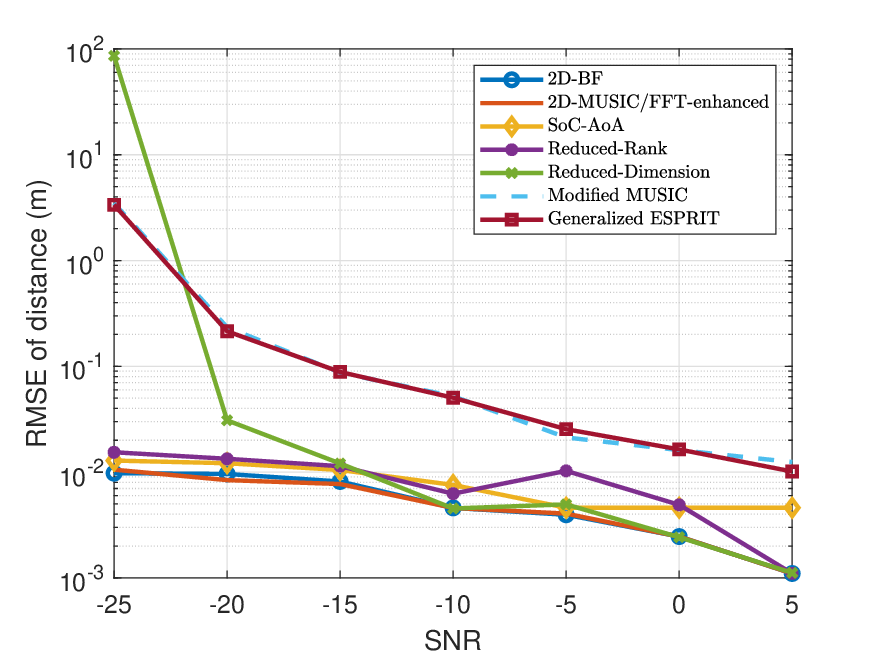}
	\caption{RMSE of distance parameter for near-field MIMO-OFDM ISAC.}
	\label{aoa_distance}
    \vspace{-0.2cm}
\end{figure}

Figs.~\ref{aoa_angle} and \ref{aoa_distance} show the RMSE of the angle and distance parameters versus SNR, respectively. The number of antenna elements is $M_r=64$ with $d=\frac{\lambda}{4}$. The target is located at $(10.2^\circ, 5.64\ \rm{m})$. As the SNR increases, the estimation accuracy of all algorithms improves. For angle estimation, RD and RR perform best, thanks to their high-sensitivity angle search procedures. On the other hand, the SoC-based algorithm performs the worst since the exploited structure of the covariance holds asymptotically. We further see that 2D BF and 2D MUSIC are sensitive to SNR, and can only achieve good estimation performance when the SNR exceeds $-5$ dB. 

\subsection{Lessons Learned}
	Compared with the far-field case, the spherical wavefronts in near-field scenarios bring new opportunities and challenges for sensing. In particular, the nonlinear phase of the steering vector enables distance estimation without the need for propagation delay estimation, but the highly coupled parameters in the angle and distance domains also increase the estimation complexity. 2D algorithms and reduced dimension/rank algorithms achieve high RMSEs for angle and distance estimation, but also suffer from higher complexity. Modified MUSIC and SoC-based algorithms have lower complexity, but their RMSE performance is degraded. FFT-enhanced algorithms are able to achieve a good balance between performance and complexity. Several challenges still need to be addressed when implementing existing algorithms, as elaborated below.

    \begin{itemize}
        \item \textbf{Element spacing constraint:} Most near-field sensing algorithms require the inter-element spacing to be smaller than a quarter wavelength. However, such dense array configurations may significantly increase the mutual coupling among antenna elements.

       \item \textbf{Degraded spatial resolution:} The reduced element spacing inevitably shrinks the overall physical aperture, thereby reduing the inherent spatial resolution from $\frac{2\lambda}{D}$ to $\frac{4\lambda}{D}$. Moreover, for statistics-based algorithms such as the SoC method, the symmetric correlation operation combines each pair of antennas into a single output, which further halves the effective aperture and consequently results in an additional loss of spatial resolution.

        \item \textbf{Prior knowledge of the number of targets:} Almost all the algorithms mentioned above require prior knowledge of the number of targets, so the ability to accurately obtain this information will greatly impact the estimation performance. For MUSIC, this can be achieved by calculating the number of significantly large eigenvalues. For RR and RD algorithms, the number of targets is crucial to properly choose the number of peaks in the angle spectrum.

        \item \textbf{Error propagation:} By decoupling the angle and distance estimation, computational complexity can be significantly reduced. However, the distance parameter estimation accuracy is highly dependent on the quality of the angle estimates. For the extreme case when several targets are located in the same direction, SoC and RD algorithms are no longer applicable, precluding accurate distance estimation.

        \item \textbf{Parameter grouping:} Parameter grouping is an additional step after separate estimation of angle and distance. However, algorithms involving a 2D search naturally lead to matched parameter estimates. For the SoC-AoA and RR algorithms, the distances are obtained by performing a 1D search for each estimated angle, thus the problem of parameter grouping is also avoided. For other algorithms, the problem and complexity associated with parameter grouping should be considered.
    \end{itemize}

\section{Extended Discussion for MIMO-OFDM ISAC} \label{openp}
In this section, we discuss open problems and outline promising avenues that warrant further investigation for MIMO-OFDM ISAC to inspire future work.
\subsection{Wideband XL-MIMO OFDM ISAC}
Deploying a huge number of antennas and transmission in the mmWave/THz frequency bands can significantly enlarge the array aperture and exploit enormous bandwidth resources for MIMO-OFDM ISAC systems, thereby enhancing the resolution in both the spatial/angular and delay domains. However, wideband XL-MIMO OFDM ISAC introduces spatial wideband effects~\cite{8354789,10736561}, also known as beam squint and Doppler squint \cite{10283625}, where the propagation delay and Doppler spread across the array aperture cannot be ignored. Spatial wideband effects occur when the frequency dependence of the array response cannot be ignored.  In such cases, the phase of the array response is a function of the subcarrier index. In addition, the use of extremely large arrays causes the target Doppler frequency to vary across the array aperture, resulting in the coupling of the phase shift components in the spatial and Doppler domains. These challenging issues will fundamentally affect target sensing in XL-MIMO OFDM ISAC systems. Thus, the sensing methods and results established in this paper for conventional systems will require reformulation and modification.

\subsection{Signal Pre-processing for MIMO-OFDM ISAC}
\textbf{Clutter rejection and clustering:} Clutter from environmental scatterers makes it difficult to distinguish sensing targets from other reflections. An effective approach for clutter rejection is to use adaptive filtering methods, such as MMSE and Kalman filtering~\cite{b77}. Additionally, techniques such as spatial filtering and Doppler processing can enhance the ability of the radar to distinguish between targets and clutter, especially in dynamic environments~\cite{b78}. Clustering methods also play a significant role in improving the SNR and facilitating more accurate detection in complex sensing environments. For instance, K-means clustering and Gaussian mixture models (GMM) can be employed to effectively separate target echoes from clutter~\cite{b79}. Emerging techniques based on machine learning are increasingly explored for clutter rejection in ISAC systems. For instance, convolutional neural networks (CNNs) have shown promise in automatically learning the features that distinguish clutter from desired signals, leading to improved classification and interference rejection~\cite{b80}. Filtering clutter based on pre-stored channel information in CKM is also an emerging and promising approach for clutter rejection. As an example, the authors of \cite{CKMclutter} proposed a new CKM approach referred to as clutter angle map (CLAM) that can effectively remove environmental clutter through pre-acquired clutter angle information.

\textbf{SI cancellation:} In BS monostatic ISAC systems, the ISAC BS needs to operate in full-duplex mode, making it vulnerable to SI due to imperfect isolation. SI significantly increases the noise level at the Sen-RX, thereby degrading sensing performance. Various methods have been proposed to address this issue, such as a combination of analog and digital cancellers \cite{xiao2022waveform}, where the analog canceller is implemented through an additional RF chain, while the digital canceller operates in the baseband of the sampled digital sensing signal.

\textbf{Synchronization:} For bistatic and multi-static ISAC systems, the ISAC-TX and RX are spatially separated and not synchronized since they have separate LO and sampling clock references. Therefore, the received signal experiences an additional STO $\tau_\Delta$, CFO $f_\Delta$, and a sampling frequency offset (SFO) $\delta_\Delta$ caused by the deviation in the sampling frequency between the DAC at the ISAC-TX and the ADC at the RX\cite{dd29}. Typically, synchronization is performed based on the LoS channel. Therefore, the RX needs to estimate and compensate for the time offset (TO) $\tau_{off}=\tau_0+\tau_\Delta$, the frequency offset (FO) $f_{off}=v_0+f_\Delta$, and $\delta_\Delta$. 
The TO must be compensated for first to determine the starting point for the OFDM signal. A coarse TO and FO can be estimated based on C\&S algorithms \cite{b72}. Furthermore, cross-correlation between the received signal and the first preamble symbol is performed for fine TO estimation \cite{b70}.
After finding the starting point of the received signal, the SFO can be estimated based on the remaining preambles via a weighted LS algorithm \cite{b73}. For more accurate SFO estimation, the entire received signal can be interpolated by a multi-rate finite impulse response (FIR) filter \cite{b74}, \cite{b75} and fed into a sample rate converter based on a polynomial filter \cite{b76}. 
Finally, the coarse FO estimate can be used to find the residual FO by constructing a delay Doppler profile (DDP) based on pilots\cite{b71}.

\subsection{Resource Allocation and Beam Management}\label{RABM}
 In practice, an ISAC system cannot continuously occupy all time-frequency resources to execute a single communication or sensing task. For example, in 5G NR TDD mode, the slots for uplink and downlink communications are alternating and non-uniform \cite{xiao2024achieving,wang2024coprime}, presenting a new challenge for sensing tasks in ISAC systems. For practical multi-user ISAC scenarios, users obtain C\&S services through orthogonal frequency division multiple access (OFDMA). Thus, different users are allocated only part of the time-frequency resources and transmit beamforming is exploited to serve multiple users to accomplish different C\&S tasks. Therefore, resource allocation and beam management are two key design issues in ISAC systems, as discussed in the following.

\textbf{Sparse resource allocation:} In conventional communication systems, the collocated allocation of time-frequency resources inherently constrains the achievable delay and Doppler resolution for individual users. In ISAC systems, sparse resource allocation \cite{xiao2024achieving,b21} emerges as a promising alternative, enhancing sensing resolution by expanding the effective time and frequency spans. Sparse allocation can be broadly categorized into uniform and non-uniform approaches. Uniform sparse allocation employs fixed-interval resource distribution, which improves sensing resolution at the expense of a significantly reduced unambiguous range. Non-uniform sparse allocation maintains the unambiguous range while enhancing resolution, albeit at the cost of elevated sidelobes in delay/Doppler periodograms, potentially obscuring weak targets. Pre-storing non-uniform sparse patterns with suppressed sidelobes via traversal search, and structured allocation based on coprime \cite{wang2024coprime}, nested \cite{5456168}, or minimum redundancy \cite{1139138}, coupled with differential coarray processing, are feasible allocation strategies. However, traversal search methods suffer from prohibitive computational complexity, while fixed allocation structures may conflict with functional reference signals in mobile communication standards. Consequently, critical challenges remain in optimizing time-frequency resources and subcarrier power allocation \cite{9945983}, necessitating further investigation into trade-offs between fairness and efficiency, as well as between C\&S performance.

\textbf{Non-orthogonal multiple access:} Non-orthogonal multiple access (NOMA) is another promising technology for enhancing the allocation of wireless resources in ISAC systems \cite{8114722,multia}. By enabling the simultaneous provision of C\&S services for multiple users within the same spatial, time, frequency, and code domains, NOMA can circumvent inherent limitations on C\&S performance imposed by resource allocation in orthogonal multiple access (OMA). Current NOMA schemes primarily include power-domain \cite{8114722} and code-domain \cite{noma1} approaches. Power-domain NOMA allocates power through superposition coding (SC) and employs successive interference cancellation (SIC) to distinguish signals from different users, while code-domain NOMA relies on dedicated non-orthogonal sparse codebooks and message passing algorithms for user signal separation. However, the C\&S performance of NOMA is constrained by the error propagation effect in SIC. Further research is warranted to model and mitigate residual inter-user or inter-C\&S interference \cite{noma2,noma3}.

 \textbf{Beam management:} Beam management \cite{10422712,8458146} can refer to beam alignment, beam sweeping, or beam tracking, based on different transmission or reception methods. Beam alignment involves aligning the beams between a transmitter and a receiver to maximize the signal transmission efficiency. Beam sweeping is the most commonly used method for achieving beam alignment, which involves transmitting or receiving beams from a predefined codebook over a given time interval to cover a given spatial area. Beam tracking aims to achieve beam alignment by predicting and tracking the positions or angles of moving users or targets. Beam management techniques alter the transmit beamforming vectors for different time-frequency resources across different CPIs and affect the duration of a CPI. The signal processing methods discussed in this paper are applicable within a given CPI for each beam management technique. However, multiple beams under different management schemes may be transmitted or received simultaneously. OMA can separate them via orthogonal time-frequency resources, whereas the use of NOMA presents additional challenges. Furthermore, multi-beam transmission requires power allocation among the beams, which is another critical consideration.

\subsection{Sparse MIMO ISAC}
Looking ahead to future 6G mobile communication networks, MIMO systems are expected to provide finer spatial resolution, not only enhancing the SE of wireless communication but also enabling more precise wireless sensing. 
To tackle the higher power consumption and signal processing complexity of MIMO systems, sparse arrays have been proposed as a way to achieve a large array aperture with a small number of antennas~\cite{sparse1}\cite{SA1}.
Sparse MIMO arrays allow for a large virtual MIMO system for sensing~\cite{5739227}, which still enables underdetermined target parameter estimation. 
However, sparse MIMO also leads to a number of challenging issues.
For example, since sparse antenna geometries result in antenna spacing that exceeds one half wavelength, undesired grating lobes are generated. When users or targets are located within these lobes, severe communication inter-user interference and radar angle ambiguity occur. For sparse MIMO sensing, differential or sum co-arrays have been demonstrated to effectively mitigate grating lobes and prevent angular ambiguity. However, the application of these techniques to communication or ISAC systems remains unclear.
Moreover, in near-field scenarios, the spatially nonstationary phase that results from spherical wave propagation does not allow formation of a virtual array as in the far-field case~\cite{coprimloca}. Therefore, developing methods to enhance DoFs in the near-field is a critical challenge.
Sparse array design itself is a challenging problem, and more work is needed to develop designs that provide an effective balance between C\&S performance.

\subsection{NLoS Sensing and Localization in ISAC}
The vision of pervasive sensing and connectivity for future 6G wireless networks relies not only on LoS signals but also on NLoS sensing and localization. Nevertheless, achieving highly accurate localization in NLoS scenarios remains an open issue. The parameter ambiguity introduced by NLoS propagation presents both challenges and opportunities in ISAC. Although it is challenging to directly perform sensing and localization based on NLoS signal components, hybrid localization schemes (e.g., RSS-ToA~\cite{sieskul2009hybrid} and AoA-ToA~\cite{fu2009cramer}) integrated with prior information (e.g., CSI~\cite{gao2024integrated} and CKM) make NLoS sensing possible. NLoS components can significantly contribute to more precise localization services and transmission coverage. Thus, IRS \cite{ris2,ris3} aided ISAC is regarded as a promising approach that can enhance C\&S performance by creating additional NLoS paths~\cite{shao2022target}. 

\subsection{Mutualism between ISAC and CKM}
 More comprehensive and environment-aware ISAC networks can be designed by establishing a feedback loop in which ISAC facilitates the construction, updating, and calibration of CKM.  CKM \cite{CKM1} is a site-specific database, marked with transmitter and/or receiver locations and containing channel-related information to enable environment-awareness.  CKM provides a structured approach for directly incorporating environmental characteristics into communication and sensing processes, allowing for a deeper understanding of the relationship between the physical environment and wireless channel properties. This breaks the heavy reliance of current ISAC systems on geometric assumptions and LoS conditions, enabling more robust and accurate communication and sensing.  For ISAC,  CKM can leverage prior knowledge of CSI to facilitate environment-aware ISAC.  This means ISAC systems can actively adapt to the surrounding environment, reducing frequent real-time channel estimation, optimizing data transmission, while enabling environmental clutter rejection, assisting in target detection and tracking,  enabling NLoS localization and sensing, and enhancing localization and sensing performance.  On the other hand, ISAC can utilize CSI captured through sensing to achieve initial CKM construction,  enabling dynamic updates and long-term calibration of CKM. The mutualism between CKM and  ISAC warrants more in-depth exploration for future wireless networks. 

\subsection{ISAC for 6G System-Level Requirements}
Beyond signal processing, several system-level requirements for 6G networks also need to be considered for ISAC system design. For example, 6G is likely to feature ultra-dense deployment of BSs and access points (APs), requiring efficient coordination to ensure seamless connectivity and consistent sensing coverage. ISAC can facilitate such coordination by providing shared sensing information across nodes, enabling cooperative beam management and multi-static sensing~\cite{han2025network,meng2024network}. Many envisioned applications, such as autonomous driving and industrial automation, impose ultra-low latency requirements, necessitating efficient ISAC designs to support real-time operation~\cite{zeng2025ultra}. By integrating sensing feedback directly into communication protocols, ISAC enables real-time perception-control loops that are difficult to achieve with conventional separated systems. Furthermore, with the anticipated deployment of massive numbers of devices in 6G, energy consumption becomes a critical bottleneck. ISAC inherently enhances efficiency by sharing spectrum, hardware, and infrastructure between communication and sensing functions, thereby avoiding redundant deployments and significantly reducing energy and cost overhead. These aspects showcase that ISAC is not merely a physical-layer technique focused on waveforms and algorithms, but also a key enabler for achieving broader 6G network objectives. Therefore, addressing these system-level requirements in a structured manner is an important direction for future ISAC research.

\subsection{Security and Privacy in MIMO-OFDM ISAC}
Although significant advantages in spectral efficiency and overall performance can be achieved by MIMO OFDM-aided ISAC systems, security and privacy concerns are becoming increasingly prominent~\cite{11124234,Su2023}. 
Communication security faces new challenges since the signal used for sensing may carry confidential information intended for communication users. Targets may thus act as eavesdroppers, threatening the confidentiality of the communication links. In an ISAC scenario, the transmitter should focus signal energy towards targets to obtain strong echoes while preventing the eavesdroppers from exploiting this signal to decode confidential information, which requires a careful trade-off~\cite{9755276}. 
Radar privacy is also a critical concern, involving not only protection of target parameter data (e.g., location, velocity) but also sensitive information related to the radar itself, such as its location and transmitted waveforms. Even when communication data is protected through encryption, eavesdroppers can still infer the AP’s location by detecting the existence of a communication link, thereby exposing the system to potential security threats. Therefore, security and privacy challenges arise in both the communication and sensing domains, which in ISAC systems are inherently coupled. This necessitates integrated protection strategies that enhance communication security while preserving sensing privacy~\cite{8935749}.

\section{Conclusion} \label{conclusion}
MIMO-OFDM is the most promising wireless technology to realize ISAC in future 6G wireless networks due to its predominant use in existing wireless systems. This paper has provided a comprehensive tutorial overview of ISAC systems enabled by MIMO-OFDM. A unified approach to modeling MIMO-OFDM ISAC systems and their enabling signal processing algorithms were presented, under both far-field and near-field channel conditions. Open problems for MIMO-OFDM ISAC that deserve further investigation were also discussed. We hope that this tutorial will provide a useful guide for researchers working on ISAC, towards efficiently integrating it into future 6G wireless networks via effective synergy with MIMO and OFDM.

\bibliographystyle{IEEEtran}
\bibliography{IEEEabrv,ISAC_tutorial}

\end{document}